%% file: ms.tex
\documentclass[modern]{aastex702}

\usepackage{hyperref}
\usepackage{color}
\usepackage{natbib}
\usepackage{graphicx}
\usepackage{xfrac}
\usepackage{xcolor}
\usepackage{amsmath}
\usepackage{mathtools}
\usepackage{afterpage}
\usepackage{longtable}

\newcommand{\fastspecfit}{\texttt{FastSpecFit}}
\newcommand{\fastspec}{\texttt{fastspec}}
\newcommand{\fastphot}{\texttt{fastphot}}

\newcommand{\kms}{\ensuremath{\mathrm{km~s}^{-1}}}
\newcommand{\flux}{\ensuremath{\mathrm{erg~s}^{-1}~\mathrm{cm}^{-2}}}
\newcommand{\fluxdensity}{\ensuremath{\mathrm{erg~s}^{-1}~\mathrm{cm}^{-2}~\text{\AA}^{-1}}}

\newcommand{\mgii}{\ion{Mg}{2}}
\newcommand{\oii}{[\ion{O}{2}]}
\newcommand{\oiii}{[\ion{O}{3}]}
\newcommand{\nii}{[\ion{N}{2}]}
\newcommand{\sii}{[\ion{S}{2}]}
\newcommand{\siii}{[\ion{S}{3}]}
\newcommand{\hii}{\ion{H}{2}}
\newcommand{\hei}{\ion{He}{1}}

\newcommand{\ha}{\ensuremath{\mathrm{H}\alpha}}
\newcommand{\hb}{\ensuremath{\mathrm{H}\beta}}
\newcommand{\hg}{\ensuremath{\mathrm{H}\gamma}}
\newcommand{\lya}{\ensuremath{\mathrm{Ly}\alpha}}

\newcommand{\niilam}{[\ion{N}{2}]~$\lambda6584$}
\newcommand{\oiiilam}{[\ion{O}{3}]~$\lambda5007$}

\newcommand{\mgiidoublet}{\ion{Mg}{2}~$\lambda\lambda2796,2803$}
\newcommand{\oiidoublet}{[\ion{O}{2}]~$\lambda\lambda3726,29$}
\newcommand{\oiiidoublet}{[\ion{O}{3}]~$\lambda\lambda4959,5007$}
\newcommand{\niidoublet}{[\ion{N}{2}]~$\lambda\lambda6548,84$}
\newcommand{\siidoublet}{[\ion{S}{2}]~$\lambda\lambda6716,31$}
\newcommand{\siiidoublet}{[\ion{S}{3}]~$\lambda\lambda9069,9532$}

\newcommand{\vdisp}{\ensuremath{\sigma_\star}}
\newcommand{\tauv}{\ensuremath{\tau_V}}
\newcommand{\Mstar}{\ensuremath{\mathcal{M}_{\star}}}
\newcommand{\msun}{\ensuremath{\mathcal{M}_{\odot}}}
\newcommand{\logmstar}{\ensuremath{\log_{10}(\Mstar\,h^{-2}/\msun)}}

\graphicspath{{./}{figures/}}
\makeatletter\def\input@path{{./}{tables/}}\makeatother

\submitjournal{\aj}

\shorttitle{FastSpecFit}
\shortauthors{Moustakas et al.}

\begin{document}

\title{FastSpecFit: Spectrophotometric Modeling of 38 Million
  Extragalactic Targets from the Dark Energy Spectroscopic Instrument
  Second Data Release}

\correspondingauthor{John Moustakas}

\author[0000-0002-2733-4559]{John Moustakas}
\email{jmoustakas@siena.edu}
\affil{Department of Physics and Astronomy, Siena University,
  515 Loudon Road, Loudonville, NY 12211, USA}

\author[0000-0002-4159-4226]{Jeremy Buhler}
\email{jbuhler@wustl.edu}
\affiliation{Department of Computer Science and Engineering,
  Washington University, One Brookings Drive, St. Louis, MO 63130, USA}

\author[0000-0002-6867-1244]{Dirk Scholte}
\email{dscholte@roe.ac.uk}
\affiliation{Institute for Astronomy, University of Edinburgh, Royal
  Observatory, Blackford Hill, Edinburgh EH9 3HJ, UK}

\author{Jessica Aguilar}
\email{jaguilar@lbl.gov}
\affiliation{Lawrence Berkeley National Laboratory, 1 Cyclotron Road,
  Berkeley, CA 94720, USA}

\author[0000-0001-6098-7247]{Steven Ahlen}
\email{ahlen@bu.edu}
\affiliation{Department of Physics, Boston University, 590
  Commonwealth Avenue, Boston, MA 02215, USA}

\author[0000-0003-2923-1585]{Abhijeet Anand}
\email{abhijeetanand2011@gmail.com}
\affiliation{Inter-University Centre for Astronomy and Astrophysics,
  Post Bag 4, Ganeshkhind, Pune 411 007, India}

\author[0000-0001-7600-5148]{Eric Armengaud}
\email{eric.armengaud@cea.fr}
\affiliation{IRFU, CEA, Universit\'{e} Paris-Saclay, F-91191
  Gif-sur-Yvette, France}

\author[0000-0001-5998-3986]{Alejandro Aviles}
\email{aviles@icf.unam.mx}
\affiliation{Instituto Avanzado de Cosmolog\'{\i}a A.~C., San Marcos 11
  - Atenas 202. Magdalena Contreras. Ciudad de M\'{e}xico C.~P.~10720,
  M\'{e}xico}
\affiliation{Instituto de Ciencias F\'{\i}sicas, Universidad Nacional
  Aut\'onoma de M\'exico, Av. Universidad s/n, Cuernavaca, Morelos,
  C.~P.~62210, M\'exico}

\author[0000-0003-4162-6619]{Stephen Bailey}
\email{stephenbailey@lbl.gov}
\affiliation{Lawrence Berkeley National Laboratory, 1 Cyclotron Road,
  Berkeley, CA 94720, USA}

\author[0000-0003-0467-5438]{Florian Beutler}
\email{florian.beutler@ed.ac.uk}
\affiliation{Institute for Astronomy, University of Edinburgh, Royal
  Observatory, Blackford Hill, Edinburgh EH9 3HJ, UK}

\author[0000-0001-9712-0006]{Davide Bianchi}
\email{davide.bianchi1@unimi.it}
\affiliation{Dipartimento di Fisica ``Aldo Pontremoli'', Universit\`a
  degli Studi di Milano, Via Celoria 16, I-20133 Milano, Italy}
\affiliation{INAF-Osservatorio Astronomico di Brera, Via Brera 28, 20122
  Milano, Italy}

\author{David Brooks}
\email{david.brooks@ucl.ac.uk}
\affiliation{Department of Physics \& Astronomy, University College
  London, Gower Street, London, WC1E 6BT, UK}

\author[0000-0003-3044-5150]{Aurelio Carnero Rosell}
\email{acarnero@iac.es}
\affiliation{Departamento de Astrof\'{\i}sica, Universidad de La
  Laguna (ULL), E-38206, La Laguna, Tenerife, Spain}
\affiliation{Instituto de Astrof\'{\i}sica de Canarias, C/ V\'{\i}a
  L\'{a}ctea, s/n, E-38205, La Laguna, Tenerife, Spain}

\author[0000-0001-7316-4573]{Francisco J. Castander}
\email{fjc@ice.csic.es}
\affiliation{Institut d'Estudis Espacials de Catalunya (IEEC), c/
  Esteve Terradas 1, Edifici RDIT, Campus PMT-UPC, 08860
  Castelldefels, Spain}
\affiliation{Institute of Space Sciences, ICE-CSIC, Campus UAB, Carrer
  de Can Magrans s/n, 08913 Bellaterra, Barcelona, Spain}

\author[0000-0001-8996-4874]{Edmond Chaussidon}
\email{echaussidon@lbl.gov}
\affiliation{Lawrence Berkeley National Laboratory, 1 Cyclotron Road,
  Berkeley, CA 94720, USA}

\author{Todd Claybaugh}
\email{tmclaybaugh@lbl.gov}
\affiliation{Lawrence Berkeley National Laboratory, 1 Cyclotron Road,
  Berkeley, CA 94720, USA}

\author[0000-0002-2169-0595]{Andrei Cuceu}
\email{acuceu@lbl.gov}
\affiliation{Lawrence Berkeley National Laboratory, 1 Cyclotron Road,
  Berkeley, CA 94720, USA}

\author[0000-0002-1769-1640]{Axel de la Macorra}
\email{macorra@fisica.unam.mx}
\affiliation{Instituto de F\'{\i}sica, Universidad Nacional
  Aut\'{o}noma de M\'{e}xico, Circuito de la Investigaci\'{o}n
  Cient\'{\i}fica, Ciudad Universitaria, Cd. de M\'{e}xico
  C.~P.~04510, M\'{e}xico}

\author[0000-0002-5665-7912]{Biprateep Dey}
\email{b.dey@utoronto.ca}
\affiliation{Department of Astronomy \& Astrophysics, University of
  Toronto, Toronto, ON M5S 3H4, Canada}
\affiliation{Department of Physics \& Astronomy and Pittsburgh
  Particle Physics, Astrophysics, and Cosmology Center (PITT PACC),
  University of Pittsburgh, 3941 O'Hara Street, Pittsburgh, PA 15260,
  USA}

\author{Peter Doel}
\email{apd@star.ucl.ac.uk}
\affiliation{Department of Physics \& Astronomy, University College
  London, Gower Street, London, WC1E 6BT, UK}

\author[0000-0003-1251-532X]{Victoria A. Fawcett}
\email{victoria.fawcett@eso.org}
\affiliation{European Southern Observatory,
  Karl-Schwarzschild-Stra{\ss}e 2, 85748 Garching bei M\"unchen,
  Germany}

\author[0000-0002-3033-7312]{Andreu Font-Ribera}
\email{afont@ifae.es}
\affiliation{Instituci\'{o} Catalana de Recerca i Estudis
  Avan\c{c}ats, Passeig de Llu\'{\i}s Companys, 23, 08010 Barcelona,
  Spain}
\affiliation{Institut de F\'{i}sica d'Altes Energies (IFAE), The
  Barcelona Institute of Science and Technology, Edifici Cn, Campus
  UAB, 08193 Bellaterra (Barcelona), Spain}

\author[0000-0002-2890-3725]{Jaime E. Forero-Romero}
\email{je.forero@uniandes.edu.co}
\affiliation{Departamento de F\'isica, Universidad de los Andes, Cra.
  1 No. 18A-10, Edificio Ip, CP 111711, Bogot\'a, Colombia}
\affiliation{Observatorio Astron\'omico, Universidad de los Andes,
  Cra. 1 No. 18A-10, Edificio H, CP 111711 Bogot\'a, Colombia}

\author[0000-0001-9632-0815]{Enrique Gazta\~{n}aga}
\email{gaztanaga@gmail.com}
\affiliation{Institut d'Estudis Espacials de Catalunya (IEEC), c/
  Esteve Terradas 1, Edifici RDIT, Campus PMT-UPC, 08860
  Castelldefels, Spain}
\affiliation{Institute of Cosmology and Gravitation, University of
  Portsmouth, Dennis Sciama Building, Portsmouth, PO1 3FX, UK}
\affiliation{Institute of Space Sciences, ICE-CSIC, Campus UAB, Carrer
  de Can Magrans s/n, 08913 Bellaterra, Barcelona, Spain}

\author{Gaston Gutierrez}
\email{gaston@fnal.gov}
\affiliation{Fermi National Accelerator Laboratory, PO Box 500,
  Batavia, IL 60510, USA}

\author[0000-0001-9822-6793]{Julien Guy}
\email{jguy@lbl.gov}
\affiliation{Lawrence Berkeley National Laboratory, 1 Cyclotron Road,
  Berkeley, CA 94720, USA}

\author[0000-0002-2312-3121]{Boryana Hadzhiyska}
\email{boryanah@ast.cam.ac.uk}
\affiliation{Institute of Astronomy \& Kavli Institute for Cosmology,
  University of Cambridge, Madingley Road, Cambridge CB3 0HA, UK}

\author[0009-0007-2936-1124]{Tyler Hagen}
\email{thagen@uwlax.edu}
\affiliation{Department of Physics, University of Wisconsin-La Crosse,
  La Crosse, WI 54601, USA}

\author[0000-0003-1197-0902]{ChangHoon Hahn}
\email{changhoon.hahn@utexas.edu}
\affiliation{Department of Astronomy, University of Texas at Austin,
  2515 Speedway, Austin, TX 78712, USA}

\author[0000-0002-6550-2023]{Klaus Honscheid}
\email{kh@physics.osu.edu}
\affiliation{Center for Cosmology and AstroParticle Physics, The Ohio
  State University, 191 West Woodruff Avenue, Columbus, OH 43210, USA}
\affiliation{Department of Physics, The Ohio State University, 191
  West Woodruff Avenue, Columbus, OH 43210, USA}

\author[0000-0001-8528-3473]{Jorge Jimenez}
\email{jjimenez@ifae.es}
\affiliation{Institut de F\'{i}sica d'Altes Energies (IFAE), The
  Barcelona Institute of Science and Technology, Edifici Cn, Campus UAB,
  08193 Bellaterra (Barcelona), Spain}

\author[0000-0002-0000-2394]{Stephanie Juneau}
\email{stephanie.juneau@noirlab.edu}
\affiliation{NSF NOIRLab, 950 N. Cherry Ave., Tucson, AZ 85719, USA}

\author[0000-0002-5652-8870]{Tanveer Karim}
\email{tanveer.karim@utoronto.ca}
\affiliation{Department of Astronomy \& Astrophysics, University of
  Toronto, Toronto, ON M5S 3H4, Canada}

\author[0000-0002-8828-5463]{David Kirkby}
\email{dkirkby@uci.edu}
\affiliation{Department of Physics and Astronomy, University of
  California, Irvine, 2115 Frederick Reines Hall, Irvine, CA 92697,
  USA}

\author[0000-0003-2644-135X]{Sergey E. Koposov}
\email{skoposov@ed.ac.uk}
\affiliation{Institute for Astronomy, University of Edinburgh, Royal
  Observatory, Blackford Hill, Edinburgh EH9 3HJ, UK}
\affiliation{Institute of Astronomy, University of Cambridge,
  Madingley Road, Cambridge CB3 0HA, UK}

\author[0000-0001-6356-7424]{Anthony Kremin}
\email{akremin@lbl.gov}
\affiliation{Lawrence Berkeley National Laboratory, 1 Cyclotron Road,
  Berkeley, CA 94720, USA}

\author[0000-0002-6731-9329]{Claire Lamman}
\email{lamman.1@osu.edu}
\affiliation{Department of Astronomy, The Ohio State University, 4055
  McPherson Laboratory, 140 W 18th Avenue, Columbus, OH 43210, USA}

\author[0000-0003-1838-8528]{Martin Landriau}
\email{mlandriau@lbl.gov}
\affiliation{Lawrence Berkeley National Laboratory, 1 Cyclotron Road,
  Berkeley, CA 94720, USA}

\author[0000-0001-7178-8868]{Laurent Le Guillou}
\email{llg@lpnhe.in2p3.fr}
\affiliation{Sorbonne Universit\'{e}, CNRS/IN2P3, Laboratoire de
  Physique Nucl\'{e}aire et de Hautes Energies (LPNHE), FR-75005
  Paris, France}

\author[0000-0002-3677-3617]{Alexie Leauthaud}
\email{alexie@ucsc.edu}
\affiliation{Department of Astronomy and Astrophysics, UCO/Lick
  Observatory, University of California, 1156 High Street, Santa Cruz,
  CA 95064, USA}

\author[0000-0003-1887-1018]{Michael E. Levi}
\email{melevi@lbl.gov}
\affiliation{Lawrence Berkeley National Laboratory, 1 Cyclotron Road,
  Berkeley, CA 94720, USA}

\author[0000-0002-4623-0683]{Yufeng Luo}
\email{yluo4@uwyo.edu}
\affiliation{Department of Physics \& Astronomy, University of
  Wyoming, 1000 E. University, Dept.~3905, Laramie, WY 82071, USA}

\author[0000-0003-4962-8934]{Marc Manera}
\email{mmanera@ifae.es}
\affiliation{Departament de F\'{i}sica, Serra H\'{u}nter, Universitat
  Aut\`{o}noma de Barcelona, 08193 Bellaterra (Barcelona), Spain}
\affiliation{Institut de F\'{i}sica d'Altes Energies (IFAE), The
  Barcelona Institute of Science and Technology, Edifici Cn, Campus
  UAB, 08193 Bellaterra (Barcelona), Spain}

\author[0000-0002-7113-0262]{Viraj Manwadkar}
\email{virajvm@stanford.edu}
\affiliation{SLAC National Accelerator Laboratory, 2575 Sand Hill Road,
  Menlo Park, CA 94025, USA}

\author[0000-0002-4279-4182]{Paul Martini}
\email{martini.10@osu.edu}
\affiliation{Center for Cosmology and AstroParticle Physics, The Ohio
  State University, 191 West Woodruff Avenue, Columbus, OH 43210, USA}
\affiliation{Department of Astronomy, The Ohio State University, 4055
  McPherson Laboratory, 140 W 18th Avenue, Columbus, OH 43210, USA}

\author[0000-0002-1125-7384]{Aaron Meisner}
\email{aaron.meisner@noirlab.edu}
\affiliation{NSF NOIRLab, 950 N. Cherry Ave., Tucson, AZ 85719, USA}

\author[0000-0002-6610-4836]{Ramon Miquel}
\email{rmiquel@ifae.es}
\affiliation{Instituci\'{o} Catalana de Recerca i Estudis
  Avan\c{c}ats, Passeig de Llu\'{\i}s Companys, 23, 08010 Barcelona,
  Spain}
\affiliation{Institut de F\'{i}sica d'Altes Energies (IFAE), The
  Barcelona Institute of Science and Technology, Edifici Cn, Campus
  UAB, 08193 Bellaterra (Barcelona), Spain}

\author[0000-0001-9070-3102]{Seshadri Nadathur}
\email{seshadri.nadathur@port.ac.uk}
\affiliation{Institute of Cosmology and Gravitation, University of
  Portsmouth, Dennis Sciama Building, Portsmouth, PO1 3FX, UK}

\author[0000-0002-4637-2868]{Enrique Paillas}
\email{enrique.paillas@udp.cl}
\affiliation{Instituto de Estudios Astrof\'isicos, Facultad de
  Ingenier\'ia y Ciencias, Universidad Diego Portales, Av. Ej\'ercito
  Libertador 441, Santiago, Chile}

\author[0000-0003-3188-784X]{Nathalie Palanque-Delabrouille}
\email{npalanque-delabrouille@lbl.gov}
\affiliation{IRFU, CEA, Universit\'{e} Paris-Saclay, F-91191
  Gif-sur-Yvette, France}
\affiliation{Lawrence Berkeley National Laboratory, 1 Cyclotron Road,
  Berkeley, CA 94720, USA}

\author[0000-0002-5854-7426]{Swayamtrupta Panda}
\email{swayamtrupta.panda@noirlab.edu}
\affiliation{NSF NOIRLab, 950 N. Cherry Ave., Tucson, AZ 85719, USA}

\author[0000-0002-0644-5727]{Will J. Percival}
\email{will.percival@uwaterloo.ca}
\affiliation{Department of Physics and Astronomy, University of
  Waterloo, 200 University Ave W, Waterloo, ON N2L 3G1, Canada}
\affiliation{Perimeter Institute for Theoretical Physics, 31 Caroline
  St. North, Waterloo, ON N2L 2Y5, Canada}
\affiliation{Waterloo Centre for Astrophysics, University of Waterloo,
  200 University Ave W, Waterloo, ON N2L 3G1, Canada}

\author[0000-0001-6979-0125]{Ignasi P\'erez-R\`afols}
\email{ignasi.perez.rafols@upc.edu}
\affiliation{Departament de F\'isica, EEBE, Universitat Polit\`ecnica
  de Catalunya, c/Eduard Maristany 10, 08930 Barcelona, Spain}

\author[0000-0001-7145-8674]{Francisco Prada}
\email{fprada@iaa.es}
\affiliation{Instituto de Astrof\'{i}sica de Andaluc\'{i}a (CSIC),
  Glorieta de la Astronom\'{i}a, s/n, E-18008 Granada, Spain}

\author[0000-0002-4940-3009]{Ragadeepika Pucha}
\email{dr.raga.pucha@gmail.com}
\affiliation{Department of Physics and Astronomy, The University of
  Utah, 115 South 1400 East, Salt Lake City, UT 84112, USA}

\author[0000-0002-3500-6635]{Corentin Ravoux}
\email{corentin.ravoux@clermont.in2p3.fr}
\affiliation{Universit\'{e} Clermont-Auvergne, CNRS, LPCA, 63000
  Clermont-Ferrand, France}

\author[0000-0002-7522-9083]{Ashley J. Ross}
\email{ross.1333@osu.edu}
\affiliation{Center for Cosmology and AstroParticle Physics, The Ohio
  State University, 191 West Woodruff Avenue, Columbus, OH 43210, USA}
\affiliation{Department of Astronomy, The Ohio State University, 4055
  McPherson Laboratory, 140 W 18th Avenue, Columbus, OH 43210, USA}

\author{Graziano Rossi}
\email{graziano@sejong.ac.kr}
\affiliation{Department of Physics and Astronomy, Sejong University,
  209 Neungdong-ro, Gwangjin-gu, Seoul 05006, Republic of Korea}

\author[0000-0002-0394-0896]{Rossana Ruggeri}
\email{rossana.ruggeri@qut.edu.au}
\affiliation{Queensland University of Technology, School of Chemistry
  \& Physics, George St, Brisbane 4001, Australia}

\author[0000-0003-4357-3450]{Am\'{e}lie Saintonge}
\email{a.saintonge@ucl.ac.uk}
\affiliation{Department of Physics and Astronomy, University College
  London, Gower Street, London, WC1E 6BT, UK}
\affiliation{Max-Planck-Institut f\"{u}r Radioastronomie, Auf dem
  H\"{u}gel 69, 53121 Bonn, Germany}

\author[0000-0002-1609-5687]{Lado Samushia}
\email{lado@phys.ksu.edu}
\affiliation{Abastumani Astrophysical Observatory, Tbilisi, GE-0179,
  Georgia}
\affiliation{Department of Physics, Kansas State University, 116
  Cardwell Hall, Manhattan, KS 66506, USA}

\author[0000-0002-9646-8198]{Eusebio Sanchez}
\email{eusebio.sanchez@ciemat.es}
\affiliation{CIEMAT, Avenida Complutense 40, E-28040 Madrid, Spain}

\author[0000-0002-0408-5633]{Christoph Saulder}
\email{csaulder@mpe.mpg.de}
\affiliation{Max Planck Institute for Extraterrestrial Physics, Gie\ss
  enbachstra\ss e 1, 85748 Garching, Germany}

\author{David Schlegel}
\email{djschlegel@lbl.gov}
\affiliation{Lawrence Berkeley National Laboratory, 1 Cyclotron Road,
  Berkeley, CA 94720, USA}

\author[0000-0002-3461-0320]{Joseph Silber}
\email{jhsilber@lbl.gov}
\affiliation{Lawrence Berkeley National Laboratory, 1 Cyclotron Road,
  Berkeley, CA 94720, USA}

\author[0000-0002-2949-2155]{Ma{\l}gorzata Siudek}
\email{msiudek@iac.es}
\affiliation{Institute of Space Sciences, ICE-CSIC, Campus UAB, Carrer
  de Can Magrans s/n, 08913 Bellaterra, Barcelona, Spain}
\affiliation{Instituto de Astrof\'{\i}sica de Canarias, C/ V\'{\i}a
  L\'{a}ctea, s/n, E-38205, La Laguna, Tenerife, Spain}

\author[0000-0003-1704-0781]{Gregory Tarl\'{e}}
\email{gtarle@umich.edu}
\affiliation{University of Michigan, 500 S. State Street, Ann Arbor,
  MI 48109, USA}

\author{Benjamin A. Weaver}
\email{benjamin.weaver@noirlab.edu}
\affiliation{NSF NOIRLab, 950 N. Cherry Ave., Tucson, AZ 85719, USA}

\author[0000-0001-5381-4372]{Rongpu Zhou}
\email{rongpuzhou@lbl.gov}
\affiliation{Lawrence Berkeley National Laboratory, 1 Cyclotron Road,
  Berkeley, CA 94720, USA}

\author[0000-0002-6684-3997]{Hu Zou}
\email{zouhu@nao.cas.cn}
\affiliation{National Astronomical Observatories, Chinese Academy of
  Sciences, A20 Datun Road, Chaoyang District, Beijing, 100101,
  P.~R.~China}

\begin{abstract}
Optical spectrophotometry provides a direct window into the physical
conditions, kinematics, and formation histories of galaxies and
quasars. \fastspecfit{} is an open-source Python software package that
forward-models each Dark Energy Spectroscopic Instrument (DESI)
spectrum jointly with its broadband photometry, convolving
stellar-population synthesis and emission-line templates with each
fiber's native spectral resolution. This forward-modeling approach
preserves DESI's uncorrelated pixel statistics, fits a typical
spectrum in $\approx0.1$~s of single-core time, and scales to tens of
millions of objects, with uncertainties estimated via Monte Carlo
resampling. \fastspecfit{} measures stellar velocity dispersions,
non-parametric star-formation histories, stellar masses,
star-formation rates, dust attenuation, K-corrections, rest-frame
magnitudes, and fluxes and equivalent widths for 46 emission lines,
including separate narrow and broad components for the Balmer
lines. We describe these algorithms in detail, validate the
measurements against independent codes, and present the resulting
value-added catalog for DESI Data Release 2, which provides
spectrophotometric and physical properties for 38 million
extragalactic targets at $0<z\lesssim4$. We also release a companion photometry-only
catalog and document the catalogs previously released for the DESI
Early Data Release and Data Release 1.
\end{abstract}


\section{Introduction}
\label{sec:intro}

The Dark Energy Spectroscopic Instrument \citep[DESI;][]{levi13a,
  desi-collaboration16a, desi-collaboration24e} is conducting an
eight-year survey (2021--2029) which will ultimately yield
spectroscopic redshifts for more than 60 million galaxies and quasars
over 17,000~deg$^2$. These data are being made public in a series of
staged releases. The Early Data Release
\citep[EDR;][]{desi-collaboration24d}, comprising 1.7~million unique
spectra from the survey validation (SV) campaign (2020 December--2021
May), was released in 2023 June. Data Release~1
\citep[DR1;][]{desi-collaboration26a}, released in 2025 March,
uniformly reprocessed the SV data and added the first 13 months of the
main-survey sample, for a total of 20.4~million unique spectra. Data
Release~2 (DR2), the subject of this paper, extends the main-survey
data through 2024 April, its first three years, includes an
unprecedented 38 million unique redshifts, and is scheduled for public
release in the first part of 2027. Each successive release has
sharpened DESI's baryon acoustic oscillation distance measurements to
sub-percent precision and currently shows a 2.8--4.2$\sigma$
preference for an evolving dark energy equation of state over the
standard $\Lambda$CDM concordance cosmological model \citep{adame25b,
  adame25d, adame25e, abdul-karim25a, abdul-karim25b}.

Each spectrum DESI acquires encodes a wealth of astrophysical
information beyond its redshift. The shape and absorption-line content
of the stellar continuum trace the underlying stellar populations,
dust content, and kinematics of a galaxy, while its emission-line
spectrum constrains the star-formation rate, chemical abundances,
ionization state, and gas kinematics, as well as the presence and
strength of an active galactic nucleus (AGN) \citep[and references
  therein]{conroy13a}. Extracting these physical properties at DESI
scale---and combining them with broadband photometry to normalize the
overall spectral energy distribution (SED)---requires a fitting code
that is fast, robust, and automated across an enormous range of
signal-to-noise ratios (S/N), redshifts, and galaxy types.

Various publicly available codes already perform this kind of
spectrophotometric modeling. Full-spectrum fitting codes---among them
\texttt{pPXF} \citep{cappellari04a, cappellari23a}, \texttt{GandALF}
\citep{sarzi06a, sarzi17a}, \texttt{STARLIGHT}
\citep{cid-fernandes05a}, \texttt{FIREFLY} \citep{wilkinson17a},
\texttt{FIT3D}/\texttt{pyPipe3D} \citep{sanchez16a, sanchez22a},
\texttt{ULySS} \citep{koleva09a}, \texttt{alf} \citep{conroy18a}, and
others---extract the kinematics and properties of the stellar
populations and line-emitting gas; SED-fitting codes such as
\texttt{CIGALE} \citep{boquien19a}, \texttt{Bagpipes}
\citep{carnall18a}, \texttt{Prospector} \citep{johnson21a}, and
\texttt{MAGPHYS} \citep{da-cunha08a} combine photometry (and, in some
cases, spectroscopy) with physically motivated stellar population
synthesis (SPS) models to infer star-formation histories, stellar
masses, and dust properties; and codes like \texttt{PyQSOFit}
\citep{guo18a} are tailored to fitting the ultraviolet (UV) and
optical spectra of quasars.

Each of these codes makes different trade-offs among flexibility,
physical realism, and computational speed, but none were designed to
be run efficiently on the tens of millions of spectra acquired by
DESI. Existing codes also do not take advantage of the DESI resolution
matrix, which compactly encodes the instrumental resolution and
guarantees uncorrelated uncertainties from
pixel to pixel \citep{bolton10a,
  guy23a}.

\fastspecfit\footnote{\url{https://fastspecfit.readthedocs.io/en/latest}}
has been developed to meet both these challenges, scaling efficiently
to tens of millions of spectra while fully exploiting DESI's native
spectral resolution. Using a forward-modeling fitting approach,
\fastspecfit{} convolves physically motivated stellar continuum and
emission-line templates with each fiber's resolution matrix and
compares the result directly to the observed spectra, preserving the
uncorrelated pixel statistics that resampling would otherwise destroy
(\S\ref{sec:modeling}). In addition, \fastspecfit{} is engineered for
speed; it fits a typical DESI spectrum in $\approx0.1$~s on a single
CPU core ($\approx5$~s including the default 50 Monte Carlo
realizations; \S\ref{ssec:performance}), making it possible to process
the full DESI dataset---and to refit it as the data-reduction pipeline
and SPS models evolve---with modest computing resources. This speed
also makes it practical to estimate parameter uncertainties from
repeated fits to Monte Carlo-resampled data rather than from
computationally expensive posterior sampling
(\S\ref{ssec:uncertainties}), and, via a hybrid message passing
interface (MPI) plus multiprocessing design, to scale from a single
serial process on a laptop to thousands of cores on a supercomputer
(\S\ref{ssec:performance}).

Beyond speed, \fastspecfit{} is designed to be flexible, applying only
the physical constraints the data strictly require and otherwise
letting the data determine the best-fitting model. The continuum model
imposes only minimal priors: non-negative stellar population
coefficients, an energy-conserving dust model, and a freely fitted
stellar velocity dispersion bounded widely enough to accommodate the
full spectral diversity of DESI targets. In exchange for speed and
robustness at low S/N, the default configuration we adopt deliberately
fixes other ingredients, including the stellar metallicity, the slope
of the attenuation curve, and the infrared (IR) dust-emission
parameters (\S\ref{ssec:templates}, \S\ref{sssec:fitting}), although
users can relax most of these choices with custom templates and
fitting options (\S\ref{sec:modes}). The emission-line model is
similarly flexible, supporting independent narrow and broad kinematic
components and capturing the broad Balmer-line emission that
distinguishes type~1 AGN whenever the data require it. Unlike codes
that target a specific class or spectral type, \fastspecfit{} applies
this same machinery uniformly to every spectrum, without assuming in
advance whether the target is quiescent, star-forming, post-starburst,
or hosts an AGN. This flexibility minimizes classification-driven
selection effects and maximizes the discovery potential of a dataset
as large and diverse as DESI, where unusual or physically ambiguous
objects emerge naturally from the fit rather than requiring
special-case handling.

Although this paper presents the first comprehensive description of
\fastspecfit{}'s algorithms, measurements from the earlier EDR and DR1
value-added catalogs (VACs; see Appendix~\ref{sec:appendix_vacs}) have
already enabled a broad range of extragalactic science by both DESI
collaboration members and the wider scientific community. We briefly
highlight a small, illustrative sample of that work
here. Emission-line fluxes, equivalent widths, and stellar masses from
\fastspecfit{} have been used to identify and characterize
changing-look AGN \citep{guo25a} and to triple the census of dwarf AGN
candidates in the local universe \citep{pucha25a}; to measure
direct-method ($T_e$-based) chemical abundances for nearly 50,000
star-forming galaxies in DESI DR2 \citep{scholte26a}, including two of
the most metal-poor star-forming galaxies ever found (J.~Moustakas
et~al., in preparation), and to characterize the atomic gas sequence
and mass-metallicity relation across five decades in stellar mass
\citep{scholte24a}; to assemble catalogs of post-starburst galaxies
\citep{soto25a} and to infer the physical properties of galaxies in
cosmic voids \citep{rincon25a}; and to measure stellar velocity
dispersions for strong gravitational lens candidates
\citep{huang25a}. The \fastspecfit{} measurements have also been
essential to measurements of the galaxy luminosity and stellar
functions \citep{wang25a, moore26a}, and to the modeling of intrinsic
alignments \citep{siegel25a} and photometric-redshift calibration
\citep{blanco25a} for the next generation of weak-lensing surveys.

Although this paper is long, we have striven to write it so it can be
read in pieces as well as start to end. In \S\ref{sec:data} we
describe the DESI spectroscopic and photometric data \fastspecfit{}
fits, together with the survey, program, and target-class structure
particular to DESI needed to interpret the resulting catalogs. The
technical core of the paper, \S\ref{sec:modeling}, presents
\fastspecfit{}'s algorithms and the physical quantities it derives in
full detail; as noted there, readers who want the output before the
machinery can skip this section. In \S\ref{sec:casestudy} we ground
that machinery in three worked examples spanning the DESI
extragalactic classes, a natural starting point for readers who want
to see spectra and their best-fitting models to help build
intuition first. In \S\ref{sec:vac} we present the DR2 catalog
itself---its construction and organization, recommended quality cuts,
computational performance, and validation against independent
codes. Readers who only want to use the \fastspecfit{} DR2 VAC will
find this section largely self-contained. In \S\ref{sec:discussion} we
step back to illustrate the catalog's scientific reach and describe
planned improvements, before concluding with a summary in
\S\ref{sec:summary}. Finally, in Appendix~\ref{sec:appendix_fastphot}
we document a companion, photometry-only VAC, and in
Appendix~\ref{sec:appendix_vacs} we place the DR2 catalog in the
context of \fastspecfit{}'s full release history, including the
earlier EDR and DR1 catalogs.

We adopt a flat $\Lambda$CDM cosmology with $\Omega_m = 0.315$,
consistent with \citet{planck-collaboration20a}, and a
\citet{chabrier03a} initial mass function (IMF) spanning
$0.08$--$120$~\msun; we define $h \equiv H_0 / (100~\kms\,\textrm{Mpc}^{-1})$
and express stellar masses and other derived quantities
in units with $h=1$.

\section{Spectroscopic and Photometric Data}
\label{sec:data}

In this section we describe the DESI spectroscopic
(\S\ref{ssec:spectra}) and broadband photometric
(\S\ref{ssec:photometry}) data that \fastspecfit{} fits, and the
target classes and redshift measurements that define the extragalactic
objects that DESI observes (\S\ref{ssec:targets}).

\subsection{DESI Spectroscopy}
\label{ssec:spectra}

DESI is a highly multiplexed fiber spectrograph mounted on the Mayall
4-m telescope at Kitt Peak National Observatory
\citep{desi-collaboration16b, desi-collaboration22a}.  A wide-field
optical corrector \citep{miller24a} provides a $3\fdg2$-diameter field
of view. Within this field, 5000 robotically actuated positioners
  \citep{silber23a} place $1\farcs5$-diameter fibers
\citep{poppett24a} onto target coordinates. These fibers feed ten three-arm
spectrographs covering 3600--9824~\AA{} in the observed frame
\citep{desi-collaboration22a}. The focal plane is divided into ten
wedge-shaped ``petals'', each of which routes its 500 fibers to one of
the ten spectrographs. A single pointing of the full 5000-fiber focal
plane defines a ``tile'', the basic unit of the survey's observing
(tiling) strategy \citep{schlafly23a}. The three cameras sense blue
(B: 3600--5930~\AA), red (R: 5600--7720~\AA), and near-IR (Z:
7470--9824~\AA) light at a spectral resolution of
$\mathcal{R}\equiv\lambda/\Delta\lambda_\mathrm{FWHM} \approx
2000-3500$, $3400-4800$, and $3800-5200$, respectively
\citep{desi-collaboration22a}.

\citet{guy23a} describe the DESI spectroscopic pipeline in detail (see
also \citealt{desi-collaboration24d, desi-collaboration26a} for
shorter summaries). \fastspecfit{} relies on three per-fiber products
from this pipeline: the one-dimensional spectrum, its corresponding
inverse-variance spectrum, and the resolution matrix $\mathbf{R}$, a
band-diagonal, sparse matrix that quantifies the effective,
wavelength-dependent instrumental resolution. $\mathbf{R}$ is a
natural output of the ``spectroperfectionism''
algorithm of \citet{bolton10a}, which the DESI pipeline uses to
extract one-dimensional spectra with uncorrelated pixels. It is also a
key ingredient of \fastspecfit's forward-modeling philosophy:
convolving a perfect, infinite-resolution model spectrum with
$\mathbf{R}$ yields the equivalent model at DESI's effective spectral
resolution (\S\ref{sec:modeling}).

The per-fiber spectra are extracted onto an 0.8\,\AA{} linearly
spaced wavelength grid (in vacuum and in the Solar System barycentric
frame of reference), flux-calibrated using a network of F-type
standard stars (typically 10 standard stars per petal), and repeat
exposures are coadded using inverse-variance
weights. \citet{guy23a} and \citet{koposov26a} report the residual
wavelength-calibration uncertainty to be $\approx0.025$~\AA{} per sky
line, or $\approx1$\,\kms{} at 7000~\AA. Meanwhile, comparison of the
calibrated spectra against hydrogen-atmosphere white-dwarf models
shows a residual flux-calibration scatter of about 2\% over most of
the 3600--9800~\AA{} range (rising to $\approx6\%$ shortward of
3700~\AA), comparable to the $\pm6\%$ scatter between synthesized DESI
spectrophotometry and imaging photometry (\S\ref{ssec:photometry}) for
bright stars \citep{guy23a}.

\subsection{Broadband Photometry}
\label{ssec:photometry}

\fastspecfit{} uses broadband optical and IR photometry from the DESI
Legacy Imaging Surveys Data Release 9 (DR9) (hereafter, Legacy
Surveys; \citealt{dey19a}, D.~J.~Schlegel et~al., in preparation),
which covers the full DESI footprint using two imaging surveys and
three cameras. South of a declination boundary of approximately
$+32\arcdeg$, the $grz$ imaging comes from the DECam Legacy Survey
\citep[DECaLS;][]{dey19a}, obtained with the Dark Energy Camera
\citep[DECam;][]{flaugher15a} on the Blanco 4-m telescope at the Cerro
Tololo Inter-American Observatory (CTIO).  North of this boundary,
$g$- and $r$-band imaging come from the Beijing-Arizona Sky Survey
\citep[BASS;][]{zou17a}, which was conducted with the 90Prime camera
\citep{williams04a} on the Bok 2.3-m telescope, and the $z$-band
imaging comes from the Mayall $z$-band Legacy Survey (MzLS), which
used the Mosaic-3 camera \citep{dey16a} on the Mayall 4-m
telescope. The Bok and Mayall telescopes are both located at Kitt Peak
National Observatory (KPNO). Following \citet{myers23a}, we track the
imaging system for each DESI target via the \texttt{PHOTSYS} flag
(\texttt{N} or \texttt{S}) to assign the appropriate filter
curves from
  \texttt{speclite}.\footnote{\url{https://speclite.readthedocs.io/en/latest}}

The Legacy Survey catalogs are generated using the Tractor
forward-modeling photometry pipeline \citep{lang16b, lang16a}, which
fits a common two-dimensional image model simultaneously across all
available $grz$ imaging to derive the integrated flux for each
source. A forced-photometry technique is used to derive fluxes in the
four WISE bands---$W1$ (3.4\,\micron), $W2$ (4.6\,\micron), $W3$
(12\,\micron), and $W4$ (22\,\micron) \citep{wright10a}---based on the
unWISE coadds\footnote{\url{https://unwise.me}} produced by
\citet{meisner21a} for $W1$ and $W2$ (based on the six-year WISE plus
NEOWISE-Reactivation imaging; \citealt{mainzer14a}) and
\citet{lang14a} for $W3$ and $W4$. \fastspecfit{} fits this total
photometry in all seven $grz$ and WISE $W1$--$W4$ bands together
with the DESI spectrum in a joint spectrophotometric fit
(\S\ref{sec:modeling}); in any band where a source is undetected, the
reported $1\sigma$ flux uncertainty is used as an upper limit.

\subsection{DESI Target Classes and Redshifts}
\label{ssec:targets}

This section briefly summarizes the DESI survey and program structure,
the extragalactic target classes, and the redshift-determination
pipeline needed to interpret the \fastspecfit{} VACs; for a more
detailed discussion of all these topics see
\citet{desi-collaboration24e, desi-collaboration24d,
  desi-collaboration26a}.

DESI observations are organized by survey and program
\citep{schlafly23a, myers23a}. The primary program is the main survey,
which began on 2021 May 14 as a five-year survey (DESI Run~1;
2021--2026) and has since been extended by an additional three years
(DESI Run~1b; 2026--2029). The main survey was preceded by several SV
phases, most notably the One-Percent Survey (SV3), which covered
roughly 1\% of the final DESI footprint to high spectroscopic
completeness \citep{desi-collaboration24e,
  desi-collaboration24d}. Within each survey, observations are
assigned to the ``bright'' program (moonlight or reduced
transparency), the ``dark'' program (nominal dark-sky conditions), or
the ``backup'' program \citep[twilight and poor-weather
  time;][]{dey26a} on the basis of sky conditions. A small fraction
of observations that do not fit any of these three categories are
assigned to an ``other'' program \citep{desi-collaboration26a}.

DESI photometrically selects primary extragalactic targets from the
Legacy Surveys (\citealt{dey19a}; D.~J.~Schlegel et~al., in
preparation) and organizes them into four classes spanning
complementary redshift ranges \citep{myers23a}. The bright galaxy
survey \citep[BGS, $0 < z \lesssim 0.6$;][]{hahn23a} targets galaxies
to $r \lesssim 20.2$~mag in the bright program; the luminous red
galaxy \citep[LRG, $0.4 < z \lesssim 1.1$;][]{zhou23a}, emission-line
galaxy \citep[ELG, $0.6 < z \lesssim 1.6$;][]{raichoor23a}, and quasar
\citep[QSO, $0.9 < z \lesssim 4$;][]{chaussidon23a} samples are all
observed in the dark program; and finally, secondary or tertiary
(special) targets are observed as part of a wide range of bespoke
science programs and can be assigned to any program \citep{myers23a,
  desi-collaboration26a}.

Spectroscopic redshifts and spectral classifications are determined by
Redrock (S.~Bailey et~al., in preparation; see also
\citealt{guy23a, anand24a}), a
principal component analysis (PCA) template-fitting code that
independently fits templates for stars, galaxies, and quasars over a
type-dependent grid of trial redshifts and records the
maximum-likelihood redshift \texttt{Z} and template class
\texttt{SPECTYPE} (\texttt{STAR}, \texttt{GALAXY}, or
\texttt{QSO}). The two primary redshift-quality diagnostics reported by
Redrock are \texttt{ZWARN}, a per-object warning bitmask where
\texttt{ZWARN=0} signals no known hardware, data-quality, or
redshift-fitting problems, and \texttt{DELTACHI2}, the difference in
$\chi^2$ between the best and second-best fits across all spectral
classes. A third diagnostic, \texttt{COADD\_FIBERSTATUS}, is a
per-object bitmask that logically combines the per-exposure
fiber-status flags contributing to the coadded spectrum and records
hardware or data-quality problems such as a broken or stuck fiber,
poor positioning, or a bad amplifier readout (D.~Green et~al., in
preparation; S.~Bailey et~al., in preparation).

Finally, three after-burner processing steps follow Redrock: an
emission-line fitter (\texttt{emlinefit}; \citealt{raichoor23a,
  desi-collaboration24d}) used to validate ELG redshifts; a
\mgiidoublet{} doublet fitter \citep{chaussidon23a}; and the
\texttt{QuasarNET} neural-network QSO classifier \citep{busca18a,
  green25a}. The latter two after-burners are essential for
improving the completeness of the QSO sample
(\S\ref{ssec:catalog}).

\section{Spectrophotometric Modeling}
\label{sec:modeling}

This section describes the technical, algorithmic structure of
\fastspec, the default spectrophotometric fitting mode within the
larger \fastspecfit{} software package. Readers interested in examples
of \fastspec{} applied to three illustrative DESI targets may safely
skip to \S\ref{sec:casestudy}, while those interested only in the
resulting DR2 VAC may skip to \S\ref{sec:vac}.

Let $F(\lambda;\theta)$ be the observed-frame UV through IR SED of a
galaxy, where $\theta$ denotes the physical properties---stellar mass,
star-formation history, dust content, kinematics, redshift, and so
on---that determine it.
Our overarching goal is to find the model
\begin{equation}
  M(\lambda;\theta) = C(\lambda;\theta_C) + E(\lambda;\theta_E),
  \label{eq:totmodel}
\end{equation}
the sum of a stellar continuum model $C$ (\S\ref{ssec:continuum}) and
an emission-line model $E$ (\S\ref{ssec:emlines}), that best
reproduces $F$ in a maximum-likelihood sense, where $\theta \equiv
\theta_C \cup \theta_E$ is the union of the continuum and
emission-line parameter sets.

A key algorithmic choice we make in \fastspec{} is to fit $C$ and $E$
sequentially rather than jointly, subtracting the best-fitting
continuum model from the data before fitting the emission
lines. Because $C$
draws on the overall continuum shape and broadband photometry while
$E$ draws on (relatively) narrow ``residual'' features, decoupling
them lets us fit the lines robustly and flexibly even when the
continuum model is a poor fit to the data due to, for example,
residual calibration errors (\S\ref{sssec:smooth}).

We prepare the spectroscopic and photometric data for fitting in two
ways. First, we correct both datasets for Galactic extinction using the
\citet{schlegel98a} dust map, recalibrated following
\citet{schlafly11a}, and the \citet{fitzpatrick99a} extinction curve
with $R_V\equiv A_V/E(B-V)=3.1$. We evaluate the extinction curve at
the vacuum wavelength of each spectral pixel and at the effective
wavelength of each of the seven $grz$ and WISE $W1$--$W4$ bandpasses.
Second, to guard against underestimated pipeline uncertainties, we add
a minimum uncertainty floor in quadrature to the formal inverse
variances. We adopt a 1\% floor for the spectroscopy and, for the
photometry, a fixed per-band floor of 2\% for $grz$ (DECam, BASS, and
MzLS) and 5\% for WISE $W1$--$W4$. We use these floored inverse
variances throughout the pipeline, including in the Monte Carlo
resampling described in \S\ref{ssec:uncertainties}.

In the remainder of this section we describe the fitting pipeline's
two primary model inputs, the SPS template library
(\S\ref{ssec:templates}) and the emission-line list
(\S\ref{ssec:emlinelist}). We then describe the three sequential
algorithmic stages \fastspec{} executes in turn---adaptive
emission-line masking via fast, preliminary fits in wavelength patches
(\S\ref{ssec:patches}), stellar continuum modeling
(\S\ref{ssec:continuum}), and emission-line modeling
(\S\ref{ssec:emlines}). We close by describing the physical quantities
\fastspec{} derives from these fits (\S\ref{ssec:quantities}), how we
estimate their uncertainties (\S\ref{ssec:uncertainties}), and the
additional fitting modes, including a ``photometry-only'' mode called
\fastphot, built on this same machinery (\S\ref{sec:modes}).

\subsection{SPS Template Library}
\label{ssec:templates}

\begin{figure*}
\centering\includegraphics[width=1.0\textwidth]{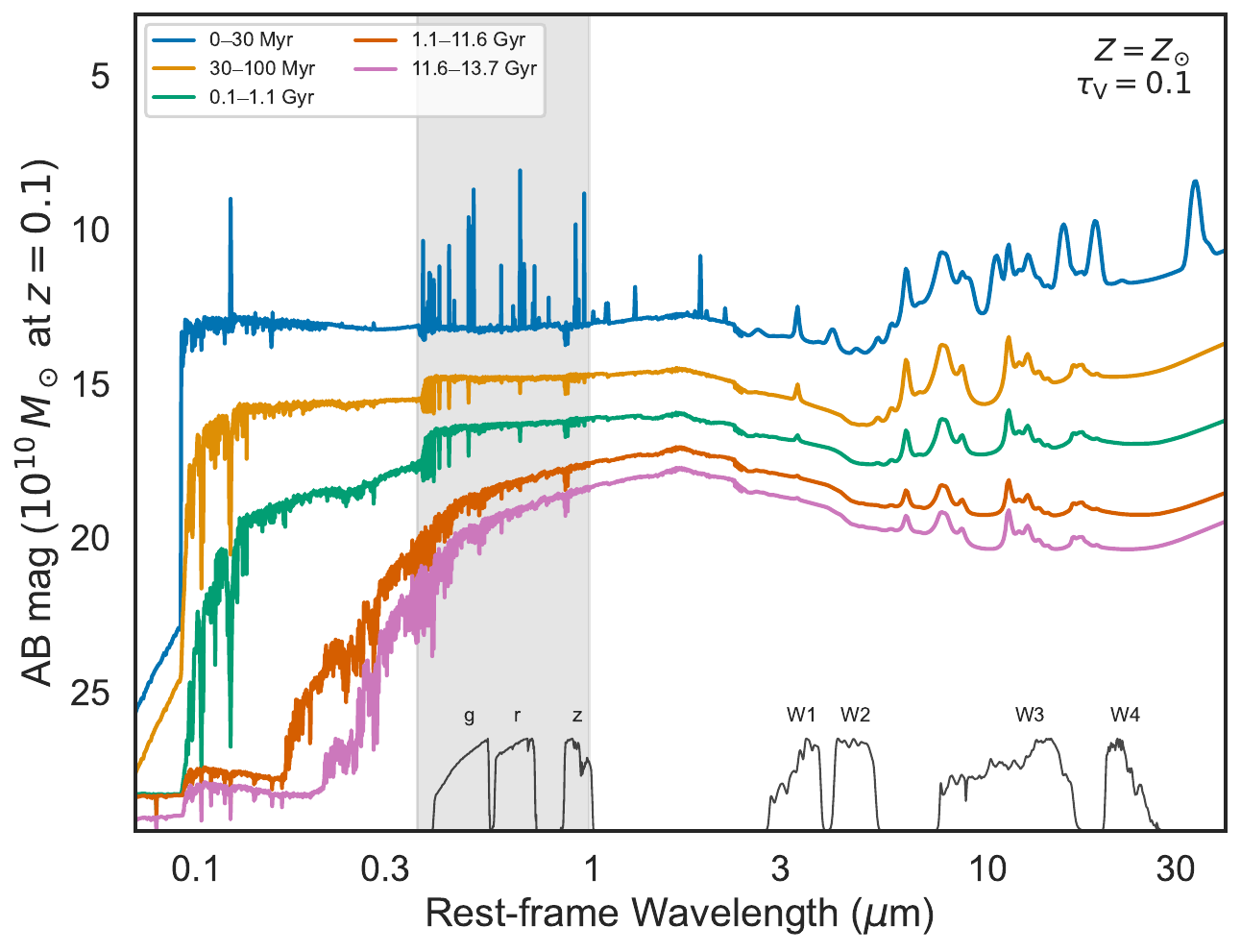}
\caption{The five SPS templates spanning \fastspecfit's lookback-time
  age bins (\S\ref{ssec:templates}), computed at solar metallicity
  ($\log_{10}Z/Z_\odot=0$) with $\tauv=0.1$ and normalized to a
  $10^{10}\,\msun$ galaxy at $z=0.1$. Each curve corresponds to one
  age bin, assuming a constant star-formation rate over its duration,
  ranging from 0--30~Myr (youngest) to 11.6--13.7~Gyr (oldest); the
  SED shifts from UV-bright continua with prominent nebular emission
  in the youngest populations to red, featureless spectra in the
  oldest. The gray band marks the DESI optical spectral range
  (0.36--0.98~$\mu$m); the curves along the bottom of the panel show
  the DECam $grz$ and WISE $W1$--$W4$ broadband filter throughputs that
  \fastspecfit{} includes in the simultaneous spectrophotometric
  fit. \label{fig:sps}}
\end{figure*}

We generate SPS templates using Flexible Stellar Population
Synthesis\footnote{\url{https://github.com/moustakas/fsps}}
\citep[FSPS;][]{conroy09a, conroy10b} via a fork of the
\texttt{python-fsps}\footnote{\url{https://github.com/moustakas/python-fsps}}
interface. We construct these models using a custom version of the
\texttt{C3K} theoretical spectral library,
\texttt{C3K\_R3K}\footnote{\url{https://github.com/moustakas/c3k}},
the MIST stellar isochrones \citep[v1.0;][]{choi16a,
  dotter16a}, and a \citet{chabrier03a} IMF from
$0.08$--$120$~\msun. The \texttt{C3K\_R3K} library spans 100~\AA{} to
1000~$\mu$m. In the rest-frame optical (3500--9800~\AA) it has
$\mathcal{R}_{\mathrm{C3K}}=3000$, corresponding to a Gaussian
line-spread function with
$\sigma_\mathrm{C3K}=c/(2.355\,\mathcal{R}_\mathrm{C3K})=42$~\kms{},
sampled in constant-velocity $\Delta v=50$~\kms{} pixels. At shorter
and longer wavelengths, our models have lower resolution and
sparser wavelength sampling, which keeps the UV-to-IR template
lightweight for synthetic photometry and dust-emission modeling
(\S\ref{sssec:fitting}) without wasting resolution where the
spectroscopic fit does not need it.

Following \citet{leja17a}, we build five ``continuity age'' basis
templates spanning fixed lookback-time bins, each assuming a constant
star-formation rate over its duration. The two youngest bins span
0--30~Myr and 30--100~Myr (with upper edges at
$\log_{10}(t/\mathrm{yr}) = 7.48$ and $8.0$, respectively), while the
next two bins are logarithmically spaced between
  100~Myr and 11.6~Gyr (0.1--1.1~Gyr and 1.1--11.6~Gyr) and the oldest
  bin spans 11.6--13.7~Gyr, the last 15\% of a 13.7~Gyr cosmic age
(Figure~\ref{fig:sps}). We generate templates at fixed solar
metallicity, $\log_{10}(Z/Z_\odot) = 0$ (with no $\alpha$-element
enhancement), with
  and without nebular emission lines \citep{byler17a}. We need both
  variants because the broadband photometry includes any nebular line
  emission, whereas we model the emission lines in the spectrum
  separately (\S\ref{ssec:emlines}). We therefore use the variant with
  emission lines in the joint spectrophotometric fit
  (\S\ref{sssec:fitting}), where masking line-contaminated pixels
  restricts its effect to the synthetic photometry, and for
  K-corrections (\S\ref{ssec:quantities}). We use the variant without
  emission lines everywhere else, including the velocity-dispersion
  fit (\S\ref{sssec:vdisp}), the continuum model we subtract before
  fitting the emission lines, and continuum-derived quantities such as
  $D_n(4000)$ and the monochromatic luminosities
  (\S\ref{ssec:quantities}).

We index the resulting rest-frame templates $S_n(\lambda)$, $n =
1,\ldots,n_\mathrm{template}$ (with $n_\mathrm{template}=5$) by
star-formation history (SFH) age bin, separately for each
nebular-emission variant, and
normalize each to the flux density produced at 10~pc by one solar mass
of stars formed at the tabulated lookback time,
\begin{equation}
  S_n(\lambda) =
  \frac{L_\odot}{4\pi(10\,\mathrm{pc})^2}\,\frac{L_{\lambda,n}}{L_\odot},
  \label{eq:fnorm}
\end{equation}
in units of
$\mathrm{erg\,s^{-1}\,cm^{-2}\,\text{\AA}^{-1}\,\msun^{-1}}$, so that
each template's fitted coefficient $c_n \geq 0$
(\S\ref{sssec:fitting}) carries units of \msun.

In addition, we generate a single infrared dust emission template
$S_\mathrm{dust}(\lambda)$ from \citet{draine07a}. This model is
parameterized by the polycyclic aromatic hydrocarbon (PAH) mass
fraction, $q_\mathrm{PAH}$, the minimum starlight intensity heating
the diffuse interstellar medium (ISM) dust component $U_\mathrm{min}$,
and the fraction $\gamma$ of dust mass exposed to a power-law
distribution of starlight intensities (up to a fixed maximum
$U_\mathrm{max}$) representing photodissociation regions. Because most
DESI targets do not have adequately deep IR photometry to constrain a
large number of dust emission parameters, we adopt fiducial values of
$q_\mathrm{PAH} = 3.5$, $U_\mathrm{min} = 1.0$, and $\gamma = 0.01$
and normalize our IR template to unit bolometric luminosity.

There is no AGN-like emission (either continuum or emission lines) in
these templates, a shortcoming which should be kept in mind throughout
this paper (\S\ref{ssec:example_qso}). However,
\S\ref{ssec:improvements} discusses some of the planned improvements
for \fastspecfit, including a QSO/AGN fitting mode.

\subsection{Emission-Line List}
\label{ssec:emlinelist}

The default \fastspecfit{} line-list, \texttt{data/emlines.ecsv},
includes 46 emission lines, spanning rest-frame
Ly$\alpha$~$\lambda$1215 through [\ion{S}{3}]~$\lambda$9532
(Table~\ref{table:emlines}). We select this list to span the full range
of galaxy and QSO spectral types over the DESI redshift range,
targeting the strongest and most diagnostically useful transitions
rather than every possible detectable line; the resulting set supports
target selection and scientific follow-up across a broad range of
object classes, from star-forming galaxies to high-redshift
quasars.

The adopted vacuum rest-frame wavelengths in Table~\ref{table:emlines}
come from the NIST Atomic Spectra
Database\footnote{\url{https://physics.nist.gov/PhysRefData/ASD/lines_form.html}}
for the Balmer series, the helium lines, the \mgiidoublet{} doublet,
and most forbidden transitions; the UV broad-line wavelengths come
from \citet{vanden-berk01a}; and wavelengths for \niidoublet,
\oiiidoublet, \nii~$\lambda5755$, and \oiii~$\lambda4363$ come from
the CHIANTI atomic
database.\footnote{\url{https://db.chiantidatabase.org}}

We organize these lines into four general categories---UV/QSO lines,
narrow Balmer/helium lines, broad Balmer/helium lines, and narrow
forbidden lines---which are used to set kinematic fitting priors
(\S\ref{sssec:kinematic}). Table~\ref{table:emlines} also lists a set
of 23 spectrally adjacent ``patches'', and a Boolean bit (``strong'')
which flags which lines, if present, are most likely to be
well-detected in a given spectrum. We describe these latter quantities
in detail in the next subsection (\S\ref{ssec:patches}).

\input{table-emlines}

\subsection{Adaptive Emission-Line Masking}
\label{ssec:patches}

In order to fit the stellar continuum robustly, we must first identify
pixels which could be affected by emission lines. The challenge is
that DESI targets span an enormous range of emission-line properties,
from purely absorption-line quiescent galaxies with no measurable
line-emission to extreme starbursts and broad-line AGN whose emission
dominates large stretches of the observed spectrum. A single, fixed
masking window sized for the strongest expected lines would over-mask
quiescent spectra and discard usable continuum information, while a
window sized for narrow lines found in typical star-forming galaxies
would under-mask galaxies and QSOs with broad emission lines and bias
the stellar continuum fit.

Therefore, before fitting the stellar continuum
(\S\ref{sssec:fitting}), we run a lightweight, per-object fitting
pass---the \textit{line masker}---that determines whether a spectrum
contains statistically significant line-emission, and, if so, the
approximate velocity widths and velocity shifts of those
lines. Crucially, the line-masker also performs an initial
determination of whether the spectrum contains broad Balmer
line-emission (although the statistically defensible decision is made
in the final fitting pass, described in \S\ref{sssec:kinematic}).

\begin{figure*}
\centering\includegraphics[width=1.0\textwidth]{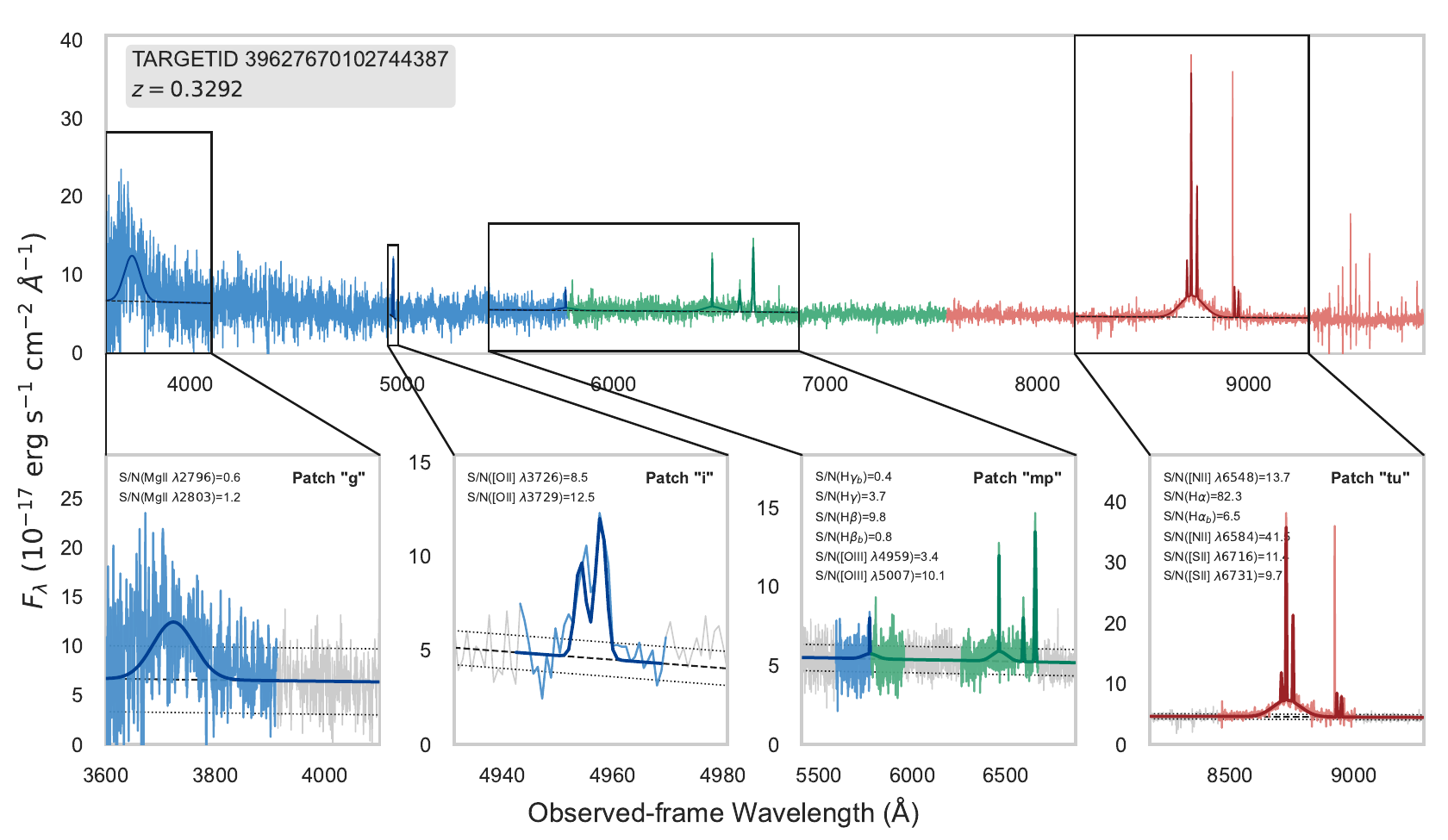}
\caption{Illustration of the line-masking algorithm for an example BGS
  target at $z=0.3292$ (\texttt{TARGETID=39627670102744387} from the
  main survey bright-time program, appearing on HEALPix
  17366). \textit{Top:} the full per-camera DESI spectrum (B, R, Z in
  blue, green, and red, respectively) together with the per-patch
  best-fitting local linear continuum and Gaussian model line-profiles
  (darker blue, green, and red in B, R, and Z, respectively) for all
  the candidate strong lines in the observed spectral
  range. \textit{Bottom:} Zoomed-in panels which more clearly show the
  per-patch fitting results. In each panel, camera-colored pixels mark
  the pixels in the final line-mask (and therefore excluded from the
  subsequent continuum-fit), while light-gray pixels are the
  surrounding continuum used for that patch's local linear fit. The
  darker curves show the combined best-fit line-plus-continuum model,
  while the dashed black line and surrounding dotted band show the
  local linear continuum and its $\pm1\sigma$ noise envelope. Finally,
  the legends indicate the S/N value inferred for each line and the
  patch label. For this particular object, a significant broad \ha{}
  Balmer line was correctly identified, resulting in patches ``m'' and
  ``p'' and patches ``t'' and ``u'' merging into larger patches ``mp''
  and ``tu'', respectively, and wider line masks. \label{fig:linemasker}}
\end{figure*}

First, we activate all the spectral patches
(Table~\ref{table:emlines}) with at least one line in the observed
spectral range and fit all the ``strong'' lines with the narrow-only
kinematic model, together with a simple linear local continuum in each
patch. The lines we initially fit are simply the subset of lines
which, if present, are most likely to yield useful kinematic
constraints. In the narrow-only kinematic model, all the forbidden
lines are tied together into a single kinematic group (shared velocity
width and velocity center) anchored on \oiiilam, while all the narrow
Balmer and \hei{} lines are anchored on \ha{} in a second
kinematic group (see \S\ref{sssec:kinematic} for more details). Then,
we define a pixel window of $\pm5\,\sigma_\mathrm{line}$ centered on
each line's expected observed-frame wavelength, where
$\sigma_\mathrm{line}$ is that line's current velocity-width
estimate. We iteratively update $\sigma_\mathrm{line}$, beginning with
150~\kms{} and 3000~\kms{} for narrow and (candidate) broad lines,
respectively, but always floored at a minimum line-width of 50~\kms,
and grow a surrounding continuum window until at least 11 unmasked
pixels are available. When this adaptive growth causes two spectrally
adjacent patches' continuum windows to overlap by more than 11 pixels,
we merge them into a single patch, sharing one local continuum and the
union of both patches' lines. We run two such iterations in total,
updating the line-pixel windows with the best-fit velocity shifts and
widths before the second pass.

Next, if any of the three strongest Balmer lines (\ha, \hb, or \hg)
fall within the observed spectral range, we repeat the procedure with
the narrow+broad model, initializing all the broad Balmer line-widths
at 1000~\kms. In this fitting mode, all the ``strong'' broad Balmer
lines are kinematically tied together, separately from the two
previously described narrow-line kinematic groups (see also
\S\ref{sssec:kinematic}). After this fitting pass, if any broad Balmer
line reaches $\mathrm{S/N} > 1.5$, we adopt the narrow+broad
line-widths for masking; otherwise, we
retain the narrow-only widths. Here, S/N is the fitted peak
amplitude divided by the robust standard deviation of the
line-subtracted residuals in that patch.

With the robustly estimated line-widths, we can now build the complete
line-mask. Iterating over all the lines in the observed spectral range
(not just the initial set of candidate strong lines), we robustly
estimate each line's S/N by comparing its peak
amplitude against the robust standard
deviation of the line-subtracted residuals in its surrounding local
continuum, and discard ``weak'' lines with $\mathrm{S/N} < 3.5$ from
further consideration, in order to avoid over-masking the continuum
spectrum. For the surviving lines, we adopt a
$\pm5\,\sigma_\mathrm{line}$ window, centered on each line's expected
wavelength and sized using its converged velocity width and velocity
shift, to build the final pixel mask. Regardless of the outcome
of the broad-line test, we floor the broad Balmer line-width at
1000~\kms{} when building this mask, so that the final emission-line
fit (\S\ref{sssec:kinematic}) can remeasure any broad component.

Figure~\ref{fig:linemasker} illustrates this process for a single BGS
target from the bright program of the main survey,
\texttt{TARGETID}=39627670102744387, at $z=0.3292$. Two pairs of
spectrally adjacent patches are merged in this example: the ``m''
(H$\gamma$) and ``p'' (\hb{} and \oiiidoublet) patches merge into a
single ``mp'' patch, and the ``t'' (\ha{} and \niidoublet) and ``u''
(\siidoublet) patches merge into ``tu'', because their adaptive
continuum windows overlap by more than 11 pixels. The line-masker also
detects broad Balmer line-emission in this target, so the
corresponding patches are masked with wider windows, using the
narrow+broad kinematic model described above.

\subsection{Stellar Continuum Model}
\label{ssec:continuum}

This section describes the steps \fastspec{} uses to find the
best-fitting continuum model, $C(\lambda;\theta_C)$. Because speed and
flexibility are paramount, the default fitting mode uses a minimal
parameter set, $\theta_C = \{\mathbf{c},\tauv,\vdisp\}$, where
$\mathbf{c}$ are the non-negative
template coefficients, $\tauv$ is the $V$-band dust optical depth, and
$\vdisp$ is the stellar velocity dispersion. We emphasize, however,
that the fitting infrastructure can be easily extended to include
additional parameters (e.g., redshift, a two-component dust
model or QSO model spectra), which we discuss in
\S\ref{ssec:improvements}.

For each object, we determine $\theta_C$ in three stages. First, we
measure $\vdisp$ from stellar absorption features
(\S\ref{sssec:vdisp}); then we derive a scalar aperture correction
that places the DESI fiber spectrum on the total-flux scale of the
broadband photometry (\S\ref{sssec:apercorr}); and finally we jointly
fit for dust attenuation $\tauv$ and the template coefficients
$\mathbf{c}$, holding $z$ fixed at its input (Redrock) value
(\S\ref{sssec:fitting}). An optional nonparametric smooth-continuum
correction then removes residual low-level systematics
(\S\ref{sssec:smooth}) before the spectrum passes to the emission-line
fitter.

\subsubsection{Stellar Velocity Dispersion}
\label{sssec:vdisp}

The line-of-sight stellar velocity dispersion \vdisp{} traces the
depth of a galaxy's gravitational potential well; it is a key
observable for several galaxy scaling relations, such as the
Fundamental Plane, the black hole mass--\vdisp{} relation, and the
halo-mass--stellar-mass relation, and a fundamental ingredient for
most theoretical models of galaxy formation \citep[and references
  therein]{renzini06a, blanton09a, kormendy13a, vegetti24a}. A
reliable measurement, however, requires adequate rest-frame wavelength
coverage and relatively high-S/N spectroscopy of the absorption-line
features most sensitive to dynamical broadening (e.g., Ca~H\&K, the
$G$-band, H$\beta$, Mg~$b$). In DESI, these conditions are generally
met only for BGS and LRG targets; however, because \fastspec{} does
not use targeting information as a prior, we attempt to measure
\vdisp{} for every spectroscopic target.

We determine whether a direct measurement of \vdisp{} is possible from
two conditions: rest-frame wavelength coverage and ``fit
significance.'' First, we require the spectrum of a given target to
span the full rest-frame range 3800--4900~\AA, bridging Ca~H\&K, the
4000-\AA{} break, and H$\beta$. If this condition is satisfied, then
we perform the actual fit using all unmasked pixels between
3800--6000~\AA{} (in the rest-frame). In practice, this requirement
means that we can only estimate \vdisp{} for galaxies up to
$z\approx1$. Second, we determine fit significance by performing a
fast $\chi^2$ grid scan of six trial velocity-broadening widths evenly
spaced between 50 and 500~\kms, simultaneously solving for \tauv{} and
the template coefficients by linear least-squares at each grid point
(\S\ref{sssec:fitting}). If this scan yields a peak-to-peak
$\Delta\chi^2 \geq 25$ (and with the minimum value away from either
boundary) then we proceed with a direct measurement. In general,
$\Delta\chi^2 < 25$ implies that the spectrum has inadequate S/N in
the stellar continuum (e.g., a faint ELG target) to accommodate
\vdisp{} as an additional free parameter.

Most DESI targets in DR2 fail one or both of the preceding conditions;
however, we would still like to use an approximately correct value of
\vdisp. In this case, we estimate \vdisp{} from an approximate
stellar-mass--velocity-dispersion relation,
\begin{equation}
  \log_{10}(\vdisp/\kms) = a + b\,(\logmstar - 11),
  \label{eq:sigma_mstar}
\end{equation}
with $(a,b) = (2.30,\,0.25)$ using a fast, preliminary estimate of the
stellar mass, \Mstar{} (\S\ref{sssec:fitting}). If our estimate of
\Mstar{} fails, we instead adopt a fallback
$\vdisp = 150$~\kms{} (\texttt{VDISP\_NOMINAL}), close to the
characteristic ``knee'' of the velocity dispersion function
\citep{bezanson11a, taylor22a}. In either fallback case we set
\texttt{VDISP\_IVAR}~$=0$ to indicate that \vdisp{} has not been
independently measured.

When a direct measurement is possible, we determine \vdisp{} by
finding the best-fitting Gaussian kernel
dispersion, $\vdisp^\mathrm{kern} \in [50,\,500]$~\kms, using
\texttt{scipy.optimize.least\_squares} with a Trust-Region Reflective
minimization algorithm \citep{branch99a}, fitting for
$\vdisp^\mathrm{kern}$ and \tauv{} simultaneously over the same range
as above. At each trial value of $\vdisp^\mathrm{kern}$ we convolve
the templates with a Gaussian broadening kernel using a Fast Fourier
Transform (FFT), which multiplies the templates' precomputed Fourier
transforms by the kernel's and is markedly faster than a direct
pixel-space convolution. We then recover the physical velocity
dispersion by adding back the intrinsic resolution of our SPS models
in quadrature,
\begin{equation}
  \vdisp = \sqrt{(\vdisp^\mathrm{kern})^2 + \sigma_\mathrm{C3K}^2},
  \label{eq:vdisptot}
\end{equation}
where $\sigma_\mathrm{C3K}=42$~\kms{} (\S\ref{ssec:templates}).

\subsubsection{Aperture Correction}
\label{sssec:apercorr}

By default, \fastspec{} fits the spectrum ($0.36-0.98$\,\micron;
\S\ref{ssec:spectra}) of each extragalactic DESI target simultaneously
with its corresponding seven-band ($grz$ and WISE $W1$--$W4$) Legacy
Surveys integrated photometry ($0.47-22$\,\micron;
\S\ref{ssec:photometry}). Fitting all the available spectrophotometry
over the broadest possible wavelength baseline helps constrain
physical properties like dust attenuation and ties the derived
stellar masses, star-formation rates, and rest-frame photometry to the
total light of each source rather than to the light captured by the
$1\farcs5$-diameter DESI fiber. However, we have to account for the
fact that the spectrum and photometry are measured over different
spatial scales before combining them.

So-called aperture effects (or aperture bias) are extremely difficult
to treat in detail because galaxies exhibit genuine radial gradients
(or, more generally, spatial inhomogeneities) in color, stellar
populations, age, metallicity, and more \citep[and references
  therein]{sanchez20a}. In addition, the amount of light the DESI
fiber captures depends on each galaxy's apparent (angular) size on the
sky, and therefore on redshift through the angular diameter distance
\citep{hogg99a}. These effects conspire to make aperture bias a strong
function of target class, affecting low-redshift BGS targets the most
and higher-redshift, angularly compact, and more
spectrophotometrically homogeneous LRG and ELG targets the least.

Nevertheless, in the current version of \fastspec{} we use the
simplest possible model to account for aperture bias: we assume that
the spectral shape within the fiber is representative of the galaxy as
a whole, independent of wavelength or the magnitude of the aperture
correction. We derive this correction by synthesizing $grz$ photometry
from a fast initial fit to the DESI spectrum to get $P_k^\mathrm{fib}$
for $k \in \{g,\,r,\,z\}$, and computing the per-band multiplicative
factor
\begin{equation}
  \eta_k = \frac{D_k}{P_k^\mathrm{fib}},
  \label{eq:apercorr}
\end{equation}
where $D_k$ is the total Tractor flux of the object in bandpass
$k$. Then, we compute the median scalar aperture correction,
$\bar{\eta}=\mathrm{median}(\eta_g,\,\eta_r,\,\eta_z)$, and apply the
resulting multiplicative factor (falling back to $\bar{\eta}=1.0$ if
any part of the preceding analysis fails) during the full
spectrophotometric fit (\S\ref{sssec:fitting}).

\subsubsection{Fitting the Stellar Continuum}
\label{sssec:fitting}

To fit the stellar continuum we use a forward-modeling approach.  We
carry each rest-frame SPS template (\S\ref{ssec:templates}) through a
sequence of physical and instrumental transformations---dust
attenuation, redshifting, and instrumental broadening (or filter
convolution)--- to
  produce a prediction that we can compare directly to the observed
  spectrophotometry. We then solve for the template weights and dust
attenuation (at fixed \vdisp; \S\ref{sssec:vdisp}) that minimize the
inverse-variance-weighted residuals.

First, we model the attenuation of each template as a
uniform foreground screen,
\begin{equation}
  S_n^\mathrm{att}(\lambda) = S_n(\lambda)\,e^{-\tauv k(\lambda)},
  \label{eq:atten}
\end{equation}

\noindent where $S_n(\lambda)$ is the unattenuated template given by
equation~(\ref{eq:fnorm}), $\tauv$ is the $V$-band optical depth, and

\begin{equation}
  k(\lambda) =
  \left(\frac{\lambda}{5500\,\text{\AA}}\right)^{n}
  \label{eq:kl}
\end{equation}

\noindent is the attenuation curve, which we model as a simple power
law with fixed index $n=-0.7$ based on the \citet{charlot00a}
birth-cloud model.  Absorbed photons are re-emitted in the IR
following energy balance. We scale the \citet{draine07a} dust emission
template, $S_\mathrm{dust}(\lambda)$, by the luminosity absorbed by
each template,
\begin{equation}
  L_{\mathrm{abs},n} =
  \int_{0.01\,\mu\mathrm{m}}^{1000\,\mu\mathrm{m}} S_n(\lambda) (1 -
  e^{-\tauv k(\lambda)})\,d\lambda,
  \label{eq:labs}
\end{equation}
so that the total (attenuated plus reradiated) continuum template
conserves bolometric luminosity,
\begin{equation}
  S_n^\mathrm{tot}(\lambda) = S_n^\mathrm{att}(\lambda) +
  L_{\mathrm{abs},n}\,S_\mathrm{dust}(\lambda).
  \label{eq:ftot}
\end{equation}

For an object at redshift $z$ with luminosity distance $D_L$ (in pc),
we relate each rest-frame template to its observed-frame counterpart
via
\begin{equation}
  S_n^\mathrm{obs}(\lambda)
  =
  T_\mathrm{IGM}(\lambda,z)\,
  \mathcal{F}_0\,\mathcal{M}_0
  \left(\frac{10\,\mathrm{pc}}{D_L}\right)^{2}
  \,\frac{S_n^\mathrm{tot}(\lambda/(1+z))}{1+z},
  \label{eq:zfactor}
\end{equation}
where $T_\mathrm{IGM}(\lambda,z)$ is the \citet{inoue14a}
intergalactic medium (IGM) transmission function which accounts for
Ly$\alpha$ forest and Lyman-limit attenuation; $\mathcal{F}_0 =
10^{17}$ is a dimensionless scale factor converting the cgs flux
densities of \S\ref{ssec:templates} to our internal flux units of
$10^{-17}$~erg~s$^{-1}$~cm$^{-2}$~\AA$^{-1}$ \citep{guy23a}; and
$\mathcal{M}_0 = 10^{10}\,\msun$ is a fixed mass normalization that
rescales the fitted template coefficients to order unity for a
well-conditioned non-negative least-squares (NNLS) solve (see below).

We restrict the dust optical depth to the range
$\tauv \in [0,\,2]$, and by default we include coefficients for all
five SPS age bins (\S\ref{ssec:templates}), even for stellar
populations nominally older than the age of the universe at the
redshift of a given object. Because our age sampling is coarse, the
boundary between physical and unphysical ages is poorly defined, and
excluding bins outright would introduce more redshift-dependent bias
than it removes \citep{leja19a}.

At this point, we have all the ingredients necessary to fit the
stellar continuum. Let $d$ and $I$ be the Galactic
extinction-corrected and floor-inflated DESI spectrum and
inverse-variance spectrum, respectively, and $m$ be the binary
emission-line mask from \S\ref{ssec:patches}. Each of these arrays
contain $n_\mathrm{pix}$ elements, with $m_i=1$ for pixels retained in
the fit and $0$ for masked pixels. Also let $D$ and $I_D$ be the
Galactic extinction-corrected and floor-inflated $n_\mathrm{band}=7$
($grz$ and WISE $W1$--$W4$) photometric fluxes and inverse variances,
respectively.

We construct the continuum model $C(\lambda;\theta_C)$ as the
non-negative linear combination of the observed-frame templates
$S_n^\mathrm{obs}(\lambda)$ from equation~(\ref{eq:zfactor}),
\begin{equation}
  C(\lambda;\theta_C) = \sum_n c_n\,S_n^\mathrm{obs}(\lambda),
  \label{eq:ccont}
\end{equation}
holding \vdisp{} fixed at the value determined in \S\ref{sssec:vdisp}
and solving only for $\tauv$ (which enters through $S_n^\mathrm{obs}$)
and the coefficients $\mathbf{c}$. To compare this model against the
spectroscopic and photometric data, we transform it in two ways: (1)
we broaden each template by \vdisp{} (\S\ref{sssec:vdisp}), convolve
it with the wavelength-dependent resolution matrix $\mathbf{R}$
(\S\ref{ssec:spectra}), and resample (using a flux-conserving
trapezoidal-resampling algorithm) onto the same $n_\mathrm{pix}$-pixel
wavelength grid as $d$, giving $C_i \equiv C(\lambda_i;\theta_C)$; and
(2) we convolve each template with the $n_\mathrm{band}$ filter
response functions (\S\ref{ssec:photometry}) to derive the synthetic
photometry $P_k$. Collecting these transformed templates as the
columns of a design matrix $\boldsymbol{\Psi}$, the continuum model at
every pixel and bandpass becomes
\begin{equation}
  \mathbf{C} = \boldsymbol{\Psi}\,\mathbf{c},
  \label{eq:designmatrix}
\end{equation}
\noindent and we determine $\tauv$ and $\mathbf{c}$ by minimizing
\begin{equation}
  \chi^2(\tauv,\mathbf{c})
  =
  \sum_{i=1}^{n_\mathrm{pix}} m_i I_i\left(d_i - \frac{C_i}{\bar\eta}\right)^2
  \;+\;
  \sum_{k=1}^{n_\mathrm{band}} I_{D,k}\left(D_k - P_k\right)^2,
  \label{eq:chi2cont}
\end{equation}
where $\bar\eta$ is the scalar aperture correction described in
\S\ref{sssec:apercorr}.

Because equation~(\ref{eq:designmatrix}) is linear in the multiple
template coefficients $\mathbf{c}$ but nonlinear only in the single
scalar $\tauv$, rather than solve for both simultaneously we exploit
the separability via variable projection (VARPRO; \citealt{golub73a}).
With this approach, we perform an outer Brent scalar minimization over
$\tauv$, with an inner NNLS solve for $\mathbf{c}$ at each trial
$\tauv$,
\begin{equation}
  \mathbf{c}^* = \arg\min_{\mathbf{c}\,\geq\,0}
  \left\|\boldsymbol{\Psi}\,\mathbf{c} - \mathbf{b}\right\|^2,
  \label{eq:nnls}
\end{equation}
where $\mathbf{b}$ is the corresponding weighted data vector, and the
mask, inverse variances, and aperture correction enter as row weights
on $\boldsymbol{\Psi}$ and $\mathbf{b}$ alike---$\sqrt{m_i I_i}/\bar\eta$
for spectral rows and $\sqrt{I_{D,k}}$ for photometric rows. Minimizing
$\|\boldsymbol{\Psi}\mathbf{c}-\mathbf{b}\|^2$ is therefore equivalent
to minimizing equation~(\ref{eq:chi2cont}) at fixed $\tauv$. Because
the inner NNLS solve is exact at every trial $\tauv$, VARPRO converges
an order of magnitude faster than direct optimization over the full
parameter vector, which would otherwise require finite-differencing
across all template coefficients.

With $\tauv$ and $\mathbf{c}$ now in hand, and $\vdisp$ fixed
from the earlier stages, every element of $\theta_C$ is determined,
giving the best-fitting stellar continuum model
$C(\lambda;\theta_C)$. From this model we derive several physical
properties of interest---stellar mass, star-formation rate, and
stellar age---as described in \S\ref{ssec:quantities}.

\subsubsection{Smooth Continuum Correction}
\label{sssec:smooth}

Despite its flexibility, our parametric continuum model $C(\lambda)$
(\S\ref{sssec:fitting}) can leave small- or even large-scale
systematic residuals in the continuum-subtracted spectrum. Depending
on their wavelength-dependence and amplitude, leaving these residuals
unaddressed can severely bias the emission-line modeling
(\S\ref{sssec:emline_model}), especially for faint lines.

The unmodeled residuals can be due to instrumental effects, template
mismatch, or both. For example, the noisy boundary between the three
DESI cameras (\S\ref{ssec:spectra}) is especially susceptible to
residual additive (bias, dark, or sky-subtraction) errors, or to
multiplicative (flat-fielding or flux-calibration) errors. Template
mismatch, meanwhile, is most severe for QSOs and broad-line AGN
(\S\ref{ssec:example_qso}). Their observed continuum typically blends
genuine host-galaxy starlight with \ion{Fe}{2} emission in the
rest-frame UV (roughly 1100--3000\,\AA; \citealt{vestergaard01a}) and
a featureless, non-thermal power-law continuum from accretion onto the
supermassive black hole \citep{vanden-berk01a}. Our SPS templates can
represent the host-galaxy light but not the other two, nor cleanly
separate them.

Keeping these effects in mind, the current version of \fastspec{}
treats all such residuals, irrespective of their origin, as a single,
low-level additive correction to $C(\lambda)$. We estimate and remove
these residuals with a nonparametric smooth continuum after
subtracting the best-fitting, line-free continuum model. Working per
camera, we slide a 75-pixel window in 125-pixel steps across the
residual spectrum, applying a $2\sigma$ iterative clip in each window
and computing the robust median and standard deviation. We discard
windows with fewer than 15 unmasked pixels. We then fit a weighted
quadratic spline (\texttt{scipy.interpolate.UnivariateSpline}, $k=2$;
\citealt{virtanen20a}) to the window medians, giving the smooth
continuum correction $C_\mathrm{smooth}(\lambda)$. For each camera we
also record \texttt{SMOOTHCORR\_B}, \texttt{SMOOTHCORR\_R}, and
\texttt{SMOOTHCORR\_Z}, the median of the smooth correction (using all
unmasked pixels) divided by the observed flux, expressed as a
percentage---a simple, per-object diagnostic of the correction's
overall amplitude relative to the source itself.

\subsection{Emission-Line Model}
\label{ssec:emlines}

After subtracting the best-fitting continuum model
$C(\lambda;\theta_C)$ (\S\ref{sssec:fitting}) and the smooth continuum
correction $C_\mathrm{smooth}(\lambda)$ (\S\ref{sssec:smooth}) from
the data $d_i$, we are left with a residual spectrum $r_i \equiv d_i -
C_i - C_{\mathrm{smooth},i}$ (in \fluxdensity) containing one or more
emission lines (\S\ref{ssec:emlinelist}). In the following three
subsections we describe how we construct and fit the emission-line
model, $E(\lambda;\theta_E)$ (\S\ref{sssec:emline_model}). The full
emission-line parameter set $\theta_E$ consists of three
parameters---an amplitude, velocity offset, and velocity width---for
each of the $n_\mathrm{line}$ lines in our list. We then describe how
we use physically motivated ``kinematic groups'' among different sets
of lines to reduce the number of free parameters in $\theta_E$,
including (optionally) decomposing the Balmer lines into broad and
narrow components (\S\ref{sssec:kinematic}), and conclude with a
summary of all the quantities we measure for each detected (or
undetected) emission line (\S\ref{sssec:lines}).

\subsubsection{Forward Model}
\label{sssec:emline_model}

We forward-model each object's emission-line spectrum as a
non-negative sum of the $n_\mathrm{line}$ emission lines from our line
list (\S\ref{ssec:emlinelist}) that fall within the observed-frame
wavelength range at the Redrock redshift $z$,
\begin{equation}
  E(\lambda;\theta_E) = \sum_{p=1}^{n_\mathrm{line}}
    \phi(\lambda;\,A_p,\lambda_p^*,\sigma_{\ell,p}),
  \label{eq:emline_sum}
\end{equation}
where we model each individual line as a log-Gaussian in observed
wavelength,
\begin{equation}
  \phi(\lambda;\,A,\lambda^*,\sigma_\ell)
  =
  A\exp\!\left(-\frac{(\ln\lambda - \ln\lambda^*)^{2}}{2\sigma_\ell^{2}}\right).
  \label{eq:intrinsic}
\end{equation}
Here, $A$ is the peak amplitude (in \fluxdensity), $\lambda^*$ is the
observed line center in \AA, including both cosmological ($z$) and
peculiar ($\Delta v$) motion of the line-emitting gas,
\begin{equation}
  \lambda^* = \lambda_\mathrm{rest}
    \left(1 + z + \frac{\Delta v}{c}\right),
  \label{eq:lstar}
\end{equation}
and $\sigma_\ell = \sigma/c$ is the line width in ln-$\lambda$ units,
where $\sigma$ (in \kms) is the physical (Doppler)
line-width.\footnote{During internal (DESI Collaboration)
  review, it was noted that this definition of $\Delta v$ is not a
  rest-frame quantity, as originally intended. Consequently, users
  should divide the reported \texttt{VSHIFT} (\S\ref{sssec:lines}) by
  $1+z$ to recover the rest-frame velocity of the line-emitting
  gas. The next version of the code will use a multiplicative
  parameterization of $\lambda^*$ so that $\Delta v$ is in the
  rest-frame, matching $\sigma$.}  With this parameterization, the
total flux of each line (in \flux) is
\begin{equation}
  \Phi = \sqrt{2\pi}\,A\,\lambda^*\,\sigma_\ell
       = \sqrt{2\pi}\,A\,\lambda^*\,\frac{\sigma}{c},
  \label{eq:flux_phys}
\end{equation}
which is independent of the DESI wavelength grid and instrumental
resolution.

Comparing this profile to a DESI spectrum requires convolving it with
the resolution matrix $\mathbf{R}$ (\S\ref{ssec:spectra}). Unlike the
stellar continuum, however, which varies smoothly enough across a
pixel that sampling the continuum model at pixel centers before
applying $\mathbf{R}$ preserves its flux (\S\ref{sssec:fitting}), it
is possible for an emission line to be narrower than a single
0.8~\AA{} DESI pixel.\footnote{DESI pixels range in velocity width
between $\approx66$ and $\approx25$~\kms{} over the observed-frame
spectral range, 3600 to 9800~\AA.} Low-mass, low-metallicity
star-forming galaxies, in particular, can exhibit strong nebular
line-emission with intrinsic line-widths of tens of \kms{} or
less---narrower than a pixel over most of the observed wavelength
range (\citealt{scholte26a}; J.~Moustakas et~al., in preparation;
V.~Manwadkar et~al., in preparation).

The obvious forward-modeling analog, $\mathbf{m} \approx
\mathbf{R}\,\boldsymbol{\phi}$, therefore fails: point-sampling
$\phi(\lambda)$ at the pixel centers captures only a small,
pixel-phase-dependent fraction of a narrow line's true area, biasing
the recovered flux low by a few percent up to $>$30\%, depending on
where the line center falls within a pixel (see Appendix~D in
\citealt{guy23a} for additional discussion). In detail, this bias
arises because $\mathbf{R}$'s rows are normalized to conserve flux
only for smooth inputs sampled at pixel centers, with no matching
column normalization to protect sub-pixel structure
\citep{guy23a}. Moreover, this bias is not
just a modeling artifact; simulations show that even a simple
(model-independent) integration of the observed pixels
($\Phi_\mathrm{box}$; \S\ref{sssec:lines}) suffers the same
flux loss for lines narrower than about one pixel (S.~Koposov
2026, private
communication).\footnote{\url{https://github.com/desihub/fastspecfit/issues/183}}

To contend with this bias, we adopt the deconvolution strategy devised
by S.~Koposov for \texttt{RVSpecFit} \citep{koposov19b, koposov26a}.
The key idea is that $\mathbf{R}$ samples a smooth underlying
resolution function, which we can represent with Gaussian radial basis
function (RBF) interpolation. Rather than sampling $\phi(\lambda)$ at
pixel centers, we pre-convolve our emission-line model with a fixed
Gaussian kernel of width $\sigma_G = 0.5$\,\AA---narrow compared to
the DESI resolution but wide enough to smooth the sub-pixel structure
of the model---and propagate the result through a correspondingly
deconvolved resolution matrix
\begin{equation}
  \mathbf{W} = \mathbf{R}\,\mathbf{M}^{-1},
  \label{eq:wmatrix}
\end{equation}
where $\mathbf{M}$ is the Gaussian RBF design matrix, $\mathbf{M}_{ij} =
G(\lambda_i - \lambda_j;\,0,\,\sigma_G)$. The predicted line model is
then $\mathbf{m} \approx \mathbf{W}\,\mathbf{F}$, where $\mathbf{F}$
is the $\sigma_G$-pre-convolved profile evaluated at pixel
centers. Because each $F_i$ is a weighted integral of $\phi(\lambda)$
rather than a point sample, it captures the correct line area
regardless of the line's width or its phase (position) relative to the
pixel grid.

Formally, $F_i = (\phi * G_{\sigma_G})(\lambda_i)$ is the convolution
of the intrinsic profile given by equation~(\ref{eq:intrinsic}) with
the unit-area Gaussian kernel, evaluated at pixel center
$\lambda_i$. Because both $\phi$ and the kernel are exact Gaussians in
ln-$\lambda$ space, their convolution is itself Gaussian, with
variances adding in quadrature to give an effective width
\begin{equation}
  \sigma_\mathrm{eff} = \sqrt{\sigma_\ell^2 + \sigma_{G,\ell}^2},
  \label{eq:sigmaeff}
\end{equation}
\noindent where $\sigma_{G,\ell} \equiv \sigma_G/\lambda^*$. In
addition, because the kernel has unit area, the convolution also
conserves the total line flux, given by
equation~(\ref{eq:flux_phys}). However, since the profile has
broadened from $\sigma_\ell$ to $\sigma_\mathrm{eff}$, its peak
amplitude must shrink by the same factor to keep that flux fixed.  The
pre-convolved profile of a single line at pixel center $\lambda_i$
therefore becomes
\begin{equation}
  F_i = A\,\frac{\sigma_\ell}{\sigma_\mathrm{eff}}
    \exp\!\left(-\frac{(\ln\lambda_i-\ln\lambda^*)^2}{2\sigma_\mathrm{eff}^2}\right),
  \label{eq:model}
\end{equation}
directly analogous to equation~(\ref{eq:intrinsic}). In addition,
because $\sigma_\mathrm{eff} \geq \sigma_{G,\ell}$, even a formally
unresolved line ($\sigma_\ell \to 0$) is represented by a profile
spanning several pixels, so its flux is correctly captured by the
pixel grid rather than lost between samples. Collecting the
pre-convolved profiles of all $n_\mathrm{line}$ lines into
$\mathbf{F}$, the pixelized counterpart of
equation~(\ref{eq:emline_sum}) finally becomes
\begin{equation}
  \mathbf{E} = \mathbf{W}\,\mathbf{F}.
  \label{eq:emline_pixel}
\end{equation}

With this pixelized model in hand, we determine the parameters
$\theta_E$ of every line by minimizing
\begin{equation}
  \chi^2(\theta_E) = \sum_{i=1}^{n_\mathrm{pix}} I_i\left(r_i - E_i\right)^2,
  \label{eq:chi2line}
\end{equation}
using the trust-region reflective algorithm
\citep[\texttt{scipy.optimize.least\_squares};][]{branch99a}, where
$r_i$ is the residual (continuum-subtracted) spectrum, $I_i$ is the
corresponding inverse variance spectrum, $E_i$ is the emission-line
model given by equation~(\ref{eq:emline_pixel}), and $n_\mathrm{pix}$
is the number of pixels in the spectrum.

Because \fastspecfit{} has to fit tens of blended lines
simultaneously\footnote{Fun fact: at $z \approx 0.076$–$0.083$, as
many as 34 of the 46 lines in the nominal \fastspecfit{} line-list
fall simultaneously within the nominal DESI wavelength range
(3600–9824~\AA).}, this minimization involves many free parameters,
and finite-difference gradients would require $\mathcal{O}(3\times
n_\mathrm{line})$ additional forward-model evaluations per
iteration. We instead speed up the minimization by supplying
\texttt{least\_squares} with an analytic Jacobian: closed-form first
derivatives of equation~(\ref{eq:model}) with respect to each free
parameter, reducing this cost to a single forward-model call per
iteration.

First, because $F_i$ is linear in $A$, the amplitude gradient is
simply
\begin{equation}
  \frac{\partial F_i}{\partial A} = \frac{F_i}{A}.
  \label{eq:dF_dA}
\end{equation}
Writing $x_i \equiv \ln\lambda_i - \ln\lambda^*$ for brevity, the
velocity-shift gradient follows from the dependence of
$\ln\lambda^* = \ln[\lambda_\mathrm{rest}(1 + z + \Delta v/c)]$ on
$\Delta v$:
\begin{equation}
  \frac{\partial F_i}{\partial \Delta v}
  =
  \frac{F_i\,x_i}{\sigma_\mathrm{eff}^2\,c\,(1+z+\Delta v/c)}.
  \label{eq:dF_dv}
\end{equation}
And finally, the line-width gradient, written in a form that remains
finite as $\sigma_\ell\to0$, accounts for the dependence of both the
amplitude factor $\sigma_\ell/\sigma_\mathrm{eff}$ and the exponent on
$\sigma_\ell = \sigma/c$:
\begin{equation}
  \frac{\partial F_i}{\partial \sigma} =
  \frac{A}{\sigma_\mathrm{eff}\,c}\,
  \exp\!\left(-\frac{x_i^2}{2\sigma_\mathrm{eff}^2} \right) \, \left[1
    - \frac{\sigma_\ell^2}{\sigma_\mathrm{eff}^2} \left(1 -
    \frac{x_i^2}{\sigma_\mathrm{eff}^2} \right)\right].
  \label{eq:dF_ds}
\end{equation}

The parameters $\theta_E$ recovered from this minimization are the
amplitude, velocity shift, and width of every individual line. In the
next section (\S\ref{sssec:kinematic}) we describe how we share
$\Delta v$ and $\sigma$ across physically motivated kinematic groups
to reduce this parameter count.

\subsubsection{Kinematic Groups and Broad-Line Decomposition}
\label{sssec:kinematic}

\fastspecfit{} reduces the dimensionality of its emission-line fitting
(\S\ref{sssec:emline_model}) by grouping lines into kinematic groups,
within which all members share a common velocity shift $\Delta v$ and
intrinsic velocity width $\sigma$. All kinematic groups, bounds, and
amplitude constraints are loaded at startup from the bundled
configuration file \texttt{data/emline-constraints.yaml}, as
summarized in Table~\ref{table:kinematic}. This parameter file also
defines two important kinematic models: an emission line model
consisting entirely of narrow lines (\emph{narrow-only}) and a model
with both narrow and broad lines (\emph{narrow+broad}).

\input{table-kinematic-bounds}

The narrow-only model ties all forbidden lines to one group anchored
to \oiiilam{} and a second group consisting of all \ion{H}{1} and
\hei{} lines anchored to \ha. Separately, the narrow+broad model
assigns the \oiiidoublet{} doublet to its own group anchored on
\oiiilam; the remaining forbidden, Balmer, and helium lines to a
second group anchored on narrow \ha; and the broad Balmer lines to a
third group anchored on broad \ha. The \oiii{} doublet gets its own
kinematic group because of the frequency with which this doublet shows
an asymmetric or complex line-profile in AGN, which can bias the
fitting of the fainter forbidden lines. Regardless, in both
  models, the lines in the global groups of
  Table~\ref{table:kinematic} have a single component, with no
  narrow+broad decomposition.

Within each kinematic group, select emission-line doublets carry an
additional amplitude constraint, set in the same constraints file as
either fixed or free. In the default configuration, the amplitude
ratios of \nii~$\lambda$6548\,/\,$\lambda$6584,
\oiii~$\lambda$4959\,/\,$\lambda$5007, and
\oii~$\lambda$7330\,/\,$\lambda$7320 are fixed to their theoretical
spontaneous-transition-probability value \citep{osterbrock89a}, while
\oii~$\lambda$3726\,/\,$\lambda$3729,
\sii~$\lambda$6716\,/\,$\lambda$6731, and
\mgii~$\lambda$2796\,/\,$\lambda$2803, carry a free (but bounded)
amplitude ratio. In the fitting engine, we parameterize each doublet
as a reference-line amplitude and a bounded (or fixed) amplitude
ratio, which lets us keep each ratio
  within its bounds during the fit. Moreover, for \oiidoublet{} and
\siidoublet{} the fitted ratio, \texttt{OII\_DOUBLET\_RATIO} and
\texttt{SII\_DOUBLET\_RATIO}, respectively, carries physical
significance as tracers of the local electron density
\citep{osterbrock89a}. In addition, this parameterization often lets
\fastspecfit{} recover a faint doublet component in noisy spectral
regions, where an independent-amplitude fit might otherwise drive it
to an unphysical part of parameter space.

\fastspecfit{} fits the narrow-only model first and then tests the
narrow+broad model only when broad-line fitting is enabled (true by
default, but see \S\ref{sec:modes}) and when at least one of the first
three Balmer lines (\ha, \hb, or \hg) falls within the observed
spectral range. When both models are fit, \fastspecfit{} selects the
preferred solution based on four conditions that must all be
satisfied:
\begin{enumerate}

\item{The change in $\chi^2$ over the pixels spanning the broad Balmer
  lines must exceed the corresponding reduction in the number of
  degrees of freedom,
\begin{equation}
  \Delta\chi^2_\mathrm{Balmer} \;>\; \Delta N_\mathrm{dof,\,Balmer},
  \label{eq:broadchi2}
\end{equation}
where $\Delta\chi^2 = \chi^2_\mathrm{narrow} - \chi^2_\mathrm{broad}$
and $\Delta N_\mathrm{dof} = N_\mathrm{dof,\,narrow} -
N_\mathrm{dof,\,broad}$, equal to the number of additional free
parameters introduced by the broad-Balmer group. In other words, the
data must justify the additional free parameters in a
maximum-likelihood sense. This criterion is only evaluated using
pixels within $\pm 2.5\sigma_\mathrm{broad}$ of the non-helium broad
Balmer lines.}
\item{The fitted velocity dispersion of the broad \ha{} component must
  exceed $\sigma_\mathrm{broad}(\ha) > 250$~\kms, in order to exclude
  spurious broad fits driven by noise or unmodeled continuum
  residuals.}
\item{The broad component must be wider than the narrow component,
  $\sigma_\mathrm{broad}(\ha) > \sigma_\mathrm{narrow}(\ha)$, ensuring
  a physical hierarchy between the two kinematic components.}
\item{The amplitude of at least one of the first two broad Balmer
  lines in the spectral range (e.g., \ha{} and \hb{} or, at higher
  redshift, \hb{} and \hg) must satisfy $\mathrm{S/N}_\mathrm{broad} >
  2.5$, where we estimate the noise robustly from the interquartile
  range of the fit residuals in the Balmer-line pixels after
  subtracting the full emission-line model.}
\end{enumerate}
If any condition fails, we retain the narrow-only result, but always
record the pixel-level $\chi^2$ comparison in the output columns
\texttt{DELTA\_LINECHI2} and \texttt{DELTA\_LINENDOF}.

Once a model has been selected, \fastspecfit{} runs an optional final
relaxation pass that frees the velocity shifts and/or line-widths of
individual lines. However, we only adopt this solution if there is a
statistically significant improvement in the fit given the increased
number of degrees of freedom, $\Delta\chi^2 > \Delta
N_\mathrm{dof}$. We record the relevant comparison parameters in the
output file as \texttt{DELTA\_KINECHI2} and \texttt{DELTA\_KINENDOF}.

In the default configuration (Table~\ref{table:kinematic}), we only
relax the final velocity shifts of lines in the narrow-only model,
allowing them to change by no more than $\pm100$~\kms{} around their
best-fit value, but keeping the line-widths constrained within each
kinematic group. This final round of flexibility accommodates minor
wavelength-calibration residuals (\S\ref{ssec:diagnostics}) and is
especially important when fitting star-forming galaxies with extremely
narrow and strong line-emission (J~Moustakas et~al., in
preparation). Doublet-tied pairs (e.g., \niidoublet) always retain a
shared velocity shift during this final optimization round, to prevent
unphysical differences between emission lines arising from the same
atomic species, and by default, we do not enable a final relaxation
pass for the three-group narrow+broad model (to prevent overfitting).

\subsubsection{Summary of Measured Emission-Line Properties}
\label{sssec:lines}

For every line in our list (Table~\ref{table:emlines}), whether
formally detected or not, we tabulate a standard set of derived
quantities in the output catalog; Table~\ref{table:linesummary}
summarizes their symbols, output-column names, and units.

We infer $A$, $\Delta v$, and $\sigma$ by optimizing the
pre-convolution Gaussian parameters given by
equation~(\ref{eq:model}), as well as $\chi^2_\mathrm{line}$ for the
fit and the number of unmasked pixels $n_\mathrm{pix}^\mathrm{line}$
contributing to it (both within $\pm3\sigma$ of the line-center). We
additionally measure the observed amplitude of each line,
$A_\mathrm{obs}$, the peak flux density of the fully pixelized,
resolution-matrix-convolved model $\mathbf{E}=\mathbf{W}\mathbf{F}$
(equation~\ref{eq:emline_pixel}) as it actually appears on the DESI
pixel grid. In general (but especially for lines narrower than one
pixel), the pre-convolution of
\S\ref{sssec:emline_model} means $A$ can substantially exceed
$A_\mathrm{obs}$; $A_\mathrm{obs}$, not $A$, is what one would measure
by eye in the spectrum and is therefore one of the quantities we can
use to determine the statistical significance of a line
(\S\ref{ssec:cuts}).

\input{table-linesummary}

Next, we compute the Gaussian-integrated flux $\Phi$ analytically from
the fitted parameters using equation~(\ref{eq:flux_phys}), again,
independent of the pixel grid or resolution matrix. As a
model-independent check on $\Phi$, we also derive the
boxcar-integrated flux $\Phi_\mathrm{box}$ by directly summing the
continuum-subtracted spectrum
\begin{equation}
  \Phi_\mathrm{box} = \sum_{i\,\in\,\mathrm{patch}} r_i\,\Delta\lambda_i,
  \label{eq:boxflux}
\end{equation}
over the unmasked pixels within $\pm3\sigma$ of the line center
(requiring at least seven pixels, and falling back to a default width
if no measurement of $\sigma$ is available for that line). Naturally,
$\Phi_\mathrm{box}$ is a biased estimate of the emission-line flux for
blended or overlapping lines (e.g., \oiidoublet) and also suffers
from the subpixel bias for very narrow lines discussed in
\S\ref{sssec:emline_model}.

For each line, we estimate the local continuum flux density $F_c$, by
computing the iteratively sigma-clipped mean of the observed spectrum,
with the best-fitting emission-line model subtracted, over two
sideband windows spanning 3--10$\sigma$ on either side of the
line-center (again requiring at least seven pixels per side). This
local, data-driven estimate is more robust to continuum-model mismatch
at the line's exact wavelength than simply interpolating $C(\lambda)$
would be. The rest-frame equivalent width (EW) then follows from
\begin{equation}
  \mathrm{EW} = \frac{\Phi}{F_c\,(1+z)}.
  \label{eq:ew}
\end{equation}

We report the $1\sigma$ upper limit on the flux of every line with a
valid continuum measurement by substituting the uncertainty in the
local continuum for the amplitude in equation~(\ref{eq:flux_phys}),
\begin{equation}
  \Phi_\mathrm{lim} = \sqrt{2\pi} \left(
  \frac{\lambda^*}{\sqrt{I_{F_c}}} \right) \frac{\sigma}{c},
  \label{eq:fluxlimit}
\end{equation}
where $I_{F_c}$ is the inverse variance of $F_c$ (discussed in
\S\ref{ssec:uncertainties}). However, because $I_{F_c}$ is the
  uncertainty on the mean continuum in the sidebands (not at the
  wavelength of the line), $\Phi_\mathrm{lim}$ is only an approximate
  flux limit. For lines that are dropped during fitting, we use a
nominal $\sigma$ of 75~\kms{} for narrow lines and 1200~\kms{} for
broad lines, so that we can still report a limit for every line in the
spectrum. We compute $\Phi_\mathrm{lim}$ as another possible metric
for determining the statistical significance of a line
(\S\ref{ssec:cuts}), and so that users can more easily fold
upper limits into their specific scientific analyses.

Separately, for four lines—\ion{C}{4}~$\lambda1549$, \mgiidoublet\ (summed
as one blended feature), \hb, and \oiiilam—we also measure three
non-parametric moments of the line profile directly from the residual
spectrum, without assuming any particular functional form. Departures
from a Gaussian profile, particularly asymmetric wings, can signal
outflows or other physical processes, most often associated with AGN;
these moments provide a model-independent starting point for a more
detailed follow-up analysis of such profiles.

We sum over pixels within $\pm5\sigma$ of the line center $\lambda^*$
(equation~\ref{eq:lstar}), using the same fallback width adopted above
when $\sigma$ is unmeasured (75~\kms{} for narrow lines, 1200~\kms{} for
broad), and requiring a window at least two pixels wide. Writing
$\lambda_i$ and $r_i$ for the observed wavelength and residual flux of
the $i$th pixel in that window, the first moment is the flux-weighted
mean wavelength,
\begin{equation}
  M_1 = \frac{\sum_i \lambda_i\,r_i}{\sum_i r_i},
  \label{eq:moment1}
\end{equation}
and the $n$th central moment is
\begin{equation}
  M_n = \frac{\sum_i (\lambda_i-M_1)^n\,r_i}{\sum_i r_i};
  \label{eq:momentn}
\end{equation}
we evaluate equation~(\ref{eq:momentn}) for the first two non-trivial
orders, $M_2$ and $M_3$. We tabulate $M_1$, $M_2$, and $M_3$ (in
observed-frame \AA, \AA$^2$, and \AA$^3$, respectively), together with
their Monte Carlo-derived inverse variances, as \texttt{MOMENT1},
\texttt{MOMENT2}, and \texttt{MOMENT3} columns prefixed by each line's
name (e.g., \texttt{OIII\_5007\_MOMENT1}).

In \S\ref{ssec:uncertainties} we describe how we derive
the companion inverse-variance columns listed in
Table~\ref{table:linesummary} via a per-object Monte Carlo procedure.

\subsection{Spectrophotometric and Physical Properties}
\label{ssec:quantities}

Because physical SPS templates underpin our best-fitting stellar
continuum model $C(\lambda;\theta_C)$, we can compute several
secondary, albeit scientifically high-impact physical and
spectrophotometric quantities. First, recall from
equation~(\ref{eq:fnorm}) that each template coefficient $c_n$ carries
units of \msun{} of stars formed in age bin $n$, rescaled by the mass
normalization $\mathcal{M}_0$ introduced in
equation~(\ref{eq:zfactor}). The current stellar mass (in
$h^{-2}\,\msun$), therefore, is simply
\begin{equation}
  \Mstar = \frac{\mathcal{M}_0}{h^2}\sum_{n=1}^{n_\mathrm{template}} c_n\,f_{\star,n},
  \label{eq:mstar}
\end{equation}

\noindent where $f_{\star,n}\in(0,1]$ is the fraction of formed mass
  that
    remains in stars and stellar remnants after mass loss from stellar
    winds and supernovae (taken directly
  from FSPS). The mean star-formation rate (SFR) over the last 100~Myr
  also follows by summing the mass formed within our first two age
  bins (0--30~Myr and 30--100~Myr; \S\ref{ssec:templates}), and
  dividing by 100~Myr,
\begin{equation}
  \mathrm{SFR} = \frac{\mathcal{M}_0}{h^{2}(10^{8}~\mathrm{yr})}\sum_{n=1}^{2} c_n,
  \label{eq:sfr}
\end{equation}
\noindent in units of $h^{-2}\,\msun$/yr.
Additional estimates of the SFR of course can be inferred from various
other tracers, such as the \ha, \oii, or UV luminosity (where they
have been reliably measured and with appropriate corrections for dust
attenuation and other effects; \citealt{moustakas06b, kennicutt98a,
  hao11a}), but \fastspecfit{} does not directly report those
quantities. Finally, we compute the mean stellar age as the
coefficient-weighted average of the age-bin midpoints $t_n$ (in Gyr),
\begin{equation}
  \langle t_\star\rangle = \frac{\sum_{n=1}^{n_\mathrm{template}}
    c_n\,t_n}{\sum_{n=1}^{n_\mathrm{template}} c_n}.
  \label{eq:age}
\end{equation}
Because the coefficients $c_n$ represent formed stellar mass rather
than surviving mass or luminosity, $\langle t_\star\rangle$
weights
older stellar populations more heavily than either a light-weighted
or a surviving-mass-weighted age, and so is at least as old as both.

Next, we compute K-corrections and rest-frame absolute magnitudes by
synthesizing photometry from the best-fitting continuum model
$C(\lambda)$ (including nebular emission; \S\ref{ssec:templates}), in
the observed and rest frames, following the standard formalism of
\citet{hogg02a} and \citet{blanton03a}. For each output bandpass we
identify the observed-frame Legacy Surveys or WISE filter whose
rest-frame effective wavelength lies closest to the target band,
restricted to filters detected at $\mathrm{S/N}>2$. This
nearest-bandpass matching minimizes the magnitude of the K-correction
at the cost of an expected discontinuity in the derived K-corrections
as a function of redshift wherever the nearest-matching observed
filter switches from one bandpass to the next.
When no bandpass satisfies the S/N threshold, we instead report the
absolute magnitude synthesized purely from the model, with its inverse
variance set to zero.

The absolute magnitude follows the standard relation \citep{blanton07a},
\begin{equation}
  m_R = M_Q + \mathrm{DM}(z) + K_{QR}(z) - 5\log_{10} h,
  \label{eq:kcorr}
\end{equation}
where $m_R$ is the observed-frame apparent magnitude in bandpass $R$
(corrected for Galactic extinction), $M_Q$ is the absolute magnitude
in the rest-frame target bandpass $Q$, $\mathrm{DM}(z)$ is the
distance modulus, and $K_{QR}(z) =
2.5\log_{10}(f_\mathrm{rest}^\mathrm{synth}/f_\mathrm{obs}^\mathrm{synth})$
is the K-correction from bandpass $Q$ to its nearest-matching observed
bandpass $R$. To build band-shifted rest-frame bandpasses
\citep{blanton03a} we blueshift the filter response itself rather than
the model,
\begin{equation}
  R_Q^{z_0}(\lambda) \equiv R_Q\bigl((1+z_0)\,\lambda\bigr),
  \label{eq:bandshift}
\end{equation}
so that convolving $R_Q^{z_0}$ with the rest-frame SED reproduces what
$R_Q$ would measure on the same galaxy at redshift
$z_0$. Table~\ref{table:absmag-bands} reports the default set of
rest-frame bandpasses and band-shifts computed by \fastspecfit{} via
the bundled \texttt{data/legacysurvey-dr9.yaml} parameter file, as
well as the
\texttt{speclite}
filter curve names and their respective rest-frame wavelengths.

\input{table-absmag-bands}

In addition to rest-frame photometry, we also measure several
rest-frame monochromatic luminosities and observed-frame continuum
flux densities using the best-fitting continuum-only model,
$C(\lambda)$, which excludes nebular emission. We report luminosities
at 1450, 1700, 3000, and 5100~\AA{} in solar units ($\lambda
L_\lambda$), and at 1500 and 2800~\AA{} in
$\mathrm{erg\,s^{-1}\,Hz^{-1}}$ ($L_\nu$), the latter chosen to
facilitate UV-continuum SFR estimates (e.g.,
\citealt{kennicutt98a}). We similarly report the observed-frame
continuum flux density (in \fluxdensity) at the rest-frame centers of
\lya, \oii, \hb, \oiiilam, and \ha, which can be used to infer the
total line-luminosity from the corresponding emission-line EW
\citep[e.g.,][]{zhu09a}. In all cases, we measure these quantities
directly from $C(\lambda)$, using a robust (sigma-clipped) linear fit
to the continuum within $\pm100$~\AA{} of each reference wavelength.

Finally, we measure three different flavors of the 4000~\AA{} break,
$D_n(4000)$ \citep{balogh99a}: \texttt{DN4000\_OBS}, computed directly
from the observed spectrum, without any correction for emission-line
contamination; \texttt{DN4000\_MODEL}, from the best-fitting stellar
continuum model (excluding the smooth continuum correction;
\S\ref{sssec:smooth}); and our fiducial \texttt{DN4000}, from the
observed spectrum after subtracting the best-fitting emission-line
model.

\subsection{Parameter Uncertainties via Monte Carlo}
\label{ssec:uncertainties}

\fastspecfit{}'s forward model chains together several nonlinear
stages: an initial grid search for \vdisp{} (\S\ref{sssec:vdisp}), the
VARPRO solve for \tauv{} and $\mathbf{c}$ (\S\ref{sssec:fitting}), and
the nonlinear emission-line kinematics (\S\ref{sssec:kinematic}).
Analytic error propagation through every downstream quantity (stellar
mass, K-corrections, EWs, etc.) is therefore impractical. Monte Carlo
resampling instead offers a fast, general way to sample the
multidimensional covariance among all fitted and derived parameters.
By perturbing the input data and refitting, each nonlinear stage of
the pipeline propagates its own uncertainty naturally, without having
to derive a bespoke error-propagation formula for every derived
quantity.

By default, \fastspecfit{} generates $n_\mathrm{monte}=50$ independent
realizations of the observed spectrum and photometry by adding
Gaussian noise consistent with the (floor-inflated;
\S\ref{sec:modeling}) inverse variance,
\begin{equation}
  d_i^{(m)} \sim \mathcal{N}\left(d_i,\,I_i^{-1}\right),
  \label{eq:mc_draw}
\end{equation}
for realizations $m=1,\ldots,n_\mathrm{monte}$, where $d_i$ indexes
both spectral pixels and photometric bandpasses. We generate these
draws once, using a reproducible per-object seed recorded in the
\texttt{SEED} output column. This seed is itself derived
deterministically from a single global run seed. We reuse the same
$n_\mathrm{monte}$ realizations, unchanged, throughout the rest of the
pipeline, so the same draws flow through the velocity-dispersion,
continuum, and emission-line fits described below. The inverse
variance of any fitted or derived quantity $\theta$ then follows from
the scatter across realizations,
\begin{equation}
  \sigma_\theta^{-2}
  = \bigl[\mathrm{Var}(\theta^{(1)},\ldots,\theta^{(n_\mathrm{monte})})\bigr]^{-1}.
  \label{eq:mc_unc}
\end{equation}
Throughout the pipeline, we report uncertainties as inverse variances;
for parameters that are not measured for a given object, or that are
held fixed at a nominal value, we assign \texttt{IVAR}=0 rather than a
Monte Carlo estimate.

On the continuum side, the Monte Carlo realizations enter in two
places. First, after determining \vdisp{} (\S\ref{sssec:vdisp}), we
refit \tauv, \vdisp, and $\mathbf{c}$ jointly on each realization to
obtain \texttt{VDISP\_IVAR}. Second, with \vdisp{} fixed at its
adopted value, we repeat the full VARPRO solve---the outer \tauv{}
minimization and inner NNLS solve for $\mathbf{c}$---on each
realization, yielding $n_\mathrm{monte}$ independent estimates
$\tauv^{(m)}$ and $\mathbf{c}^{(m)}$ that directly give
\texttt{TAUV\_IVAR}. Because every stellar-population and photometric
quantity in \S\ref{ssec:quantities} is a fixed function of
$\mathbf{c}$ (equations~\ref{eq:mstar}--\ref{eq:bandshift}), we
re-evaluate those relations on each $\mathbf{c}^{(m)}$ and propagate
the resulting scatter to determine \texttt{LOGMSTAR\_IVAR},
\texttt{SFR\_IVAR}, \texttt{AGE\_IVAR}, \texttt{DN4000\_MODEL\_IVAR},
the \texttt{ABSMAG*\_SYNTH\_IVAR}, and the rest-frame luminosity and
continuum-flux inverse variances.

The emission-line Monte Carlo is deliberately cheaper. The full
kinematic model-selection procedure (\S\ref{sssec:kinematic}) chooses
between narrow-only and narrow+broad models and optionally relaxes
velocity-width bounds; repeating it for every Monte Carlo realization
would be expensive. Instead, we fix the kinematic model and
velocity-width bounds at their fiducial values and refit only the line
parameters, warm-started at the fiducial best-fit solution with
loosened convergence tolerances. The assumption underpinning these
choices is that each realization is a small perturbation of the data
rather than an independent object, so the preferred kinematic model
should rarely change between realizations. We build each realization's
continuum-subtracted spectrum from the corresponding continuum-fitting
realization, $d^{(m)} - C^{(m)}(\lambda) -
C_\mathrm{smooth}(\lambda)$, holding the smooth continuum correction
$C_\mathrm{smooth}(\lambda)$ (\S\ref{sssec:smooth}) fixed at its
fiducial value across all realizations rather than recomputing it per
draw. This procedure yields Monte Carlo uncertainties for every fitted
line parameter (\S\ref{sssec:lines} and
Table~\ref{table:linesummary}).

\subsection{Additional Fitting Modes}
\label{sec:modes}

Everything described in this section so far uses \fastspecfit{}'s
default fitting code, \fastspec, which jointly fits each object's DESI
spectrum with its corresponding optical/IR broadband photometry using
a nominal set of modeling priors. However, we designed \fastspecfit{}
with flexibility and extensibility in mind, letting users choose
different input data and model parameters while sharing a single,
speed-oriented fitting engine.

In practice, \fastspecfit{} exposes this flexibility through
command-line options that specify what data to fit. For example, the
\texttt{--ignore-photometry} option drops the broadband photometry
from the \fastspec{} fit, effectively disabling the aperture
correction (\S\ref{sssec:apercorr}). For a scientific study focused on
lower-redshift BGS targets, this option is arguably more relevant.
Other comparable options such as \texttt{--no-smooth-continuum}
disable the smooth continuum correction (\S\ref{sssec:smooth}), while
\texttt{--no-broadlinefit} skips the narrow+broad kinematic fitting
(\S\ref{sssec:kinematic}), an option that may be useful when studying,
for example, \hii-region-dominated spectra.

A second major fitting mode is \fastphot, which fits only the
broadband photometric SED with the templates convolved to a nominal
velocity dispersion, $\vdisp = 150$~\kms. As for \fastspec, in this
mode the redshift is an input quantity, taken from the DESI pipeline's
Redrock output for whichever DESI coaddition type---per-exposure,
per-night, cumulative (multi-night), or per-HEALPix---is detected
automatically from the input Redrock file \citep{guy23a,
  desi-collaboration26a}. However, the user can override the pipeline
redshifts via the \texttt{--input-redshifts} argument if externally
supplied redshifts (e.g., from visual inspection) are available.

A third fitting mode is called \texttt{stackfit}, which fits arbitrary
user-supplied stacked spectra (e.g., binned by redshift, stellar mass,
or another physical property) rather than spectra of individual DESI
targets \citep[e.g.,][]{scholte24a}. In this mode, the user is
responsible for providing a ``stacked'' (mean) resolution matrix and
for recognizing that the assumption of uncorrelated pixels no longer
holds, but at DESI scale, stacked spectra can reveal significantly
fainter spectroscopic features. With \texttt{stackfit}, the fit is
purely spectroscopic (i.e., \texttt{--ignore-photometry} is enforced).

Finally, three bundled configuration files control the physical inputs
to every fitting mode, and each can be swapped out for a custom
version. The photometric-catalog file
(\texttt{data/legacysurvey-dr9.yaml} by default; see
\S\ref{ssec:quantities}) can be overridden with \texttt{--fphotofile},
letting \fastspecfit{} fit photometry from surveys other than the
Legacy Surveys. Users can similarly override the emission-line
parameter file (\texttt{data/emlines.ecsv} by default; see
\S\ref{ssec:emlinelist}) with \texttt{--emlinesfile} to add, remove,
or retune individual lines, or the kinematic-constraints file
(\texttt{data/emline-constraints.yaml} by default; see
\S\ref{sssec:kinematic}) with \texttt{--constraintsfile} to redefine
the kinematic groups, velocity-width bounds, and amplitude ratios that
link lines during fitting. Users can even build a custom set of SPS
templates with \texttt{bin/build-templates} (\S\ref{ssec:templates})
and use them in any fitting mode via the \texttt{--templates}
command-line argument.

More details regarding these fitting modes and all the available
options can be found in the ever-evolving online documentation for
\fastspecfit.\footnote{\url{https://fastspecfit.readthedocs.io}}
Looking ahead, we describe additional fitting modes and capabilities
currently under development in \S\ref{ssec:improvements}.

\section{Illustrative Examples}
\label{sec:casestudy}

We illustrate \fastspecfit's ability to model DESI spectrophotometry
of extragalactic targets with three example fits: a low-redshift
($z\approx0.1$) BGS galaxy dominated by an evolved stellar population
but hosting a narrow-line AGN (\S\ref{ssec:example_bgs}); an ELG at $z
\approx 1$ with prominent \oiidoublet, \oiiidoublet, and higher-order
Balmer lines (\S\ref{ssec:example_elg}); and a high-redshift
($z\approx2$) QSO with broad \lya, \ion{C}{4}~$\lambda1549$, and
\mgiidoublet{} emission (\S\ref{ssec:example_qso}). Together, these
examples highlight both the dynamic range of the code and some of its
known limitations, most notably the use of purely stellar templates to
model QSO continua.

\subsection{Low-Redshift BGS Galaxy}
\label{ssec:example_bgs}

\begin{figure*}
\centering\includegraphics[width=1.0\textwidth]{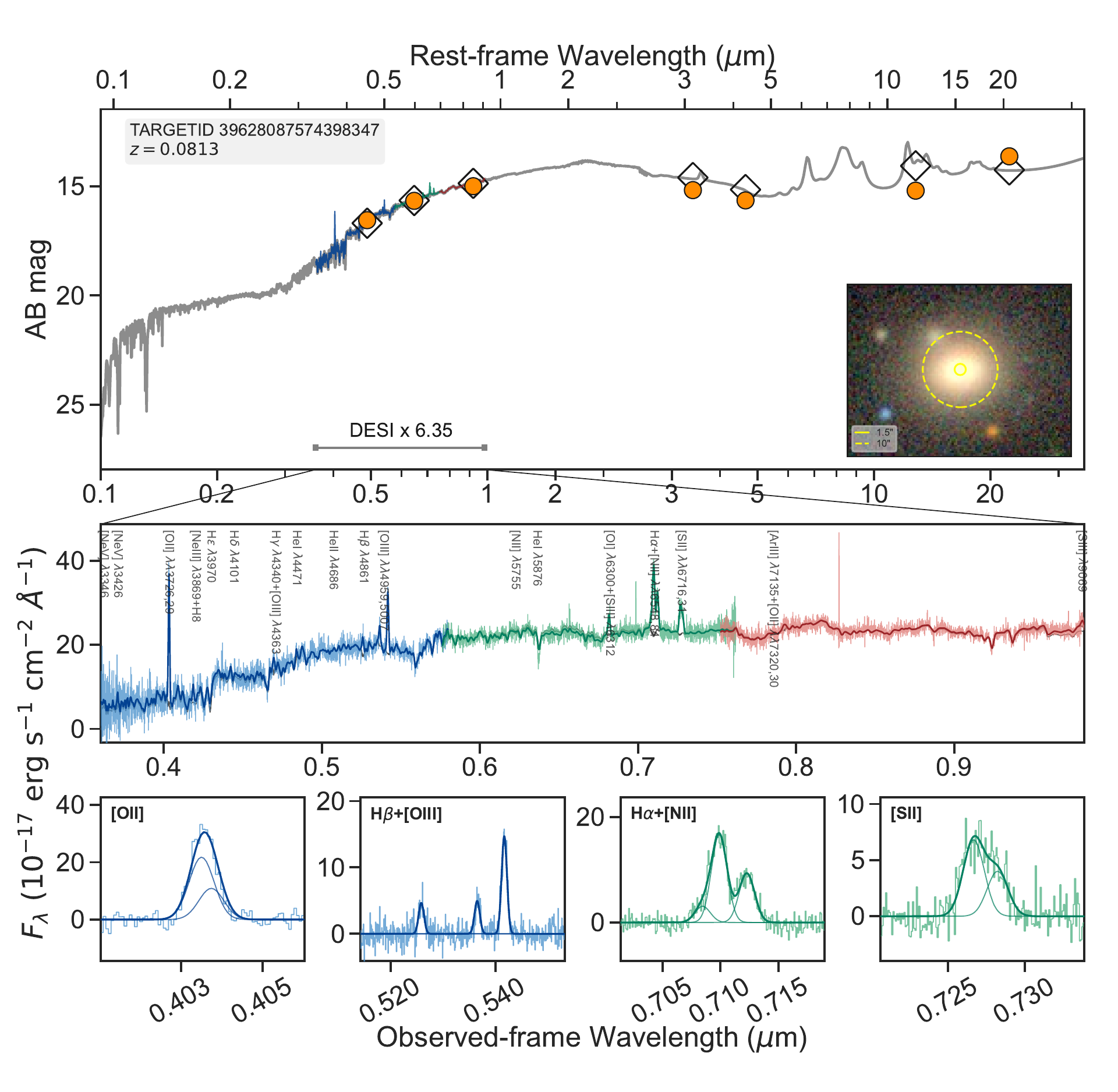}
\caption{\fastspecfit{} output for a low-redshift BGS galaxy at
  $z=0.0813$ (\texttt{TARGETID=39628087574398347}; main survey, bright
  program, HEALPix~27851). Top: broadband SED showing the observed
  Legacy Surveys $grz$ and WISE $W1$--$W4$ photometry with $1\sigma$
  uncertainties (filled orange circles; here, the uncertainties are
  smaller than the symbols), the best-fitting stellar
  continuum model (gray curve), and the synthesized model photometry
  (open diamonds). The top axis gives the rest-frame wavelength, and
  the gray bar below the photometry marks the observed DESI wavelength
  range and the median aperture correction, $\bar\eta=6.35$. The inset
  in the lower right shows the Legacy Surveys $grz$ color image with
  the $1\farcs5$-diameter DESI fiber (solid) and a $10\arcsec$
  reference aperture (dashed) overlaid. Middle: observed-frame DESI
  spectrum (light colors, one per camera) with the best-fitting
  stellar continuum (black), the stellar continuum plus
  smooth-continuum correction (dashed gray), and the total model
  including emission lines (bold, per-camera colors). Gray
  labels mark every emission line in the observed wavelength range
  included in the fit. Bottom: zooms on four representative line
  complexes---\oiidoublet, \hb+\oiiidoublet, \ha+\niidoublet, and
  \siidoublet---showing the spectrum after subtracting both the
  stellar continuum and the smooth-continuum correction, the
  individual best-fitting Gaussian components (thin curves), and the
  total emission-line model (bold curve).\label{fig:example_bgs}}
\end{figure*}

Figure~\ref{fig:example_bgs} shows the DESI spectrum and Legacy
Surveys color-cutout of a low-redshift ($z=0.0813$) BGS target
(\texttt{TARGETID=39628087574398347}) with clear line-emission and a
stellar continuum dominated by an old, evolved stellar population. The
Tractor fit to the Legacy Surveys imaging (\S\ref{ssec:photometry})
yields an apparent (Galactic extinction-corrected) $r$-band magnitude
of 15.67~mag, an $n=3.70$ \citet{sersic68a} morphological type, and a
half-light semi-major axis of $3\farcs89$. Given this galaxy's angular
extent and the $1\farcs5$ DESI fiber, we derive a median aperture
correction of $\bar\eta=6.35$, ranging between $5.82$ and $7.05$
across the three $grz$ bands (\S\ref{sssec:apercorr}).
An aperture correction this large is not surprising given the galaxy's
redshift and angular size, but it is several times larger than the
typical value for BGS targets.

From the fit to the stellar continuum we derive a stellar velocity
dispersion of $\vdisp=287\pm11$~\kms, and from the joint
spectrophotometric fit, including the infrared WISE bands, we infer a
luminosity and rest-frame color of $M_{0.1r}=-21.14$ and
$^{0.1}(g-r)=0.944$, respectively, a stellar mass of
$\log_{10}(M_\star/M_\odot)=11.24$, and an approximate age of $\langle
t_\star\rangle=12.7$~Gyr, all of which are consistent with the galaxy's optical
morphology \citep{blanton09a}. Formally, we also derive a $V$-band
optical depth of $\tauv=0.73$; however, this surprisingly large dust
attenuation value most likely reflects the age-dust-metallicity
degeneracy rather than indicating a truly dusty massive galaxy. Our
relatively simple SPS modeling choices---restricted set of model ages,
fixed solar metallicity, simple IR spectrum,
etc. (\S\ref{ssec:templates})---can be susceptible to these kinds of
degeneracies for certain classes of galaxies, so the SPS properties
inferred by \fastspecfit{} should be interpreted with care.

The fit to the emission-line spectrum builds out the rest of the
physical picture for this galaxy. Despite a relatively quiescent
stellar continuum, we detect multiple forbidden and permitted
recombination lines at high statistical significance, including \ha,
\hb, \hg, \oiidoublet, \oiiidoublet, \niidoublet, and \siidoublet. We
find no evidence for a broad Balmer-line component and derive average
narrow-line kinematics of $\sigma_\mathrm{narrow}=281\pm2$~\kms{} and
$\Delta v_\mathrm{narrow}=+76\pm30$~\kms{}
(\S\ref{sssec:kinematic}). Not surprisingly for this type of object,
the $\log_{10}($\oiiilam$/\hb)=+0.53$ and
$\log_{10}($\niilam$/\ha)=-0.25$ line-ratios place this galaxy clearly
into the type~2 Seyfert/AGN region of the Baldwin, Phillips, and
Terlevich \citep[BPT;][]{baldwin81a} diagram (\S\ref{ssec:reach}).

\subsection{Intermediate-Redshift Emission-Line Galaxy}
\label{ssec:example_elg}

\begin{figure*}
\centering\includegraphics[width=1.0\textwidth]{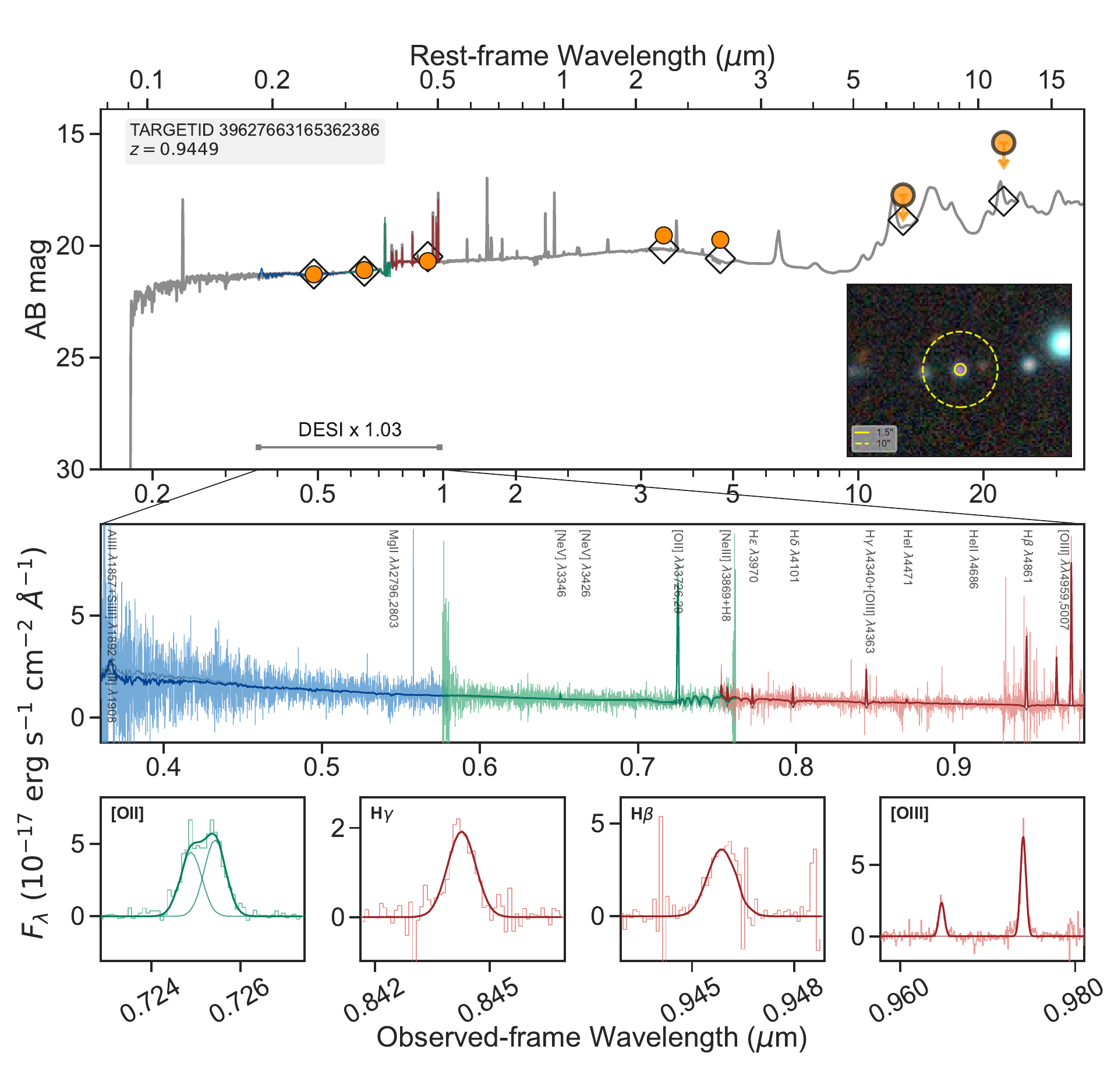}
\caption{Same as Figure~\ref{fig:example_bgs}, but for an
  intermediate-redshift ELG at $z=0.9449$
  (\texttt{TARGETID=39627663165362386}; main survey, dark program,
  HEALPix~29973) with a median aperture correction of
  $\bar\eta=1.03$. In the top panel, open circles with downward carets
  indicate broadband non-detections (upper limits). The bottom row
  zooms on \oiidoublet, \hg, \hb, and
  \oiiidoublet.\label{fig:example_elg}}
\end{figure*}

Figure~\ref{fig:example_elg} shows the broadband photometric SED,
Legacy Surveys color image, and DESI spectrum of an example
intermediate-redshift ($z=0.9449$) ELG target
(\texttt{TARGETID=39627663165362386}). This galaxy is considerably
fainter than the BGS target shown in Figure~\ref{fig:example_bgs}; its
Galactic extinction corrected $r$-band magnitude is 21.09~mag,
although it remains well-detected all the way out to
$\approx5$~\micron{} in the two shortest WISE bands. The best-fitting
Tractor morphology is a point-spread function (\texttt{PSF}); in other
words, the galaxy is unresolved in the Legacy Surveys imaging. Not
unexpectedly, the multiplicative aperture correction is small, ranging
between 1.013 and 1.033 with a median value of $\bar\eta=1.026$
(\S\ref{sssec:apercorr}).

Although the stellar continuum is only marginally detected
($\mathrm{S/N}\approx2.5$--$3.7$~pixel$^{-1}$ over the three DESI
cameras), its shape is consistent with that of a blue, star-forming
galaxy. The best-fitting \fastspecfit{} model yields
$M_{0.1r}=-21.59$, $^{0.1}(g-r)=0.053$,
$\log_{10}(M_\star/M_\odot)=9.51\pm0.21$, and $\langle
t_\star\rangle=0.06\pm3.9$~Gyr. We also infer an SED-based SFR of
$39\pm1$~\msun{}~yr$^{-1}$ and a dust attenuation value of
$\tauv=0.37\pm0.02$. All these properties are broadly consistent with
the optical morphology and color of the galaxy, and are fairly typical
for DESI ELG targets.

Despite the relatively low continuum S/N, all the strongest nebular
emission lines lying within the DESI spectral range are well-detected,
including \oiidoublet, \oiiidoublet, \hb, and \hg, as shown in the
bottom row of Figure~\ref{fig:example_elg}. From the emission-line
fits, we measure a mean narrow-line width and velocity center of
$\sigma_\mathrm{narrow}=101\pm5$~\kms{} and $\Delta
v_\mathrm{narrow}=-3\pm10$~\kms, respectively
(\S\ref{sssec:kinematic}) and cleanly resolve the \oiidoublet{}
doublet. The relative strength of the two components of the \oii{}
doublet depends on the collisional de-excitation rate of their upper
levels \citep{osterbrock89a}, and so the doublet ratio we measure,
$\lambda3726/\lambda3729=0.841\pm0.053$, implies an electron density
of $n_e=275^{+94}_{-89}$~cm$^{-3}$ for the star-forming gas
\citep{luridiana15a}. Taken together, the stellar mass, rest-frame
color, star-formation rate, and emission-line properties we measure
are all consistent with a young, low-mass, actively star-forming
galaxy.

\subsection{High-Redshift Quasar}
\label{ssec:example_qso}

\begin{figure*}
\centering\includegraphics[width=1.0\textwidth]{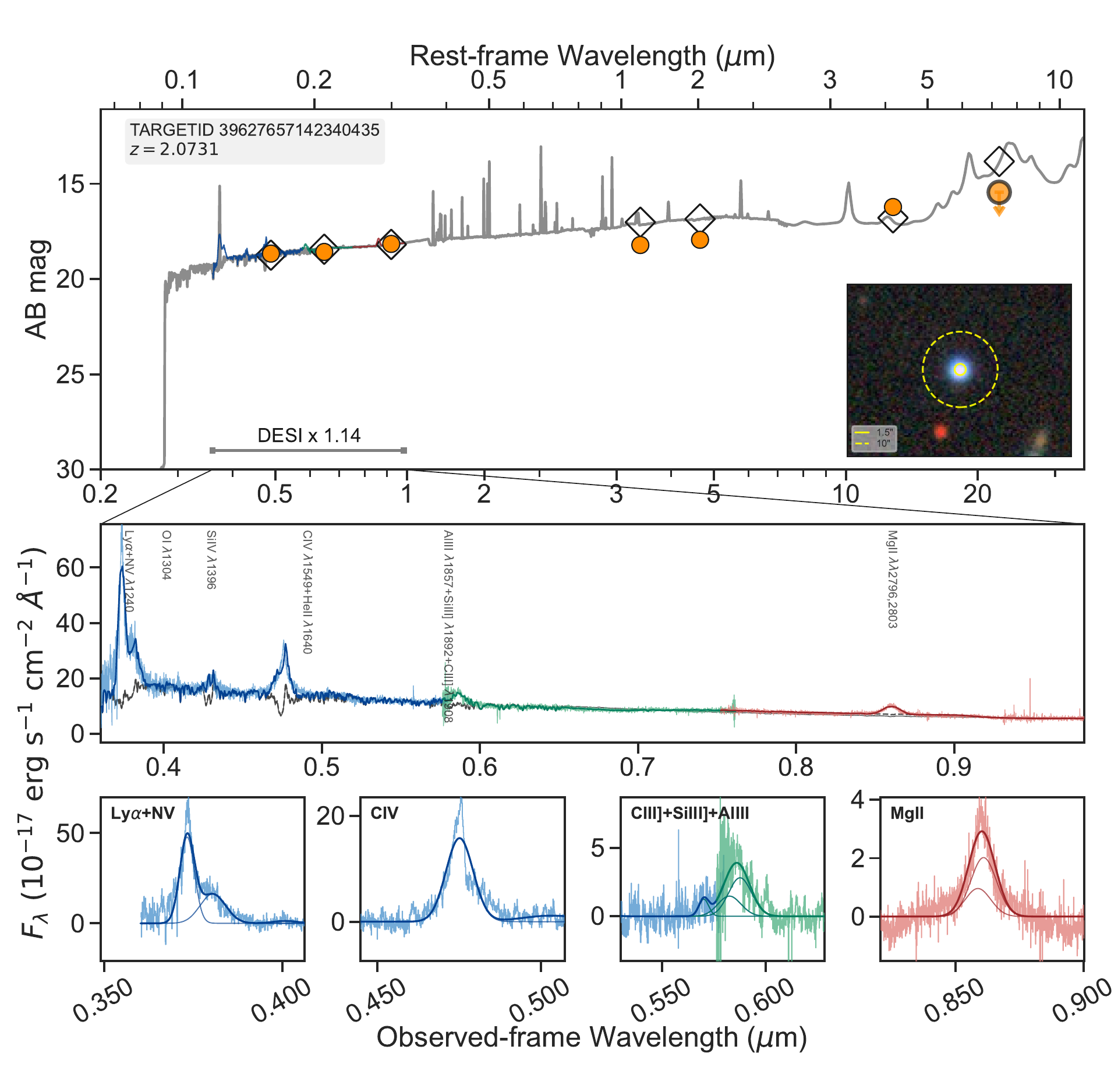}
\caption{Same as Figure~\ref{fig:example_bgs}, but for a
  high-redshift QSO at $z=2.0731$ (\texttt{TARGETID=39627657142340435};
  main survey, dark program, HEALPix~29972) with a median aperture
  correction of $\bar\eta=1.14$. As in Figure~\ref{fig:example_elg},
  open circles with downward carets in the top panel indicate broadband
  non-detections (upper limits). The bottom row zooms on four broad UV
  line complexes, \lya+\ion{N}{5}~$\lambda1240$,
  \ion{C}{4}~$\lambda1549$,
  \ion{C}{3}]~$\lambda1908$+\ion{Si}{3}]~$\lambda1892$+\ion{Al}{3}~$\lambda1857$,
  and \mgiidoublet.\label{fig:example_qso}}
\end{figure*}

Figure~\ref{fig:example_qso} shows the broadband SED, Legacy Surveys
color image, and DESI spectrum of our third and final example, a
high-redshift ($z=2.0731$) QSO target
(\texttt{TARGETID=39627657142340435}) with a Galactic
extinction-corrected $r$-band magnitude of 18.58~mag. As expected for
this class of target, the best-fitting Tractor morphology is a point
source (\texttt{PSF}) and the aperture correction is small,
$\bar\eta=1.14$ (\S\ref{sssec:apercorr}). At this redshift, the DESI
spectrum spans rest-frame wavelengths of $\approx1170$--$3200$~\AA,
revealing a blue, power-law continuum detected at high S/N
($\mathrm{S/N}\approx17$--$21$~pixel$^{-1}$ over the three DESI
cameras). Superposed on this continuum are several prominent, broad UV
emission-line complexes, including \lya+\ion{N}{5}~$\lambda1240$,
\ion{C}{4}~$\lambda1549$,
\ion{C}{3}]~$\lambda1908$+\ion{Si}{3}]~$\lambda1892$+\ion{Al}{3}~$\lambda1857$,
    and \mgiidoublet.

Although the fit to the stellar continuum is formally well-behaved, it
is not physically meaningful for this object. Our SPS templates are
purely stellar (\S\ref{ssec:templates}), so they cannot represent the
non-thermal power-law emission from the accretion disk, the
\ion{Fe}{2} pseudo-continuum in the rest-frame UV
(\S\ref{sssec:smooth}), or the hot-dust emission from the torus that
typically dominates the mid-IR SED of QSOs \citep{nenkova08a,
  nenkova08b}. Consequently, the continuum-derived physical properties
reported by \fastspec{} (stellar mass, SFR, age, dust attenuation,
etc.) should not be used for QSOs. The K-corrections and rest-frame
luminosities (\S\ref{ssec:quantities}), meanwhile, are less sensitive
to this template mismatch because they depend mainly on how well the
model reproduces the shape of the observed SED, not on its physical
interpretation. At $z\approx2$, however, the rest-frame optical bands
fall in the gap between the observed $z$ and $W1$ bands, so even these
quantities depend on how the stellar templates interpolate across this
gap and should not be used for precision work.

Because none of \ha, \hb, or \hg{} falls within the observed spectral
range, \fastspec{} does not test the narrow+broad model for this
target (\S\ref{sssec:kinematic}). Instead, it models the UV emission
lines using the global UV/AGN kinematic groups
(Table~\ref{table:kinematic}), which allow line widths up to
10,000~\kms{} and velocity shifts up to $\pm2500$~\kms{} regardless of
the kinematic model (\S\ref{sssec:kinematic}). Averaged over these
lines, we measure $\sigma_\mathrm{UV}\approx2700\pm1500$~\kms{} and
$\Delta v_\mathrm{UV}\approx-640\pm1300$~\kms, where the quoted
uncertainties represent the spread among the measured lines, not the
statistical uncertainties. These large variations in line-center (away
from the Redrock redshift) and line-width mostly reflect the complex
kinematics of the broad-line region, which affects different UV lines
in a way that depends on ionization state and QSO luminosity
\citep[e.g.,][]{shen16a}. However, the bottom row of
Figure~\ref{fig:example_qso} also shows that a single Gaussian is a
poor representation for the strong, often asymmetric UV lines,
especially \lya, whose blue side is clearly affected by absorption in
the \lya{} forest. Consequently, users who require precise UV
broad-line kinematics for DESI QSOs should interpret these values as
approximate widths and centroids rather than a complete decomposition
of the individual line-profiles. We are addressing both the continuum
and emission-line limitations illustrated by this example in a fast
QSO-fitting mode currently under development
(\S\ref{ssec:improvements}).

\section{DR2 \fastspec{} VAC}
\label{sec:vac}

In this section we present the \fastspecfit{} \texttt{Loa/v2.0} VAC,
which will be publicly released as part of DESI DR2. In DESI
vernacular, public data releases include one or more spectroscopic
productions, or ``specprods'', which consist of a well-defined input
set of spectra (usually defined by observing date range) processed
with a tagged version of the DESI pipeline
\citep{desi-collaboration26a}. DR2 includes the \texttt{Loa} specprod,
named after Mauna Loa on the island of Hawai`i (DESI specprods are
always named after mountains or mountain ranges). In production, a
tagged version of \fastspecfit{} takes the data from a specprod as its
input and produces a complete VAC. Here, we present the \texttt{v2.0}
version of the \texttt{Loa} \fastspecfit{} VAC. In
Appendix~\ref{sec:appendix_fastphot} we describe a complementary
\fastphot{} VAC produced with the same version of the code and the
same input DESI redshifts, and in Appendix~\ref{sec:appendix_vacs} we
document previously released \fastspecfit{} VACs, including those from
the EDR (\texttt{Fuji} specprod) and DR1 (\texttt{Iron} specprod).

In \S\ref{ssec:catalog} we describe the sample-selection criteria for
the \texttt{Loa/v2.0} VAC, our correction of a subset of Redrock QSO
redshifts, and the VAC's file organization. We recommend potential
quality cuts in \S\ref{ssec:cuts}, discuss the code's computational
performance at DR2 scale in \S\ref{ssec:performance}, and validate
\fastspecfit{} measurements against independent codes and catalogs in
\S\ref{ssec:validation}. Finally, in \S\ref{ssec:diagnostics} we use
\fastspecfit{} emission-line velocity-shift residuals to verify the
DESI wavelength calibration.

\subsection{Sample Selection and Catalog Production}
\label{ssec:catalog}

The \fastspecfit{} VAC includes all extragalactic objects in
\texttt{Loa} which satisfy the following minimal selection criteria:
(1) non-sky target (\texttt{OBJTYPE}~=~\texttt{TGT}; see
\citealt{myers23a}); (2) valid spectrum
(\texttt{ZWARN}~\&~\textit{NODATA}~=~0, where the \textit{NODATA} bit
($2^9$) identifies spectra in which all extracted inverse-variance values are
zero; see S.~Bailey et~al., in preparation); and (3) spectroscopic
redshift $z > 10^{-3}$ (applied after correcting a small fraction of
QSO redshifts; see below). We do not impose any other cuts on
targeting bits or other redshift-warning bits, ensuring that the VAC
includes the widest possible range of objects across every DESI target
class. And although the lower redshift limit potentially excludes a
handful of very low-redshift dwarf galaxies (V.~Manwadkar et~al., in
preparation), we impose the cut to ensure a well-defined luminosity
distance and distance modulus, which are required by our modeling
(\S\ref{sec:modeling}). For a complementary VAC of all stellar
($z\approx0$) targets in DESI, we refer the reader to
\citet{koposov26a}.

\input{table-sample}

Unfortunately, the Redrock redshift-fitting pipeline assigns an
incorrect redshift to a small but astrophysically important fraction
of QSO targets \citep{chaussidon23a, adame25e}. The default
\fastspecfit{} pipeline corrects these redshifts using the
\texttt{QuasarNet} afterburner catalog \citep{green25a}, which
  reruns Redrock with a \texttt{QuasarNet} redshift prior and is
distributed alongside each Redrock output file
\citep{desi-collaboration24d, desi-collaboration26a}.\footnote{This
option can be easily disabled by invoking the
\texttt{--ignore-quasarnet} command-line argument; see
\S\ref{sec:modes} for more options.}  For primary QSO targets and QSO
targets observed as part of the WISE variability secondary program
\citep{chaussidon23a, myers23a}, we replace the original
Redrock redshift with this result when \texttt{QuasarNet}
independently classifies the object as a QSO and when the maximum
line-confidence across six diagnostic transitions (Ly$\alpha$,
\ion{C}{4}, \ion{C}{3}, \mgii, \hb, and \ha) exceeds 0.99. For these
objects, we use the updated \texttt{Z} and \texttt{ZWARN} values and
copy the original Redrock redshift and \texttt{ZWARN} columns to
\texttt{Z\_RR} and \texttt{ZWARN\_RR},
respectively. Table~\ref{table:sample} summarizes the total number of
objects in the VAC observed as part of the DESI main, special, and SV
programs, totaling over 38~million objects, including the
approximately 165,000 targets, or $0.4\%$ of the full sample with
corrected redshifts.

We produce the \texttt{Loa/v2.0} VAC using \fastspecfit{}
version~\texttt{3.6.1} and the \texttt{ftemplates-chabrier-2.2.0.fits}
SPS template library
(\S\ref{ssec:templates}).\footnote{\url{https://fastspecfit.readthedocs.io/en/latest/loa.html}}
We run on the HEALPix-coadded DESI spectra and organize the output
catalogs identically to how the input spectra and redshift catalogs
are organized in DESI DR1 and DR2 (see \S3.3 in
\citealt{desi-collaboration26a} and DESI Collaboration et~al., in
preparation).\footnote{\url{https://fastspecfit.readthedocs.io/en/latest/vacs.html\#data-organization}}
Each \fastspec{} output file contains well over 1000 columns organized
across five FITS extensions:
\begin{itemize}
  \item \texttt{HDU0} (\texttt{PRIMARY}): empty image with processing
    metadata keywords, including the software version, Monte Carlo
    random seed, and boolean fitting flags.
  \item \texttt{HDU1} (\texttt{METADATA}): target metadata mirroring
    the DESI Redrock catalog, including sky coordinates, the input
    redshift and \texttt{ZWARN} bitmask, targeting bits, and
    extinction-corrected Legacy Survey photometry.
  \item \texttt{HDU2} (\texttt{SPECPHOT}): derived spectrophotometric
    quantities, including template coefficients, stellar-population
    parameters ($\tau_V$, $\vdisp$, age, $\logmstar$, SFR, and stellar
    metallicity; \S\ref{ssec:templates}), $D_n(4000)$, synthesized
    photometry, K-corrections, absolute magnitudes, monochromatic
    luminosities, and continuum flux densities at key emission-line
    centers.
  \item \texttt{HDU3} (\texttt{FASTSPEC}): emission-line fitting
    results, including per-camera median S/N and smooth continuum
    corrections, aperture correction factors, doublet line ratios, and
    per-line measurements (amplitude, flux, equivalent width, velocity
    shift, line width, and upper limits) for all modeled emission
    lines.
  \item \texttt{HDU4} (\texttt{MODELS}): 3D image array
    [$N_\mathrm{pix} \times 3 \times N_\mathrm{obj}$] with the
    best-fitting model spectra on a common linear wavelength grid
    (3600--9824\,\AA, $0.8$\,\AA\,pixel$^{-1}$, vacuum
    wavelengths). The three planes store the full spectrophotometric
    model, the smooth continuum correction alone, and the
    emission-line model alone, respectively, coadded across the three
    cameras. We omit this extension from the merged catalogs described
    below to keep individual file sizes manageable.
\end{itemize}
\noindent For the most complete, column-by-column description of the
data model, pinned to the \fastspecfit{} version used to produce this
VAC, we refer to the online
documentation.\footnote{\url{https://fastspecfit.readthedocs.io/en/latest/fastspec.html}}

Because there are more than 52,000 individual files (one per HEALPix),
most users will find it convenient to work with the merged catalogs
included as part of the VAC, which combine all per-HEALPix files for a
given \texttt{SURVEY} and \texttt{PROGRAM} into a single file:

\medskip
\noindent\texttt{VACDIR/catalogs/fastspec-SURVEY-PROGRAM.fits.gz}
\medskip

\noindent where \texttt{VACDIR} is the top-level VAC
directory. However, the bright and dark programs of the main survey
contain too many objects to be merged into a single catalog, so those
have been split into 12 separate \texttt{nside}~=~1
catalogs:\footnote{There are 12 \texttt{nside}~=~1 HEALPixels over the
full sky, each one covering 1.04~sr or $\approx3438$~deg$^{2}$ of
non-overlapping area.}

\medskip
\noindent\texttt{VACDIR/catalogs/fastspec-loa-main-\{bright,dark\}-nside1-hpNN.fits}
\medskip

\noindent where \texttt{NN} runs from \texttt{00} to \texttt{11}.

As a concrete example, the Python code snippet below shows how to
retrieve the \fastspec{} measurements for a single target given only
its sky position. It uses \texttt{fastspecfit.util.radec2pix} to
locate the \texttt{nside}~=~1 HEALPix file covering that position,
matches the nearest catalog entry, and reports its redshift, stellar
mass, and \ha{} equivalent width---drawn, respectively, from the
\texttt{METADATA}, \texttt{SPECPHOT}, and \texttt{FASTSPEC} extensions
described above.

\begin{verbatim}
from astropy.table import Table
from astropy.coordinates import SkyCoord
import astropy.units as u
from fastspecfit.util import radec2pix

# for main-bright/main-dark, the merged VAC is pre-split into 12
# nside=1 HEALPixels; radec2pix locates the file for a given target
ra, dec = 143.508, 55.241   # example coordinates (I Zw 18)
hpix = radec2pix(nside=1, ra=ra, dec=dec)
fastfile = f'fastspec-loa-main-bright-nside1-hp{hpix:02d}.fits'

meta = Table.read(fastfile, hdu='METADATA')
specphot = Table.read(fastfile, hdu='SPECPHOT')
fast = Table.read(fastfile, hdu='FASTSPEC')

# nearest catalog entry to the requested position
coord = SkyCoord(ra * u.deg, dec * u.deg)
catalog = SkyCoord(meta['RA'] * u.deg, meta['DEC'] * u.deg)
indx = coord.separation(catalog).argmin()

print(f"z={meta['Z'][indx]:.4f}  "
      f"log(M*/Msun)={specphot['LOGMSTAR'][indx]:.2f}  "
      f"EW(Ha)={fast['HALPHA_EW'][indx]:.1f} Ang")
\end{verbatim}

\noindent Additional examples and tutorials for working with the data
can be found in the \fastspecfit{}
documentation.\footnote{\url{https://fastspecfit.readthedocs.io/en/latest/tutorials.html}}

\subsection{Recommended Quality Cuts}
\label{ssec:cuts}

Which quality cuts to apply depends on the science being done, so
below we describe several \fastspecfit{} diagnostics rather than
recommend a single threshold. With that in mind, the starting point
for any sample selection is a target-class-dependent redshift-quality
cut using the quantities and bitmasks reported by Redrock
(\S\ref{ssec:targets}). In the remainder of this paper, we adopt the
following ``good redshift'' criteria for the three primary DESI galaxy
classes. First, we define
\begin{equation}
  \begin{split}
    \mathcal{Q} \equiv {} & [\mathtt{COADD\_FIBERSTATUS} \;\&\;
    \sim (\mathit{RESTRICTED} \,|\, \mathit{VARIABLE}) = 0] \;\&\; \\
    & (\mathtt{SPECTYPE} \texttt{!=} \texttt{"STAR"}) \;\&\; (\mathtt{Z} > 0.001),
  \end{split}
  \label{eq:goodz_common}
\end{equation}
\noindent where italics denote bit names rather than catalog columns,
the bracketed term is a bitwise \texttt{AND} restricting
\texttt{COADD\_FIBERSTATUS} to the \textit{RESTRICTED} and
\textit{VARIABLE} \texttt{FIBERSTATUS}
bits\footnote{\url{https://desidatamodel.readthedocs.io/en/latest/bitmasks.html\#fiberstatus-bit-definitions}}---informational
flags that do not by themselves indicate a data-quality problem---and
two additional cuts intended to reject stellar contaminants. The
per-class criteria are then
\begin{align}
  \mathrm{BGS:}\quad & (\mathtt{ZWARN}=0) \;\&\; (\mathtt{DELTACHI2} > 40) \;\&\;
  (\mathtt{Z} < 0.8) \;\&\; \mathcal{Q} \label{eq:goodz_bgs}\\
  \mathrm{LRG:}\quad & (\mathtt{ZWARN}=0) \;\&\; (\mathtt{Z} < 1.5) \;\&\;
  (\mathtt{DELTACHI2} > 15) \;\&\; \mathcal{Q} \label{eq:goodz_lrg}\\
  \mathrm{ELG:}\quad &
  \log_{10}\!\left[\mathrm{S/N}(\text{\oiidoublet})\right] > 0.9 -
  0.2\,\log_{10}(\mathtt{DELTACHI2}) \;\&\; \mathcal{Q}
  \label{eq:goodz_elg}
\end{align}
\noindent where the ELG cut is taken from \citet{raichoor23a} but
using the \oii{} fluxes measured by \fastspecfit{}
(\S\ref{ssec:validation}). Of course, although we adopt these
particular criteria to generate some of the figures in this paper,
per-class cuts continue to be refined by the DESI team, and so readers
building their own samples should check the DESI documentation and
associated data-release papers for the most recent
recommendations.\footnote{\url{https://data.desi.lbl.gov/doc}}

For individual emission lines, we recommend selecting on
line-amplitude S/N, \texttt{AMP}$\times\sqrt{\texttt{AMP\_IVAR}}$,
rather than integrated-flux S/N,
\texttt{FLUX}$\times\sqrt{\texttt{FLUX\_IVAR}}$. Amplitude directly
tests whether a line rises above the local continuum noise, while flux
is a derived quantity (given by equation~\ref{eq:flux_phys}) whose
precision also depends on the fitted line width and velocity
shift. \fastspecfit{} applies this same amplitude-based criterion
internally to decide which pixels to mask around each line
(\S\ref{ssec:patches}) and to test for a significant broad-line
component (\S\ref{sssec:kinematic}). Once a line is judged
significant, any analysis that uses its flux directly should instead
propagate \texttt{FLUX\_IVAR}. Our Monte Carlo procedure
(\S\ref{ssec:uncertainties}) computes \texttt{FLUX\_IVAR} by jointly
refitting the amplitude, width, and velocity shift on every
realization, so it captures their correlated contribution to the flux
uncertainty in a way \texttt{AMP\_IVAR} does not.

For continuum-derived quantities, we do not recommend a single
threshold. \fastspecfit{} instead measures dozens of independent
diagnostics, continuum and per-line $\chi^2$ values, aperture
corrections, velocity dispersion, and more, precisely so that users
can build a sample selection driven by their own science case rather
than inheriting ours. \texttt{RCHI2\_PHOT}, the reduced $\chi^2$ of
the broadband photometric fit, is one such example and is a natural
cut for any analysis that leans on the stellar-population parameters
derived from the SED (stellar mass, SFR, K-corrections). Similarly,
analyses of stellar velocity dispersions should restrict to
$\texttt{VDISP\_IVAR}>0$, which selects objects with a direct
measurement rather than a value estimated from the \Mstar--\vdisp{}
relation or fixed at \texttt{VDISP\_NOMINAL}
(\S\ref{sssec:vdisp}). Finally, we caution against using
continuum-derived physical properties for QSOs and broad-line AGN
(\S\ref{ssec:example_qso}).

\subsection{Computational Performance}
\label{ssec:performance}

Because \fastspecfit{} fits tens of millions of DESI spectra, each
refit 50 (or more) times with Monte Carlo resampling
(\S\ref{ssec:uncertainties}), computational speed is a first-order
design requirement. We strive to meet this requirement by optimizing
the per-object fitting code and by exploiting the problem's
embarrassing parallelism.

First, we cache reused arrays, especially template FFTs, the
per-object redshift and IGM prefactors, and the $\chi^2(\vdisp)$ grid
(\S\ref{sssec:vdisp}) across the Monte Carlo realizations
(\S\ref{ssec:uncertainties}). We implement the numerically intensive
kernels---dust attenuation, IGM transmission, resolution-matrix
convolution, and the emission-line forward model and its
Jacobian---using \texttt{Numba} just-in-time (JIT) compilation to
native machine code \citep{lam15a}. Moreover, we run an explicit
warm-up pass before the parallel production run, compiling every
kernel once and writing the compiled artifacts to a shared, persistent
on-disk \texttt{Numba} cache; every subsequent process, across all
ranks and nodes, loads the compiled kernels from cache rather than
recompiling them. With Monte Carlo resampling disabled, these
optimizations enable \fastspec{} to fit a typical DESI spectrum in
$\approx0.1$~s, and \fastphot{} (\S\ref{sec:modes}) to fit the
broadband photometry of a typical object in $\approx10^{-2}$~s. With
Monte Carlo,
because each realization repeats most of the same fitting stages, the
total runtime scales roughly linearly with the number of Monte Carlo
iterations requested (\S\ref{ssec:uncertainties}).

In production, we run \fastspecfit{} on the NERSC Perlmutter
supercomputer, whose 3072 CPU nodes each provide 128 physical cores
(two 64-core AMD EPYC 7763
processors).\footnote{\url{https://docs.nersc.gov/systems/perlmutter/architecture}}
When fitting the full DR2 sample---38,245,183 objects in 52,325 HEALPix
files---the code processes a typical DESI spectrum (with 50 Monte
Carlo realizations) in a median 4.8~s (mean 6.0~s) of single-core
time. For comparison, \fastphot{} processes the same sample, also with
50 Monte Carlo realizations, in a median 0.42~s of single-core time.
We use MPI for Python \citep[\texttt{mpi4py};][]{dalcin21a} to
distribute independent HEALPix files across ranks and Python
multiprocessing to parallelize per-object fits within each rank.
Per-object runtime varies between a median 3.9~s/object/core for
main-survey dark-time targets (LRGs, ELGs, QSOs) versus
8.1~s/object/core for main-survey bright-time targets (BGS). This gap
in time is primarily due to BGS targets more frequently triggering a
direct velocity-dispersion measurement, which adds a $\chi^2$-scan and
nonlinear least-squares solve to the per-object cost
(\S\ref{sssec:vdisp}).

\subsection{Validation}
\label{ssec:validation}

To validate \fastspecfit's measurements, we compare our results
against several independent codes and catalogs. Naturally, these
comparisons are not exhaustive, and because we generate the largest
possible comparison samples, it is possible for differences to be
larger or smaller for narrower subsets of objects. Nevertheless, the
following figures demonstrate that \fastspec{} produces strong
emission-line fluxes, stellar masses, and velocity dispersions that
match reasonably well to existing independent measurements.

\begin{figure*}
\centering\includegraphics[width=\textwidth]{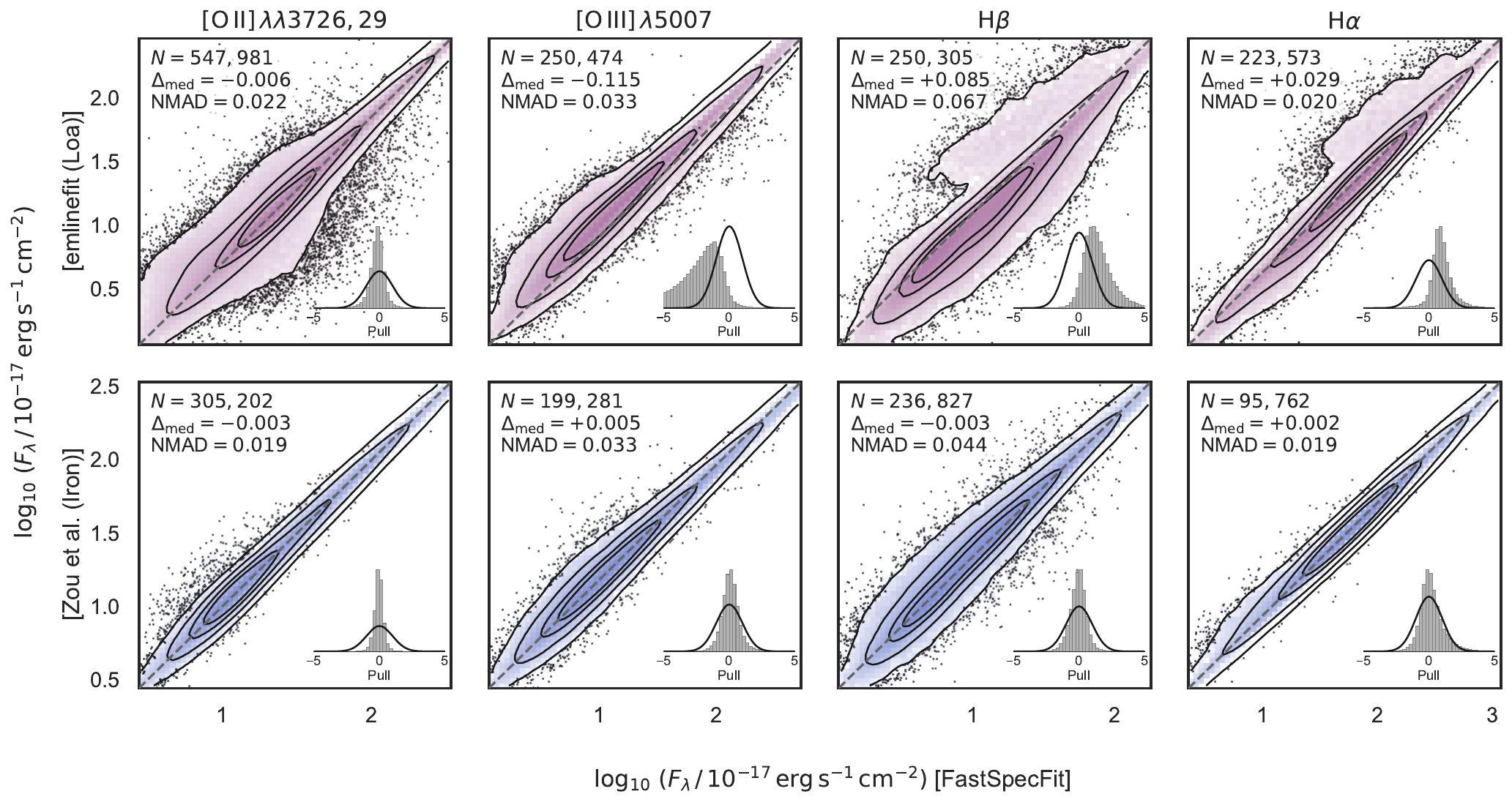}
\caption{Comparison of \fastspec{} emission-line fluxes for four
  strong nebular emission lines (\oiidoublet, summed, \oiiilam, \hb,
  and \ha) against (top) the DESI DR2/\texttt{Loa} \texttt{emlinefit}
  afterburner measurements \citep{raichoor23a}, and (bottom)
  emission-line fluxes from the H.~Zou et~al. (in preparation)
  DR1/\texttt{Iron} VAC \citep[see also][]{zou24b, sui26a}. Within each panel,
  we only plot objects with $\mathrm{S/N}>3$ in both catalogs, and the
  shaded histogram in the inset shows the pull distribution
  (equation~\ref{eq:pull}) compared to a unit Gaussian (black
  curve). We show the logarithmically-scaled two-dimensional number
  density as a color map, and the contours enclose 50\%, 75\%, 95\%,
  and 99.5\% of the joint distribution; outside the outermost contour,
  we show individual galaxies as markers. The dashed line is the
  one-to-one relation. Finally, each panel lists the matched sample
  size $N$, the median difference in logarithmic flux,
  $\Delta_{\mathrm{med}}$ (\fastspec{} minus external) in dex, and the
  \texttt{NMAD} scatter in dex. \label{fig:compare-emlines-external}}
\end{figure*}

\begin{figure*}
\centering\includegraphics[width=1.0\textwidth]{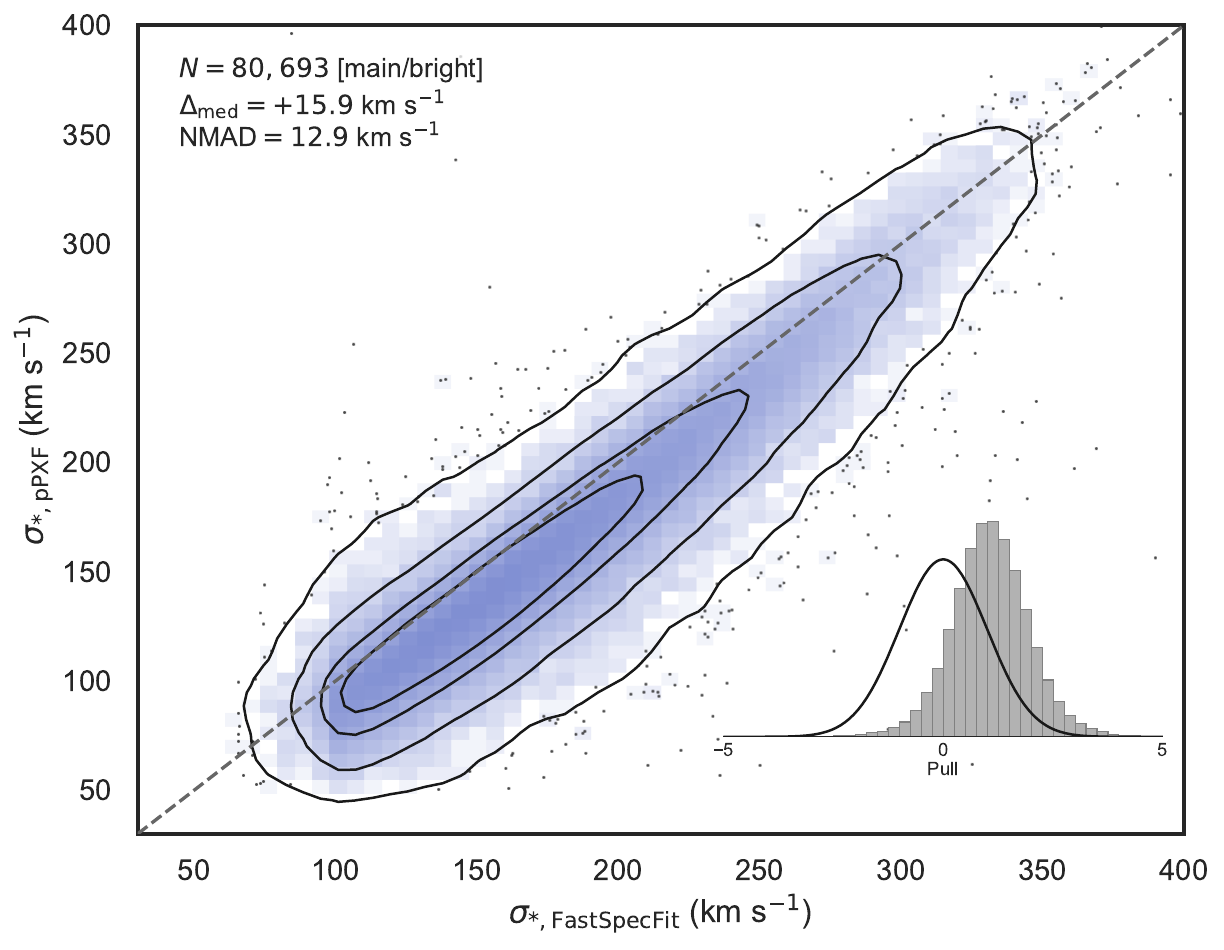}
\caption{Comparison of \fastspec{} stellar velocity dispersions with
  those measured by \citet{ross25a} using \texttt{pPXF}
  \citep{cappellari23a}, for a matched sample of main-survey,
  bright-time galaxies (\S\ref{sssec:vdisp}). As in
  Figure~\ref{fig:compare-emlines-external}, the color map, contours,
  and outlying markers show the two-dimensional density distribution,
  the dashed line is the one-to-one relation, and the inset shows the
  pull distribution (equation~\ref{eq:pull}) compared to a unit
  Gaussian (black curve). We find a median offset and \texttt{NMAD}
  scatter of $15.9\pm12.9$~\kms{} (\fastspec{} yields systematically
  larger \vdisp{} values) and a median pull of
  $1.06\pm0.82$. \label{fig:vdisp}}
\end{figure*}

For the subsequent comparisons, we match our VAC to each external
catalog by \texttt{TARGETID} when a common DESI identifier exists
\citep{myers23a}, otherwise using a $1\farcs5$ positional
threshold. In addition, we confirm each match by requiring redshift
agreement, $c|\Delta z|<1000~\kms$, which is necessary even when
comparing two DESI samples because of possible differences between
specprods and because of our override of some QSO redshifts
(\S\ref{ssec:catalog}).  For each matched pair we also compute the
pull,

\begin{equation}
\mathrm{pull} = \frac{x_{\mathrm{ref}} - x_{\mathrm{ext}}}
  {\sqrt{\delta x_{\mathrm{ext}}^2 + \delta x_{\mathrm{ref}}^2}},
\label{eq:pull}
\end{equation}

\noindent where $x$ and $\delta x$ are the measured quantity---line
flux in Figure~\ref{fig:compare-emlines-external}, velocity dispersion
in Figure~\ref{fig:vdisp}---and its uncertainty in the external and
\fastspec{} (reference) catalogs, respectively.

In Figure~\ref{fig:compare-emlines-external} we compare our
emission-line fluxes for \oiidoublet{} (summed, with the uncertainties
added in quadrature), \oiiilam, \hb, and \ha{} against the
\texttt{Loa} (DR2) \texttt{emlinefit} afterburner measurements
\citep{raichoor23a} and against the \texttt{Iron} \citep[DESI
  DR1;][]{desi-collaboration26a} \textit{Stellar Mass and Emission
  Line
  Catalog}\footnote{\url{https://data.desi.lbl.gov/doc/releases/dr1/vac/stellar-mass-emline}}
(\texttt{v1.0}) from H.~Zou et~al. (in preparation; see also
\citealt{zou24b, sui26a}). The line-fluxes in all
these catalogs have been corrected for foreground Galactic
extinction. We require $\mathrm{S/N}>3$ for each line in both
catalogs, and combine the SV3 bright and dark programs to maximize the
matched sample. (Combining the main bright and dark samples gives
nearly identical statistics, simply with many more objects.)

The agreement with \texttt{emlinefit} is very good for \oii, with a
median difference and \texttt{NMAD}\footnote{The normalized median
absolute deviation,
$\mathtt{NMAD}\equiv1.4826\,\mathrm{median}(|x-\mathrm{median}(x)|)$,
a robust estimator of the standard deviation for a distribution that
may contain outliers. Here and throughout, we quote comparisons as
$\Delta_\mathrm{med}\pm\mathtt{NMAD}$.} scatter of
$-0.006\pm0.022$~dex. For \ha{} and \hb{} we find median differences of
$0.029\pm0.020$~dex and $0.085\pm0.067$~dex, respectively---tight,
but offset by roughly one \texttt{NMAD} (\fastspec{} is
brighter). These systematic differences are most likely due to
\texttt{emlinefit}'s treatment of the continuum around each line as a
constant pedestal with no correction for stellar absorption, which
would cause the Balmer lines to be underestimated (positive pull;
Figure~\ref{fig:compare-emlines-external} insets). The differences
with \oiii, meanwhile, are notably larger, with an offset of
$-0.115\pm0.033$~dex and a distinctly non-Gaussian pull. Because the
\texttt{emlinefit} continuum window for \oiii{} (nominally
$\pm100$~\AA{} in the rest frame around each line) extends blueward to
the immediate vicinity of \hb, it is possible that the same unmodeled
stellar absorption depresses \oiii{}'s local continuum estimate and
biases its line flux high.

Meanwhile, the agreement with the VAC from H.~Zou et~al.\ (in
preparation) is excellent in terms of median offset
($|\Delta_\textrm{med}|<0.005$~dex) and scatter
($\mathtt{NMAD}<0.04$~dex) for all four lines. The pull distributions for all four
lines are narrower than a unit Gaussian, as expected because both
catalogs measure the same spectra and therefore share the same
noise. Consequently, this comparison validates the line fluxes, but it
does not test their uncertainties.

Next, Figure~\ref{fig:vdisp} compares the \fastspec{} $\vdisp$
measurements (\S\ref{sssec:vdisp}) against those from the DESI DR1
(\texttt{Iron}) Fundamental Plane study of \citet{ross25a} for the
same set of objects observed at $0.0033<z<0.1$ in the bright program
of the main survey (see also \citealt{said25a}). \citet{ross25a} fit
each DESI spectrum with \texttt{pPXF} \citep[v8.2.4;][]{cappellari23a}
using the empirical Indo-US Coud\'e Feed stellar library
\citep[$\sigma_{\mathrm{Indo-US}}=30$~\kms;][]{valdes04a}, restricted
to the DESI B camera ($\approx$3600--5550~\AA,
$\sigma_{\mathrm{B}}\approx 35-65$~\kms; see \S\ref{ssec:spectra}). By
comparison, our theoretical \texttt{C3K\_R3K} templates have
$\sigma_\mathrm{C3K}=42$~\kms{} (\S\ref{ssec:templates}) and we fit
\vdisp{} over a slightly different rest-frame wavelength range,
3800--6000~\AA{} (\S\ref{sssec:vdisp}).

\begin{figure*}
\centering\includegraphics[width=1.0\textwidth]{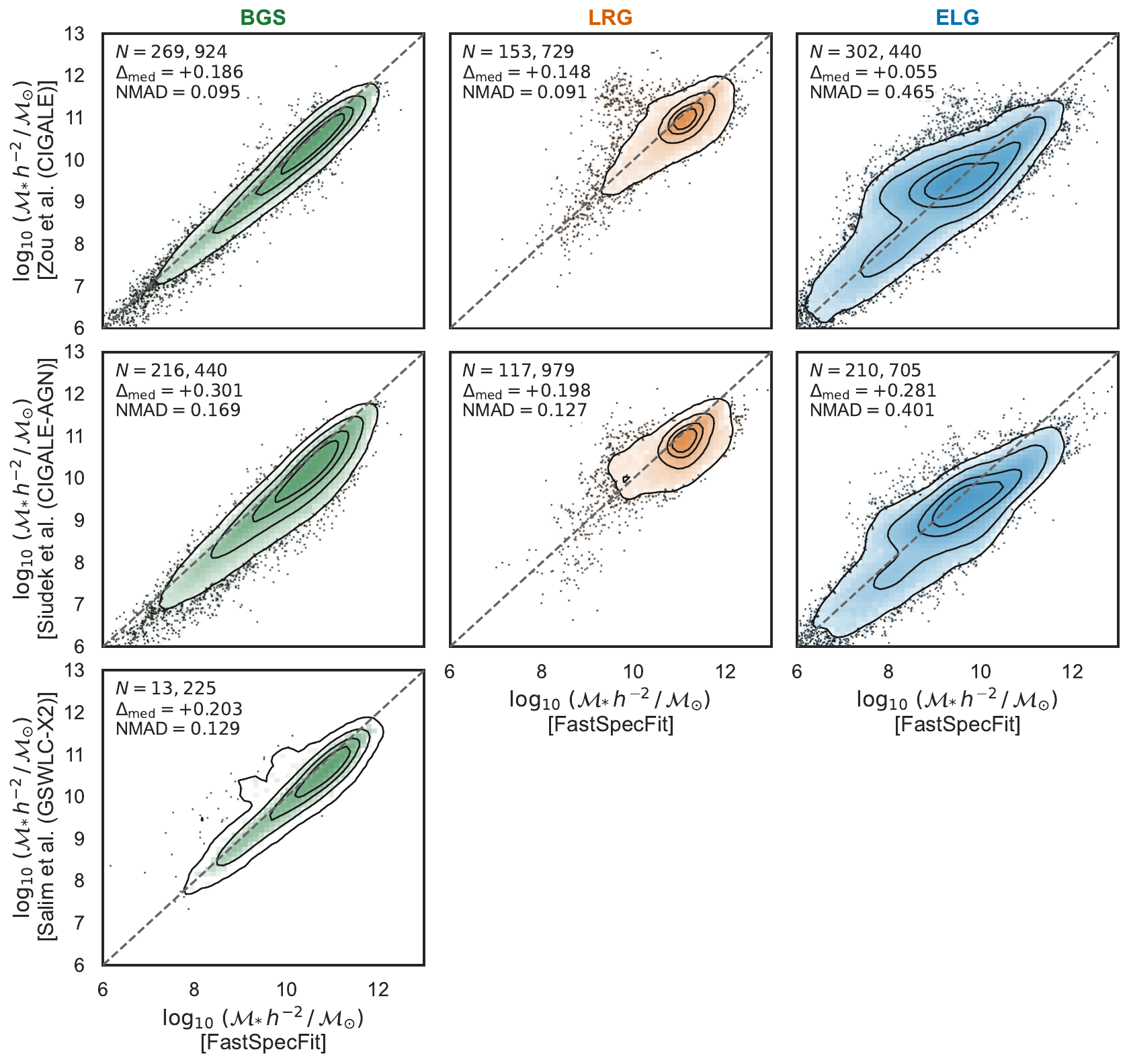}
\caption{Comparison of \fastspec{} stellar masses ($h=1$, Chabrier
  IMF) against three independent, CIGALE-based catalogs, organized by
  DESI target class along columns. In each row, we compare our
  measurements against (top) the DR2 (\texttt{Loa}) VAC of H.~Zou
  et~al. (in preparation; see also \citealt{zou24b, sui26a}), (middle)
  the DR1 (\texttt{Iron}) CIGALE-AGN VAC of \citet{siudek24a}, and
  (bottom) the GSWLC-X2 catalog of \citet{salim16a, salim18a}. As in
  Figure~\ref{fig:compare-emlines-external}, the color map, contours,
  and markers show the two-dimensional density distribution, and the
  dashed line is the one-to-one relation.  Overall, we find median
  offsets of $\Delta_\mathrm{med}=0.15$--$0.30$~dex (\fastspec{}
  systematically higher) with $\mathtt{NMAD}=0.09$--$0.17$~dex scatter
  for BGS and LRG targets, with the scatter increasing to
  $\mathtt{NMAD}\approx0.4$~dex for ELG targets. The median offsets
  and scatter are consistent with expectations from different
  (non-parametric versus parametric) SFH priors \citep{leja19b,
    leja20a} and, more generally, from the stellar-library,
  dust-attenuation, and other modeling choices that differ across
  independent SED-fitting codes. \label{fig:compare-mstar-external}}
\end{figure*}

For $N=80,693$ objects in common, we find a median offset and
\texttt{NMAD} scatter of $\Delta_\mathrm{med}=15.9\pm12.9$~\kms{}
(\fastspec{} systematically higher), with a pull of
$1.06\pm0.82$, indicating that the offset is comparable to the
combined per-object uncertainties. The residual is statistically significant but
relatively small on an instrumental scale---about a third of a native
DESI pixel ($0.8$~\AA, or $\approx53$~\kms{} at 4500~\AA). However,
its precise origin is not clear from this comparison alone. For
context, \citet{ross25a} also compare their \vdisp{} measurements
against \citet{thomas13a}, who measure velocity dispersions for a
sample of Sloan Digital Sky Survey \citep[SDSS;][]{york00a} galaxies
using the MILES empirical stellar library \citep{sanchez-blazquez06a}
and an earlier version of \texttt{pPXF}; they find very good
agreement, with an offset of only $\approx$1--2~\kms{} over most of
the $\sigma$ range. Certainly, there are real differences between the
stellar libraries used in these analyses relative to the theoretical
stellar library we adopt, especially with respect to non-solar
[$\alpha$/Fe] abundance patterns, which may have an effect on the
derived velocity dispersion \citep[e.g.,][]{conroy14a, pernet24a}. On
the other hand, \fastspec{} incorporates the wavelength-dependent
instrumental resolution (via the DESI resolution matrix) directly into
a full forward model, with no pixel resampling; we have not tested how
approximating (or neglecting) the resolution matrix, or resampling
onto a common wavelength grid, changes the derived velocity
dispersion. In conclusion, we find that \fastspec{} yields
precise velocity dispersions over the full dynamic range of \vdisp{}
values spanned by low-redshift galaxies, albeit with an as-yet
unexplained systematic difference relative to existing measurements.

Next, in Figure~\ref{fig:compare-mstar-external} we compare the
stellar masses from \fastspec{} against three independent,
CIGALE-based \citep{boquien19a} SED-fitting catalogs, organized by
DESI target class (BGS, LRG, and ELG; \S\ref{ssec:targets}). The top
row compares our measurements against the DR2 (\texttt{Loa})
SED-fitting VAC from H.~Zou et~al. (in preparation; see also
\citealt{zou24b, sui26a}), while the middle and bottom rows compare
our measurements to the DR1 (\texttt{Iron}) \textit{AGN Host Galaxies
  Physical Properties} VAC from
\citet[\texttt{v1.2};][]{siudek24a}\footnote{\url{https://data.desi.lbl.gov/doc/releases/dr1/vac/cigale}}
and the GSWLC-X2\footnote{\url{https://salims.pages.iu.edu/gswlc}}
catalog of \citet{salim16a, salim18a}, respectively. For the first two
DESI catalogs, we compare against our combined SV3 bright plus dark
catalogs, while for the GSWLC-X2 we compare just against SV3/bright
(there are a negligible number of matches from SV3/dark); as before,
matching to the main-survey \fastspec{} catalogs increases the number
of objects in the comparison sample but not the statistical results or
our conclusions.

All three of these external catalogs perform photometric SED fits, but
the Zou et~al. VAC augments its $grz$ and WISE $W1$--$W2$ photometry
with 10-band top-hat filter photometry synthesized from the DESI
optical spectrum and scaled for aperture bias using an achromatic
aperture correction analogous to our own (\S\ref{sssec:apercorr}). In
addition, all three catalogs adopt a parametric SFH---a delayed-$\tau$
model, with an optional recent burst for \citet{siudek24a} and
\citet{salim18a}, in contrast to \fastspec{}'s non-parametric,
five-bin star-formation history
(\S\ref{ssec:templates}). \citet{siudek24a} also include an AGN
component \citep{fritz06a} in their SED fits. In constructing these
comparisons, we correct for differences in IMF and cosmology.

Within the BGS and LRG subsamples, the \fastspec{} stellar masses are
systematically higher than each CIGALE-based catalog by
$\Delta_\mathrm{med}=0.15$--$0.30$~dex, with tight scatter
($\mathtt{NMAD}=0.09$--$0.17$~dex). Both the sign and the magnitude of
this offset agree well with previous comparisons between
non-parametric and parametric SFH-derived masses, which attribute
$\approx0.1$--$0.3$~dex systematic differences to the additional
flexibility of a non-parametric model accommodating more mass in old,
high-mass-to-light-ratio stellar populations \citep{leja19b, leja20a,
  lower20a}. The scatter in the difference
  for ELG targets is higher, $\mathtt{NMAD} \approx 0.4$~dex, likely
reflecting the fainter, lower-S/N
broadband photometry available for these star-forming targets,
especially in the WISE IR bands. More generally, offsets of
  this size are typical of SED-based stellar masses, which carry
  systematic uncertainties of $\approx0.2$--$0.3$~dex depending on the
  adopted SPS models, priors, and fitting technique
  \citep{moustakas13a, conroy13a}. Overall, the agreement between
\fastspec{} and three independent, CIGALE-based stellar-mass catalogs
is reasonable, particularly given the different
data (photometry only versus spectrophotometry), templates, and
star-formation-history parameterizations involved.

\begin{figure*}
\centering\includegraphics[width=1.0\textwidth]{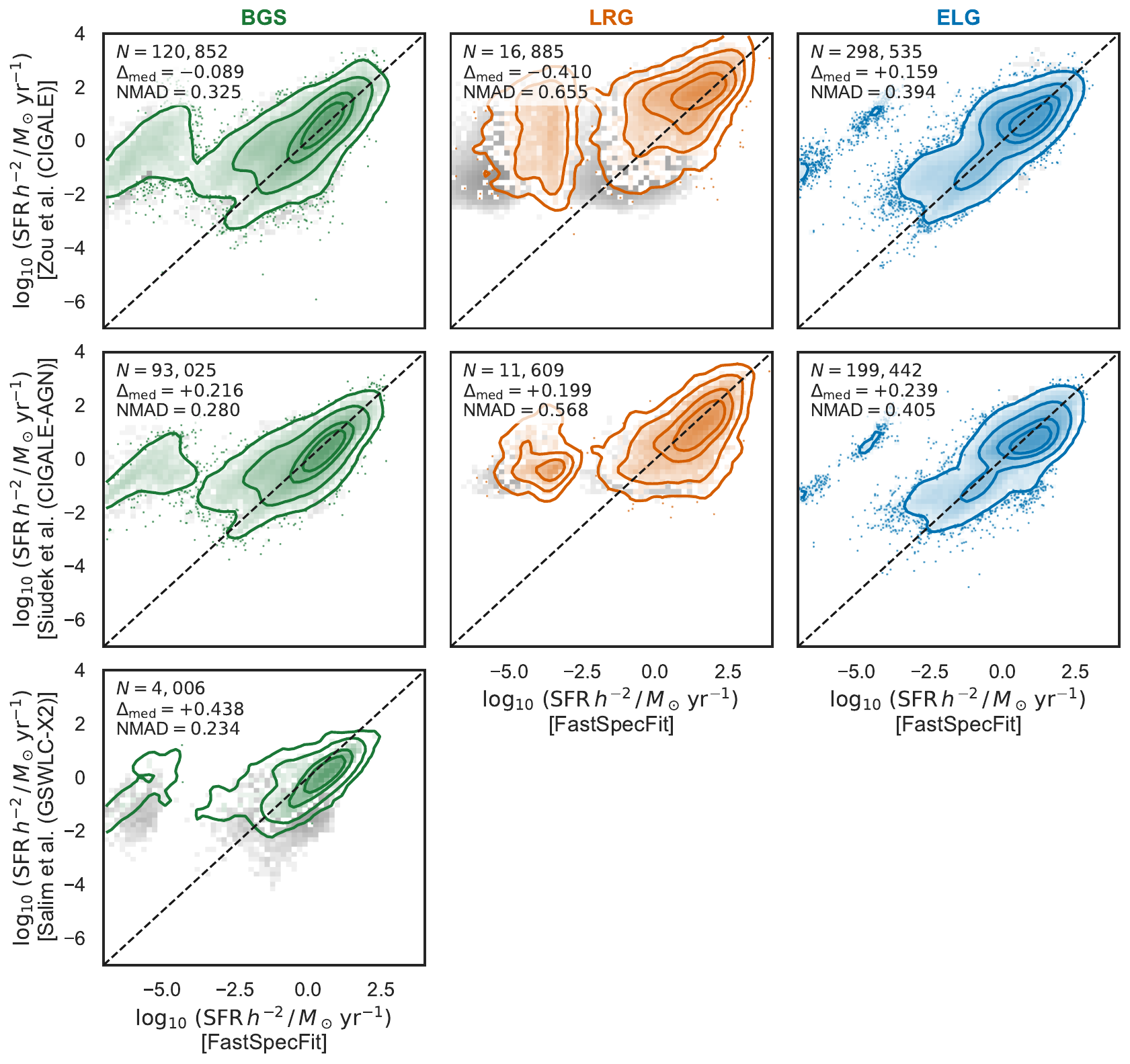}
\caption{Comparison of \fastspec{} SFRs ($h=1$, Chabrier IMF)
    against the same three CIGALE-based catalogs shown in
    Figure~\ref{fig:compare-mstar-external}, organized by DESI target
    class along columns. We classify galaxies as star-forming or
    quiescent using a rest-frame $^{0.1}(g-r)$ color cut
    (equation~\ref{eq:grcut}). For star-forming galaxies, the color
    map, contours, and markers are as in
    Figure~\ref{fig:compare-emlines-external}, while quiescent
    galaxies are shown as a gray density map only. The dashed line is
    the one-to-one relation. Each panel lists the number of
    star-forming galaxies $N$, the median difference
    $\Delta_\mathrm{med}$ (\fastspec{} minus external) in dex, and the
    \texttt{NMAD} scatter in dex, all computed using the star-forming
    galaxies only.
  \label{fig:compare-sfr-external}}
\end{figure*}

Finally, in Figure~\ref{fig:compare-sfr-external} we compare
  our SPS-derived SFRs against the same three catalogs. Unlike the
  stellar masses, the four sets of SFRs are not defined
  identically. \fastspecfit{} reports the mean SFR over the last
  100~Myr (equation~\ref{eq:sfr}), as does GSWLC-X2 \citep{salim16a},
  whereas the Zou et~al. and \citet{siudek24a} catalogs report the
  instantaneous SFR and the mean SFR over the last 10~Myr,
  respectively (DESI Collaboration et~al., in preparation). Moreover,
  only GSWLC-X2 constrains the SFR with observed-frame UV
  \citep[GALEX;][]{martin05a} photometry.

Because of the difficulty of constraining the SFRs of quiescent
  galaxies, we follow DESI Collaboration et~al. (in preparation) and
  compute the comparison statistics for star-forming galaxies only
  using the rest-frame $^{0.1}(g-r)$ color. Separately for BGS, LRG,
  and ELG targets, we measure the color of the red-sequence ridge in
  bins of stellar mass and redshift, and fit these measurements with a
  plane.
  We then classify a galaxy as star-forming if it is more than 0.2~mag
  bluer than this ridge-line,
\begin{equation}
  ^{0.1}(g-r) < 0.706 + 0.067\,(\logmstar-10) - 0.130\,(z-0.1).
  \label{eq:grcut}
\end{equation}
This cut lies $\approx0.07$~mag blueward of the minimum between
  the red sequence and the blue cloud (where we can measure it), so it
  deliberately excludes green-valley galaxies at the cost of some
  genuinely star-forming galaxies. In the sample matched to the Zou
  et~al. catalog, it classifies 49\%, 13\%, and 99\% of the BGS, LRG,
  and ELG targets as star-forming, slightly below the fractions that
  DESI Collaboration et~al. (in preparation) find for BGS and LRG
  targets (51\% and 18\%).

For star-forming galaxies, our SFRs generally track the
  external measurements over nearly four decades. For BGS and ELG
  targets, the median offsets relative to Zou et~al. and
  \citet{siudek24a} range from $-0.09$ to $+0.24$~dex with
  $\mathtt{NMAD}=0.28$--$0.41$~dex. For the relatively small fraction
  of star-forming LRGs, meanwhile, the scatter is much larger,
  $\mathtt{NMAD}=0.57$--$0.66$~dex, with large median offsets that are
  opposite in sign ($-0.41$ and $+0.20$~dex), indicating that none of
  these catalogs constrains their SFRs well (see also DESI
  Collaboration et~al., in preparation).

Compared to the GSWLC-X2 catalog, which is only available for
  BGS targets, the \fastspecfit{} SFRs are systematically higher by
  nearly a factor of three, $\Delta_\mathrm{med}=0.44\pm0.23$~dex. To
  determine which set of SFRs is more accurate, we compute an
  independent SFR for the subset of these galaxies with well-measured
  ($\mathrm{S/N}>3$) \ha{} and \hb{} line-emission. We correct the
  \ha{} flux for aperture effects and dust attenuation using the
  Balmer decrement, and convert it to SFR following \citet{hao11a}.
  GSWLC-X2 agrees with these \ha-based SFRs to within
  $\approx0.1$~dex, whereas the \fastspecfit{} SFRs exceed them by
  $\approx0.57$~dex, with no dependence on the size of the aperture
  correction. The Zou et~al. and \citet{siudek24a} SFRs also exceed
  the \ha-based SFRs, by 0.34--0.41~dex, so this excess is not unique
  to \fastspecfit.

Our stellar masses for these galaxies are also higher than
  GSWLC-X2, by 0.20~dex (Figure~\ref{fig:compare-mstar-external},
  bottom row), so the corresponding offset in specific SFR,
  $\mathrm{SFR}/\Mstar$, is about half as large,
  $\approx0.20$~dex. What distinguishes GSWLC-X2 from the other three
  catalogs, however, is its UV photometry, which leads us to posit
  that the SFR discrepancy we find is due to the well-known degeneracy
  between dust attenuation and stellar age
  \citep[e.g.,][]{conroy13a}. At optical and near-IR wavelengths, a
  fit can reproduce the same red continuum with either an older, less
  attenuated stellar population or a younger, dustier one, and the
  latter solution implies a higher SFR once corrected for
  dust. Rest-frame UV photometry, which is dominated by young, massive
  stars, breaks this degeneracy. Consistent with this picture, the SFR
  offset relative to GSWLC-X2 correlates with the difference in
  inferred attenuation (Spearman $\rho=0.50$). Moreover, the Zou
  et~al. SFRs, which have the largest median attenuation of the three
  catalogs ($\tauv=0.64$, versus 0.39 for \fastspec{} and 0.28 for
  GSWLC-X2), track this difference even more closely ($\rho=0.82$). We
  conclude that, without rest-frame UV constraints, the SPS-derived
  SFRs from \fastspecfit{} (and from other optical and near-IR SED
  fits) should be used with caution, particularly in an absolute
  sense. The relative SFRs, on the other hand, are more robust, given
  that they track the external measurements over nearly four orders of
  magnitude. For star-forming galaxies with well-measured \ha{} and
  \hb{} emission, dust-corrected SFRs computed from our emission-line
  fluxes provide a more reliable absolute calibration.

\subsection{Pipeline Diagnostics}
\label{ssec:diagnostics}

\begin{figure*}
\centering\includegraphics[width=1.0\textwidth]{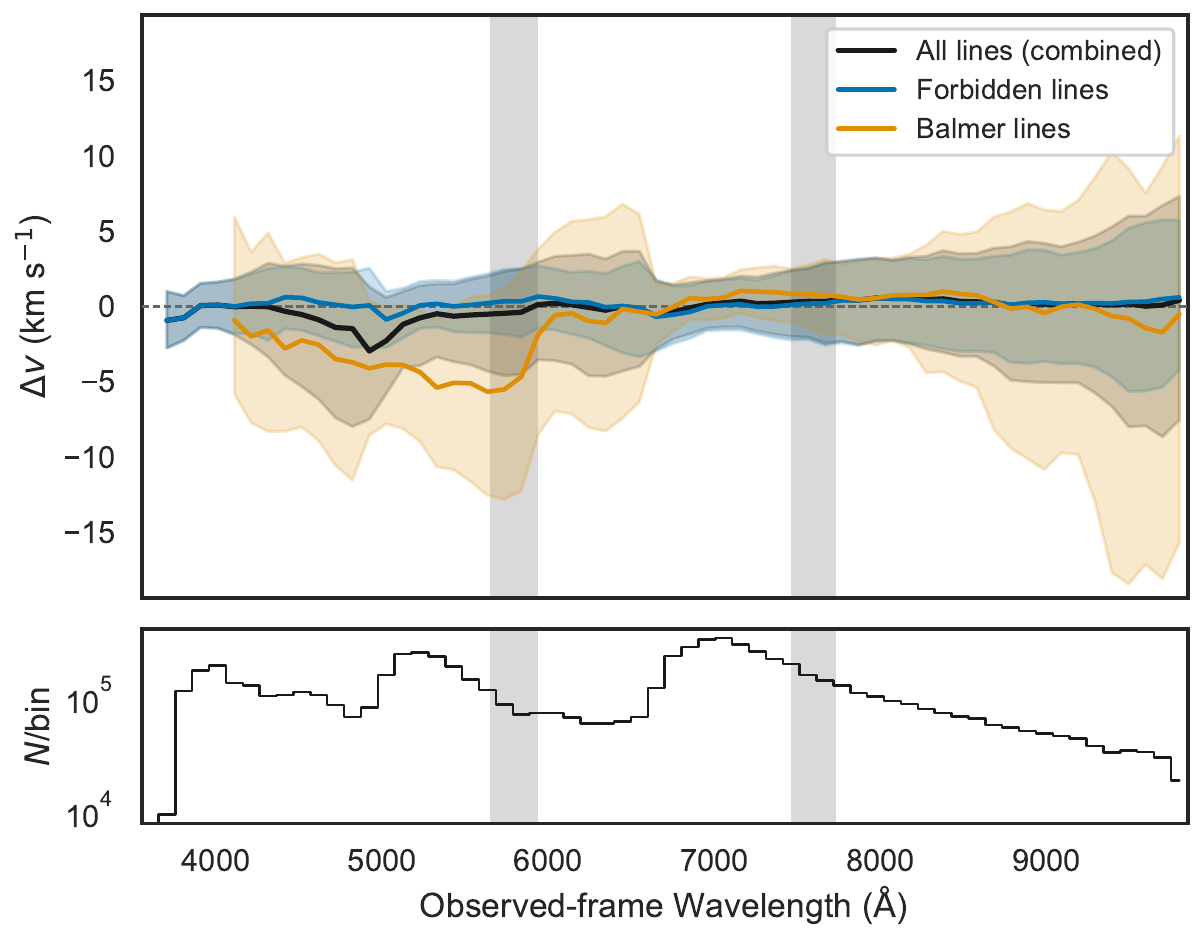}
\caption{Running median velocity shift $\Delta v$ (top) and the number
  of contributing measurements (bottom) as a function of
  observed-frame wavelength for DR2 main-survey galaxies with a
  relaxed, per-line kinematic solution (\S\ref{sssec:kinematic}). We
  compute statistics in equal-width ($\approx100$~\AA) bins spanning
  the full DESI wavelength range (3600--9824~\AA). The blue and orange
  curves show the derived trends using forbidden and Balmer lines,
  respectively, while the black curve combines all the lines; the
  shaded horizontal bands show the interquartile range of the
  individual measurements within each bin. Gray vertical bands mark
  the B/R and R/Z camera-boundary transitions, and the thin horizontal
  dashed line marks zero offset. Although the Balmer lines show a
  $\approx-5$~\kms{} offset between roughly 4800 and 5800~\AA, we
  attribute this shift to imperfect stellar absorption subtraction of
  the higher-order Balmer lines rather than systematic
  wavelength-calibration errors. Using the forbidden lines, we
  measure a systematic error of $<1$~\kms{} ($\lesssim$few\%
  of a pixel) over the full DESI spectral range, including at the
  inter-camera boundaries. \label{fig:wave_resid}}
\end{figure*}

One of the dividends of fitting every DESI spectrum with a physically
motivated forward model is that the residuals carry information about
possible systematic calibration errors in the reduced data. To
illustrate this kind of application, in this section we use
\fastspec{} measurements to characterize wavelength-solution errors
revealed by correlated per-camera emission-line velocity
shifts.

\fastspecfit{} does not fit the three DESI cameras (B, R, Z)
independently. Narrow emission lines instead share a common kinematic
anchor during the primary fit---{\oiiilam} for forbidden lines, \ha{}
for narrow Balmer and \hei{} lines (\S\ref{sssec:kinematic}). However,
in the final fitting pass, we free the velocity shift, \texttt{VSHIFT}
of individual lines and doublets and re-optimize the solution within a
$\pm100$~\kms{} window of the tied value, with the relaxed solution
adopted only if it improves $\chi^2$ (\S\ref{sssec:kinematic}).

Because the rest wavelength and systemic redshift $z$ together
determine each line's nominal observed-frame wavelength, this
population of quasi-independent \texttt{VSHIFT} measurements,
aggregated across many lines and redshifts, serves as a proxy for
residual wavelength-solution errors as a function of observed-frame
wavelength even though no single camera is ever fit in isolation.

To investigate this effect, we select main-survey galaxies passing our
baseline redshift-quality criteria (\S\ref{ssec:cuts}) for which the
relaxed kinematic pass is adopted and no broad Balmer component was
fit, so that every narrow line's \texttt{VSHIFT} is a genuine
measurement rather than a copy of its tied anchor. We select any line
detected with flux $\mathrm{S/N}>7$ and a formal \texttt{VSHIFT}
uncertainty below 25~\kms. We exclude the $0.02$\% of fits pinned at
the optimizer's hard $\pm500$~\kms{} bound, which biases their formal
uncertainties toward implausibly small values. These cuts yield a
unique set of 16 usable emission lines (11 forbidden lines and 5
Balmer lines), spanning \oiidoublet{} through \siii~$\lambda9532$.

Figure~\ref{fig:wave_resid} shows the resulting running median
\texttt{VSHIFT} as a function of observed-frame wavelength for
forbidden lines (blue), Balmer and helium lines (orange), and all
lines combined (black), as well as interquartile range of individual
measurements within each bin (shaded band). The forbidden-line median
is consistent with zero to within $\pm1$~\kms{} across the full DESI
wavelength range, with no discontinuity at the B/R or R/Z camera
boundaries (vertical shaded gray band). The Balmer lines, by contrast,
show a $\approx-5$~\kms{} offset between roughly 4800 and 5800~\AA,
the observed-frame window populated by \hb{} and higher-order Balmer
lines at low redshift, returning to zero once \ha{} dominates the
sample redward of $\approx6600$~\AA. However, because this offset
appears only in the Balmer lines and not in the forbidden lines at the
same observed wavelength, we attribute it to imperfect
continuum-subtraction in and around the higher-order Balmer lines
rather than a wavelength-calibration error, which would affect all
species equally at fixed observed wavelength. Given DESI's per-pixel
velocity sampling of 25--66~\kms{} over this wavelength range
(\S\ref{ssec:emlines}), the $<1$~\kms{} forbidden-line residual
amounts to no more than a few percent of a single pixel, indicating
that the wavelength solution is calibrated well below the pixel scale
across the full DESI optical range.

One final notable finding: We exclude the \siiidoublet{} doublet from
Figure~\ref{fig:wave_resid} because both lines show a consistent
$+11$--$12$~\kms{} offset which is not corroborated by any other lines
in that part of the observed-frame spectrum. Therefore, we interpret
the offset as a rest-wavelength error rather than a
wavelength-calibration residual. Taking the median \texttt{VSHIFT} of
each line as a pure rest-wavelength correction, i.e., setting $\Delta
v=0$ in equation~(\ref{eq:lstar}) and solving for
$\lambda_\mathrm{rest}$, implies revised rest wavelengths of
$9071.4$~\AA{} and $9533.6$~\AA{} for \siii{}~$\lambda9069$ and
\siii{}~$\lambda9532$, respectively, $\approx0.3$--$0.4$~\AA{} redward
of the NIST (Ritz\footnote{A Ritz wavelength is computed from the
difference between two independently measured energy levels via the
Ritz combination principle. This technique is used when the transition
itself has not been directly, reliably observed in the laboratory.})
values used in this version of \fastspecfit.

\section{Discussion}
\label{sec:discussion}

The \fastspecfit{} DR2 VAC provides simultaneous measurements of
stellar continuum, emission-line, and broadband SED properties for
more than 38~million extragalactic DESI targets spanning a broad range
of intrinsic physical properties and redshift. Although a specific
scientific investigation is outside the scope of the current paper, in
\S\ref{ssec:reach} we briefly illustrate the scientific reach of the
\fastspecfit{} VAC with two example figures on the stellar mass and
AGN content of the DESI sample. We then conclude in
\S\ref{ssec:improvements} with a short discussion of some of the
planned updates and improvements to the code.

\subsection{Example Scientific Applications}
\label{ssec:reach}

\begin{figure*}
\centering\includegraphics[width=1.0\textwidth]{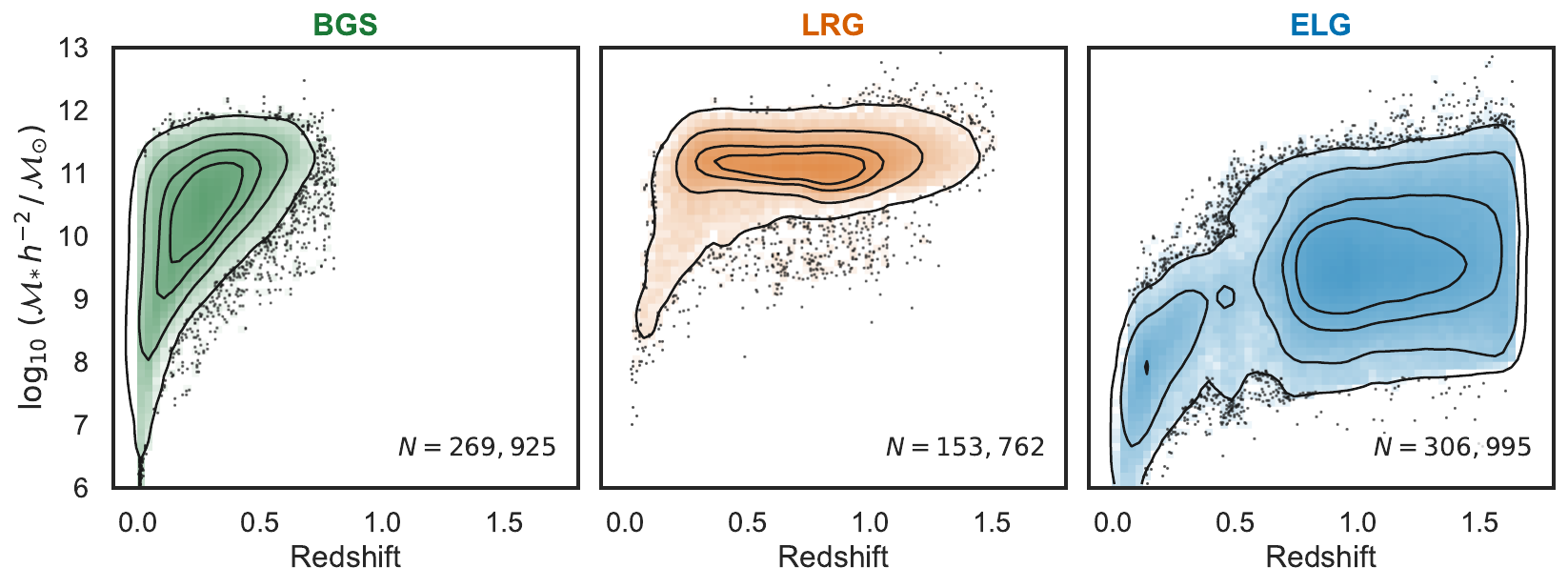}
\caption{Stellar mass versus redshift for BGS (left), LRG (center),
  and ELG (right) targets observed during the DESI SV3 bright and dark
  programs. The color scale and contours in each panel show the
  two-dimensional number density of galaxies passing per-class
  redshift quality cuts (see \S\ref{ssec:cuts}); contours enclose 50, 75, 95,
  and 99.5\% of the sample, and individual galaxies outside the
  outermost contour are plotted as points.
  \label{fig:mstar_z}}
\end{figure*}

We begin by showing the distribution of stellar mass versus redshift
for BGS, LRG, and ELG targets in Figure~\ref{fig:mstar_z}. We limit
the sample to the SV3 bright (BGS) and SV3 dark (LRG, ELG) programs
for this illustrative figure, but the results using the DR2
main-survey data are equivalent. For each panel, we only retain
objects which pass the target-class-dependent criteria listed in
\S\ref{ssec:cuts}, as well as requiring $\mathtt{LOGMSTAR\_IVAR}>0$.

This figure shows that the three target classes trace complementary
regions of stellar mass and redshift. The BGS sample spans the widest
mass range at $z \lesssim 0.6$, and includes galaxies from well below
the knee of the stellar mass function to the most massive systems in
the local universe \citep{moustakas13a, hahn24a}; the LRG sample
includes, by design, the most massive, passively evolving galaxies
over $0.3 \lesssim z \lesssim 1.5$ \citep{zhou23a}; and the ELG sample
preferentially consists of low-to-intermediate mass star-forming
systems at $0.6 \lesssim z \lesssim 1.6$, with a secondary tail of
dwarf galaxies, $\logmstar\lesssim9$, to $z\approx0$ (V.~Manwadkar
et~al., in preparation; \citealt{raichoor23a}). Together, these three
samples provide expansive coverage of the galaxy population from $z
\approx 0$ to $z \approx 1.6$, making the DR2 catalog a uniquely
comprehensive resource for studies of galaxy evolution over the past
$\approx$10~Gyr.

\begin{figure*}
\centering\includegraphics[width=1.0\textwidth]{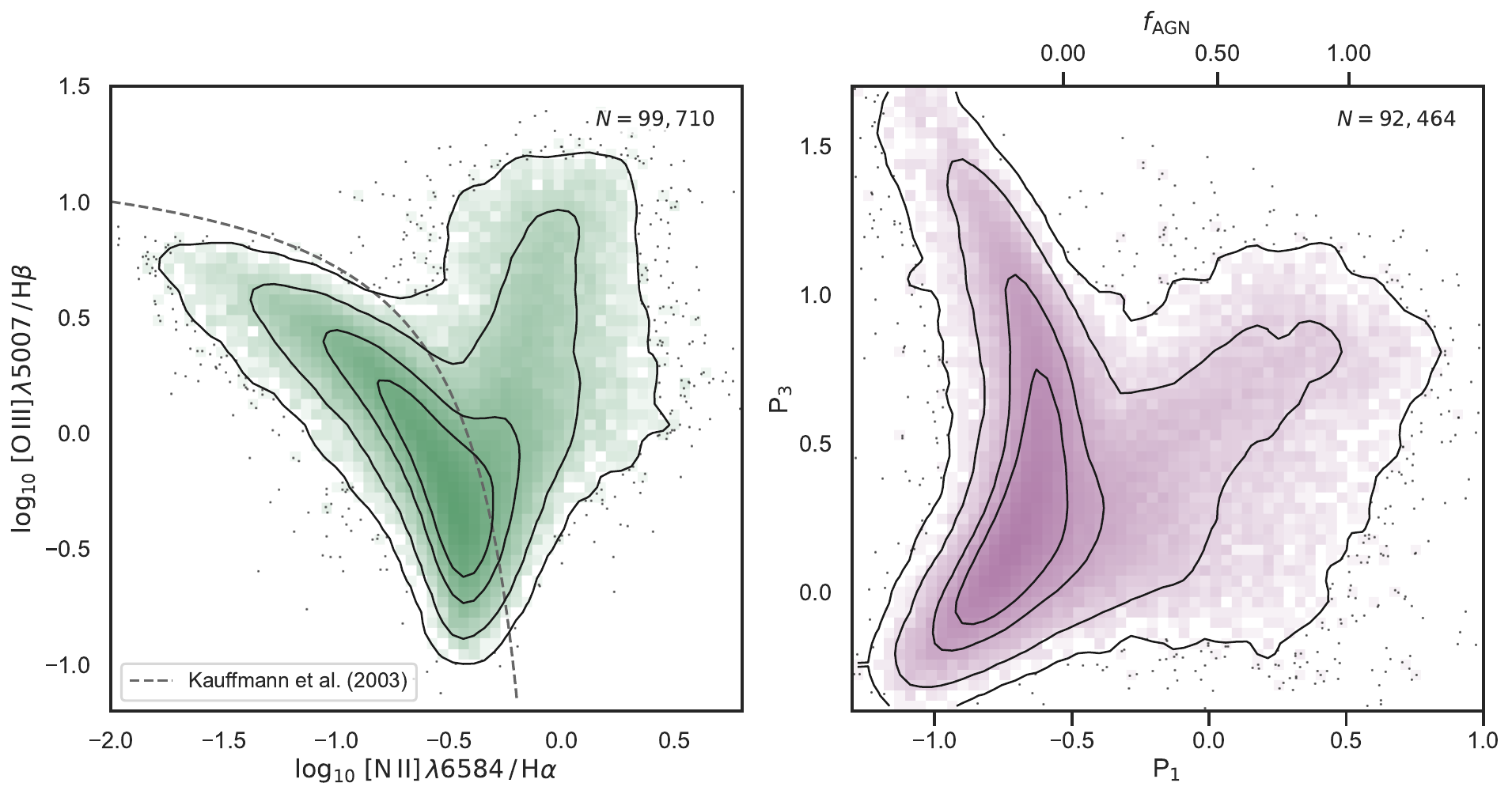}
\caption{AGN diagnostic diagrams for BGS galaxies in the SV3 bright
  program. In both panels, we require flux $\mathrm{S/N}>3$ in all the
  emission lines used to construct each diagram, and the background
  color map and contours show the number density of objects. (Left)
  Traditional BPT diagram \citep{baldwin81a}, where the dashed curve
  shows the empirical star-forming/AGN demarcation boundary from
  \citet{kauffmann03c}. (Right) $P_1$--$P_3$ optical diagnostic
  diagram of \citet{ji20a}, which separates star-forming galaxies (low
  $P_1$) from AGN (high $P_1$). $P_1$ and $P_3$ are defined in
  equations~(\ref{eq:p1}) and (\ref{eq:p3}). The top axis shows the
  AGN fraction $f_\mathrm{AGN}$ mapped from $P_1$ following
  \citet{jin21a}. \label{fig:bpt}}
\end{figure*}

Next, in Figure~\ref{fig:bpt} we construct two optical emission-line
diagnostic diagrams from the \fastspecfit{} VAC which can be used to
identify and characterize AGN. For simplicity, we show just BGS
targets from the SV3 bright program with a flux $\mathrm{S/N}>3$ on
all the relevant emission lines. For a more comprehensive sample of
AGN in the BGS sample and other target classes, see \citet{juneau25a},
\citet{siudek25a}, and S.~Juneau et~al., in preparation. The left-hand
panel of Figure~\ref{fig:bpt} shows the traditional BPT
\citep{baldwin81a} diagram, \oiiilam/\hb{} versus \niilam/\ha, with
the empirical star-forming/AGN demarcation curve from
\citet{kauffmann03c} shown as a dashed curve. Galaxies above and below
this boundary have line-emission consistent with ionization from an
AGN and star formation, respectively. Meanwhile, the star-formation
sequence itself is a sequence in gas-phase abundance, with a
second-order dependence on excitation or ionization parameter
\citep[e.g.,][]{kewley02a}.

The right-hand panel of Figure~\ref{fig:bpt} visualizes the same BGS
sample in the $P_1$--$P_3$ optical diagnostic diagram of
\citet{ji20a}, where
\begin{align}
  P_1 &\equiv  0.63\,\log(\text{\nii}/\ha) + 0.51\,\log(\text{\sii}/\ha) + 0.59\,\log(\text{\oiii}/\hb) \label{eq:p1}
\\
  P_3 &\equiv -0.46\,\log(\text{\nii}/\ha) - 0.37\,\log(\text{\sii}/\ha) + 0.81\,\log(\text{\oiii}/\hb) \label{eq:p3}
\end{align}
This panel contains $\approx8\%$ fewer objects than the left-hand
panel, due to the additional S/N cut on the \siidoublet{} doublet. In
this projection, $P_1$ maps onto a continuous AGN fraction,
$f_\mathrm{AGN}$ \citep{ji20a, jin21a}, the fractional contribution of
AGN photoionization to the total line emission. At fixed $P_3$,
\citet{ji20a} show that contours of constant ionization parameter run
nearly parallel to the $P_1$ axis, largely decoupled from variations
in gas-phase abundance, with second-order variations due to
differences in nitrogen abundance. Together, these diagrams show that
the \fastspec{} emission-line measurements recover the well-known
structure of the star-forming and AGN populations, making the DR2
catalog a powerful resource for statistical studies of nuclear
activity and ionized gas in galaxies.

\subsection{Planned Improvements}
\label{ssec:improvements}

\fastspecfit{} continues to be actively developed based on feedback
from a small but growing community of users both within and outside of
the DESI collaboration. We strive to fix reported bugs rapidly and to
implement user requests on a somewhat slower, but regular basis. In
addition, there is ongoing effort to improve and expand the
documentation and to add to the number of user-facing tutorials for
working with the code and VACs.

The next major development cycle will focus on making \fastspecfit{}'s
default inputs and assumptions easier for users to modify, building on
infrastructure that is already partly in place.  \fastspecfit{}
already supports a flexible line list and a fairly flexible
kinematic-constraints file (\S\ref{ssec:emlinelist},
\S\ref{sssec:kinematic}). Users can build and load (at runtime)
custom, project-specific files to add or remove lines and adjust
amplitude constraints without touching the fitting
code. \citet{scholte26a}, for example, extend the default line-list to
measure a larger number of electron-temperature-sensitive auroral
lines in a large sample of low-redshift star-forming galaxies. The
kinematic model choices, however, are more rigid; they currently only
support the narrow-only and narrow+broad options documented in
\S\ref{sssec:kinematic}. To address this shortcoming, we plan to
generalize the configuration file so that users can assign an
arbitrary mixture of Gaussian components to any line, not just the
Balmer series. With this change, for example, the \oiiidoublet{}
doublet and other forbidden lines can be modeled with multiple
kinematic components (with flexible constraints between them) rather
than a single Gaussian line \citep[e.g.,][]{harrison14a, pucha25a}.

Two additional fitting modes (\S\ref{sec:modes}) are also under active
development. The first, \texttt{fastqso}, is being designed to better
model AGN/QSO spectra (\S\ref{ssec:example_qso}). Instead of a
non-negative sum of SPS models, its continuum model will consist of a
UV power law, a velocity-broadened \ion{Fe}{2} (1100--3100~\AA)
template, a tunable amount of host-galaxy (stellar) light, and an
infrared torus component. Combined with the more flexible
emission-line modeling described above, \texttt{fastqso} will yield
significantly better rest-frame photometry, emission-line properties,
and more. The second fitting mode under development,
\texttt{fastbayes} (for ``fast Bayesian''), is being scoped as a
complementary alternative to \texttt{fastphot}. Instead of the current
maximum-likelihood method (\S\ref{sssec:fitting}), \texttt{fastbayes}
will interpolate and likelihood-weight a large precomputed grid of SPS
models to estimate the posterior probability distribution of stellar
mass, SFR, dust attenuation, and other physical properties. Because
the grid is precomputed rather than sampled on the fly, this approach
will return full posterior distributions in a fraction of the time
required by standard Bayesian fitting codes
\citep[e.g.,][]{johnson21a}. We defer a more complete description of
these and other developments to forthcoming papers.

\section{Summary}
\label{sec:summary}

\fastspecfit{} is a fast, flexible, and physically motivated code for
modeling DESI stellar continuum and emission-line spectra jointly with
broadband photometry. By convolving stellar-population and
emission-line templates with each fiber's native resolution matrix
(\S\ref{ssec:continuum} and \S\ref{ssec:emlines}) rather than
resampling the data, \fastspecfit{} preserves DESI's uncorrelated
pixel statistics while fitting a typical spectrum in $\approx0.1$~s on
a single CPU core. Its non-parametric SFH, energy-conserving dust
model, and independently fitted narrow and broad emission-line
kinematics let the data---rather than a pre-assumed galaxy
type---determine the best-fitting model, so the same machinery applies
uniformly across the full diversity of DESI extragalactic targets,
from quiescent LRGs to star-forming ELGs to broad-line QSOs
(\S\ref{sec:casestudy}). Its speed and hybrid MPI/multiprocessing
design make it practical to estimate parameter uncertainties from
repeated Monte Carlo fits (\S\ref{ssec:uncertainties}) and to scale
from a single laptop process to thousands of supercomputer cores
(\S\ref{ssec:performance}).

We use \fastspecfit{} to produce the DR2 VAC (\texttt{Loa/v2.0}), the
primary subject of this paper: approximately 38~million extragalactic
targets spanning the DESI commissioning, survey-validation,
special-program, and main-survey campaigns (\S\ref{ssec:catalog}),
including corrected redshifts for roughly 165,000 QSOs. We validate
the resulting stellar masses, velocity dispersions, and emission-line
fluxes against independent codes and catalogs
(\S\ref{ssec:validation}), and we illustrate the catalog's scientific
reach with the stellar mass--redshift distribution of the three
primary galaxy target classes and with optical emission-line
diagnostics of nuclear activity (\S\ref{ssec:reach}). Alongside the
\fastspec{} VAC we also produce a companion, photometry-only
\fastphot{} VAC using the same input dataset and template set
(Appendix~\ref{sec:appendix_fastphot}). We describe companion catalogs
for the Early Data Release and Data Release~1, produced with earlier
template and \fastspecfit{} versions, in
Appendix~\ref{sec:appendix_vacs}.

Several improvements are planned for future \fastspecfit{} releases,
including more flexible, multi-component emission-line kinematics and
two new fitting modes, \texttt{fastqso} for QSOs and broad-line AGN
and \texttt{fastbayes} for fast posterior estimation
(\S\ref{ssec:improvements}). The code and its measurements are already
enabling a broad range of extragalactic science using the public EDR
and DR1 VACs (\S\ref{sec:intro}), and a growing number of studies are
already drawing on the DR2 catalog described here, including
direct-method chemical abundance measurements for star-forming
galaxies \citep{scholte26a} and a search for the most metal-poor
star-forming galaxies yet found (J.~Moustakas et~al., in preparation);
we expect this number to grow substantially once the DR2 VAC is
publicly released in 2027.

\begin{acknowledgements}
JM gratefully acknowledges funding support from the U.S. Department of
Energy, Office of Science, Office of High Energy Physics under Award
Number DE-SC0020086, and DS acknowledges support from a UKRI Frontier
Research Guarantee Grant (PI Cullen; grant reference EP/X021025/1). We
also thank Ben Johnson for assistance constructing custom FSPS models,
Benjamin J. Weiner for the suggestion of fitting emission-line
doublets as an amplitude and a line-ratio, and Ashod Khederlarian for
contributing a tutorial to an early version of \fastspecfit.

This research used data obtained with the Dark Energy Spectroscopic
Instrument (DESI). DESI construction and operations is managed by the
Lawrence Berkeley National Laboratory. This material is based upon
work supported by the U.S. Department of Energy, Office of Science,
Office of High-Energy Physics, under Contract No. DE–AC02–05CH11231,
and by the National Energy Research Scientific Computing Center, a DOE
Office of Science User Facility under the same contract. Additional
support for DESI was provided by the U.S. National Science Foundation
(NSF), Division of Astronomical Sciences under Contract
No. AST-0950945 to the NSF’s National Optical-Infrared Astronomy
Research Laboratory; the Science and Technology Facilities Council of
the United Kingdom; the Gordon and Betty Moore Foundation; the
Heising-Simons Foundation; the French Alternative Energies and Atomic
Energy Commission (CEA); the National Council of Humanities, Science
and Technology of Mexico (CONAHCYT); the Ministry of Science and
Innovation of Spain (MICINN), and by the DESI Member Institutions:
\url{https://www.desi.lbl.gov/collaborating-institutions}. The DESI
collaboration is honored to be permitted to conduct scientific
research on I’oligam Du’ag (Kitt Peak), a mountain with particular
significance to the Tohono O’odham Nation. Any opinions, findings, and
conclusions or recommendations expressed in this material are those of
the author(s) and do not necessarily reflect the views of the
U.S. National Science Foundation, the U.S. Department of Energy, or
any of the listed funding agencies.

The Legacy Surveys consist of three individual and complementary
projects: the Dark Energy Camera Legacy Survey (DECaLS; Proposal ID
\#2014B-0404; PIs: David Schlegel and Arjun Dey), the Beijing-Arizona
Sky Survey (BASS; NOAO Prop. ID \#2015A-0801; PIs: Zhou Xu and Xiaohui
Fan), and the Mayall $z$-band Legacy Survey (MzLS; Prop. ID
\#2016A-0453; PI: Arjun Dey). DECaLS, BASS and MzLS together include
data obtained, respectively, at the Blanco telescope, Cerro Tololo
Inter-American Observatory, NSF’s NOIRLab; the Bok telescope, Steward
Observatory, University of Arizona; and the Mayall telescope, Kitt
Peak National Observatory, NOIRLab. Pipeline processing and analyses
of the data were supported by NOIRLab and the Lawrence Berkeley
National Laboratory (LBNL). The Legacy Surveys project is honored to
be permitted to conduct astronomical research on I’oligam Du’ag (Kitt
Peak), a mountain with particular significance to the Tohono O’odham
Nation.

NOIRLab is operated by the Association of Universities for Research in
Astronomy (AURA) under a cooperative agreement with the National
Science Foundation. LBNL is managed by the Regents of the University
of California under contract to the U.S. Department of Energy.

This project used data obtained with the Dark Energy Camera (DECam),
which was constructed by the Dark Energy Survey (DES)
collaboration. Funding for the DES Projects has been provided by the
U.S. Department of Energy, the U.S. National Science Foundation, the
Ministry of Science and Education of Spain, the Science and Technology
Facilities Council of the United Kingdom, the Higher Education Funding
Council for England, the National Center for Supercomputing
Applications at the University of Illinois at Urbana-Champaign, the
Kavli Institute of Cosmological Physics at the University of Chicago,
Center for Cosmology and Astro-Particle Physics at the Ohio State
University, the Mitchell Institute for Fundamental Physics and
Astronomy at Texas A\&M University, Financiadora de Estudos e
Projetos, Fundacao Carlos Chagas Filho de Amparo, Financiadora de
Estudos e Projetos, Fundacao Carlos Chagas Filho de Amparo a Pesquisa
do Estado do Rio de Janeiro, Conselho Nacional de Desenvolvimento
Cientifico e Tecnologico and the Ministerio da Ciencia, Tecnologia e
Inovacao, the Deutsche Forschungsgemeinschaft and the Collaborating
Institutions in the Dark Energy Survey. The Collaborating Institutions
are Argonne National Laboratory, the University of California at Santa
Cruz, the University of Cambridge, Centro de Investigaciones
Energeticas, Medioambientales y Tecnologicas-Madrid, the University of
Chicago, University College London, the DES-Brazil Consortium, the
University of Edinburgh, the Eidgenossische Technische Hochschule
(ETH) Zurich, Fermi National Accelerator Laboratory, the University of
Illinois at Urbana-Champaign, the Institut de Ciencies de l’Espai
(IEEC/CSIC), the Institut de Fisica d’Altes Energies, Lawrence
Berkeley National Laboratory, the Ludwig Maximilians Universitat
Munchen and the associated Excellence Cluster Universe, the University
of Michigan, NSF’s NOIRLab, the University of Nottingham, the Ohio
State University, the University of Pennsylvania, the University of
Portsmouth, SLAC National Accelerator Laboratory, Stanford University,
the University of Sussex, and Texas A\&M University.

BASS is a key project of the Telescope Access Program (TAP), which has
been funded by the National Astronomical Observatories of China, the
Chinese Academy of Sciences (the Strategic Priority Research Program
“The Emergence of Cosmological Structures” Grant \# XDB09000000), and
the Special Fund for Astronomy from the Ministry of Finance. The BASS
is also supported by the External Cooperation Program of Chinese
Academy of Sciences (Grant \# 114A11KYSB20160057), and Chinese National
Natural Science Foundation (Grant \# 12120101003, \# 11433005).

The Legacy Survey team makes use of data products from the Near-Earth
Object Wide-field Infrared Survey Explorer (NEOWISE), which is a
project of the Jet Propulsion Laboratory/California Institute of
Technology. NEOWISE is funded by the National Aeronautics and Space
Administration.

The Legacy Surveys imaging of the DESI footprint is supported by the
Director, Office of Science, Office of High Energy Physics of the
U.S. Department of Energy under Contract No. DE-AC02-05CH11231, by the
National Energy Research Scientific Computing Center, a DOE Office of
Science User Facility under the same contract; and by the
U.S. National Science Foundation, Division of Astronomical Sciences
under Contract No. AST-0950945 to NOAO.

Portions of the \fastspecfit{} code base, beginning with version
\texttt{3.3.0}, were developed with the assistance of Anthropic's
Claude (via Claude Code, models Sonnet~4.6 and 5). This assistance
included documentation, unit and integration tests, packaging and
automated testing and release workflows, parallel-processing and
production scripts, and parts of the continuum and emission-line
fitting code centered around optimization and speed. All AI-generated
code was reviewed, tested, and validated by the authors, who take full
responsibility for its correctness. In addition, Claude Code (models
Sonnet~5 and Opus~5.5) was used in the preparation of this manuscript
to generate outlines; translate the technical implementation in the
code into equations; develop and refactor the figure-generation and
analysis scripts and tables (all using author-provided data); suggest
candidate references (all validated by the authors); check the
manuscript for internal consistency (including quoted numerical
values, notation, and cross-references); suggest revisions to the text
for clarity and conciseness; and reformat the manuscript source for
submission. The authors reviewed and revised all AI-generated text and
take full responsibility for the content of this paper.

\end{acknowledgements}

\software{
  \texttt{Astropy} \citep{astropy-collaboration22a},
  \texttt{corner.py} \citep{foreman-mackey16a},
  \texttt{FastSpecFit} \citep{moustakas23b},
  \texttt{healpy} \citep{zonca19a},
  \texttt{mpi4py} \citep{dalcin21a},
  \texttt{NumPy} \citep{harris20a},
  \texttt{Numba} \citep{lam15a},
  \texttt{Matplotlib} \citep{hunter07a},
  \texttt{SciPy} \citep{virtanen20a},
  \texttt{seaborn} \citep{waskom21a},
  \texttt{speclite} \citep{kirkby24a}
}

\begin{appendix}

\section{DR2 \fastphot{} VAC}
\label{sec:appendix_fastphot}

This appendix briefly documents the \fastphot{} \texttt{Loa/v2.0} VAC,
a closely related but distinct catalog to the \fastspec{} VAC
introduced in \S\ref{sec:vac}. As described in \S\ref{sec:modes},
\fastphot{} is an alternative fitting mode available within the larger
\fastspecfit{} software package which models just broadband
photometry, but at the fixed DESI spectroscopic redshift. The VAC has
been generated using the identical input DESI dataset (\texttt{Loa} specprod),
\fastspecfit{} software tag (\texttt{3.6.1}), and template set
(\texttt{ftemplates-chabrier-2.2.0.fits}); the sample selection,
corrected QSO redshifts, and final object counts are identical to
those given in Table~\ref{table:sample}.

Because \fastphot{} fits only broadband photometry, its per-HEALPix
files contain three FITS extensions rather than \fastspec{}'s
five---\texttt{PRIMARY}, \texttt{METADATA}, and
\texttt{SPECPHOT}---mirroring the corresponding \fastspec{} extensions
but omitting the emission-line \texttt{FASTSPEC} table and the
\texttt{MODELS} best-fitting-spectra extension entirely, since there
is no spectrum to model or emission lines to fit
(\S\ref{ssec:catalog}). Its per-HEALPix and merged catalogs, on the
other hand, follow the same
\texttt{SURVEY}/\texttt{PROGRAM}/\texttt{HEALPIX} naming convention as
\fastspec, but with a ``fastphot'' rather than ``fastspec''
prefix. For a detailed column-by-column description of the \fastphot{}
data model, pinned to the \fastspecfit{} version used to produce this
VAC, please see the online
documentation.\footnote{\url{https://fastspecfit.readthedocs.io/en/latest/fastphot.html}}

\begin{figure*}
\centering\includegraphics[width=1.0\textwidth]{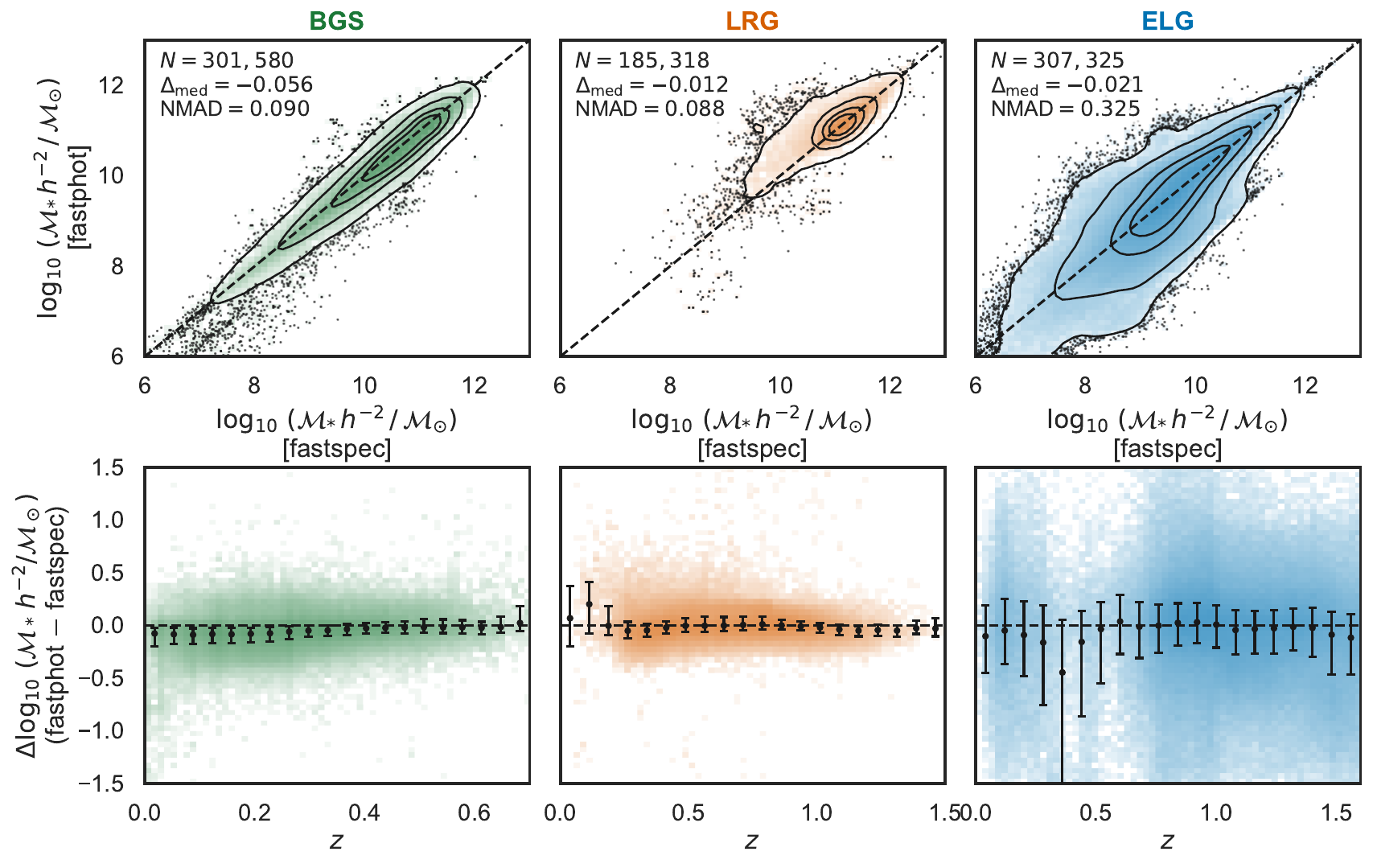}
\caption{Comparison of \fastspecfit{} stellar masses derived from
  spectrophotometric fitting (\fastspec) and from photometry-only
  fitting (\fastphot{}), organized by target class.  \textit{Top row:}
  \fastphot{} $\logmstar$ versus \fastspec{} $\logmstar$ ($h=1$,
  Chabrier IMF) for BGS (left), LRG (center), and ELG (right)
  subsamples.  As in Figure~\ref{fig:compare-mstar-external}, the color
  map, contours, and outlying markers show the two-dimensional density
  distribution, and the dashed line is the one-to-one relation; each
  panel lists the matched sample size $N$, the median mass offset
  $\Delta_\textrm{med}$ (\fastphot{} minus \fastspec) in dex, and the
  \texttt{NMAD} scatter.
  \textit{Bottom row:} Residual $\Delta\logmstar$ (\fastphot{} minus
  \fastspec) as a function of redshift for the same three subsamples.
  Black points mark the running median in bins of redshift, with error
  bars enclosing the interquartile range.  The dashed horizontal line
  marks zero offset.  We attribute the systematic offset seen in BGS
  targets at low redshift to imperfect aperture corrections for the
  most extended, nearby galaxies, whose half-light radii subtend a
  significant fraction of the DESI fiber
  diameter.\label{fig:compare_mstar_fastphot}}
\end{figure*}

Besides being substantially faster to produce (median runtime is
0.42~s/object/core, an order of magnitude faster than \fastspec;
\S\ref{ssec:performance}), \fastphot{} offers a purely photometric
alternative for science cases that do not require the DESI spectrum
itself. For example, its K-corrections and global physical properties,
including the inferred stellar mass, come from the integrated
broadband photometry, without the simplistic and uncertain
aperture-correction \fastspec{} uses (\S\ref{sssec:apercorr}).

To explore just one example, in
Figure~\ref{fig:compare_mstar_fastphot} we compare the stellar masses
we derive using \fastspec{} and \fastphot{} split by DESI target
class. Overall, the agreement is very good, with median offsets and
\texttt{NMAD} scatter of $-0.056\pm0.091$~dex, $-0.012\pm0.088$~dex,
and $-0.019\pm0.324$~dex for BGS, LRG, and ELG targets,
respectively. Examining the residuals with redshift, BGS targets at
$z\lesssim0.4$ exhibit the most pronounced differences in exactly the
sense that you might expect from aperture effects. The systematically
higher \fastspec{} stellar masses are possibly due to the larger
mass-to-light ratio (older stellar ages) exhibited
in the centers of galaxies and traced by the central DESI
spectrum. Even so, the median offset does not exceed $|\Delta_\textrm{med}|\approx0.05$~dex
for any of the three target classes,
underscoring the overall agreement between the two fitting modes.

\section{Value-Added Catalog Revision History}
\label{sec:appendix_vacs}

\fastspecfit{} has been in near-continuous development since late-2020
(not coincidentally, the beginning of DESI on-sky observations), and
so the code base and its template library have evolved progressively
and, in some cases, substantially over time. Concurrently, each new
DESI specprod has incorporated both minor and major improvements in
the data-reduction pipeline, as well as adding spectra for millions of
new objects. In order to meet the significant demand for astrophysical
measurements of DESI targets beyond just their redshift---both within
and outside of the DESI collaboration---\fastspecfit{} has been on a
fairly rapid development cycle, resulting in 12 unique \fastspec{}
VACs and four unique \fastphot{} VACs (including the ones documented
in this paper) since 2022 August. In \S\ref{sec:intro} we cite just a
small fraction of the publications which have resulted from these
datasets.

\input{table-vac-history}

In Table~\ref{table:vac-history} we document this history by listing
the current set of \fastspec{} and \fastphot{} VACs, the date they
were released, and the templates and version of \fastspecfit{} which
were used, organized by DESI specprod and data release. For additional
details regarding each of these releases, especially substantive
algorithmic changes, major bug fixes, and much more, we refer to the
detailed \fastspecfit{} change log and release
notes.\footnote{\url{https://github.com/desihub/fastspecfit/blob/main/doc/changes.rst}}

In the remainder of this appendix, we quantify the statistical
  differences across a subset of these releases for a handful of key
  quantities measured by \fastspecfit. We match \texttt{Loa/v2.0} to
  the three earlier \fastspec{} VACs which have been public the
  longest, \texttt{Iron/v3.0}, \texttt{Iron/v2.1}, and
  \texttt{Fuji/v3.2}. We omit \texttt{Iron/v4.0}, which we produced
  with the same code and templates as \texttt{Loa/v2.0}, as well as
  the earlier \texttt{Iron} and \texttt{Fuji} VACs, which were only
  available internally to DESI collaboration members before being
  superseded. We also restrict the comparison to SV3, whose targets
  appear in every release and whose coadded spectra include the same
  exposures in each specprod. After matching on \texttt{TARGETID}, we
  require each object to satisfy the good-redshift criteria for its
  target class (\S\ref{ssec:cuts}) in both catalogs. Among these
  objects, fewer than 0.2\% of the BGS, LRG, and ELG targets have
  redshifts which differ by more than 1000~\kms{} between
  releases. For QSO targets, however, this fraction rises to 12\%
  relative to \texttt{Iron} and 19\% relative to \texttt{Fuji}, which
  we attribute to changes in both the \texttt{QuasarNet}
  classifications and Redrock redshifts between specprods and in how
  \fastspecfit{} updates its QSO redshifts (\S\ref{ssec:catalog}). We
  exclude objects whose redshift changed from the following
  comparisons.

\input{table-vac-compare}

In Table~\ref{table:vac-compare} we list the median difference
  and NMAD of several key quantities for each target class. Relative
  to \texttt{Iron/v3.0}, the median offsets are smaller than 0.01~dex
  (or 0.01~mag) for every quantity except the SFR. For BGS and LRG
  targets, the scatter in stellar mass, absolute magnitude, velocity
  dispersion, and emission-line flux is also below 0.01~dex (or
  0.01~mag). The small positive offset in \vdisp{} (0.004--0.007~dex)
  is consistent with the correction for the intrinsic resolution of
  the C3K templates (equation~\ref{eq:vdisptot}), which we introduced
  in \fastspecfit{}~3.5.0. Relative to \texttt{Iron/v2.1} and
  \texttt{Fuji/v3.2}, both produced with \fastspecfit{}~2.5.0, the
  scatter is larger (typically 0.02--0.06~dex), but the median offsets
  remain below 0.015~dex. The notable exception is the stellar mass of
  ELGs, which scatters by 0.21--0.28~dex relative to these two catalogs
  with no significant median offset. We attribute this scatter to the
  substantial changes in our SPS templates since 2.5.0
  (Table~\ref{table:vac-history}), combined with the low
  S/N of most ELG continua. Moreover, because
  \texttt{Iron/v2.1} and \texttt{Fuji/v3.2} share the same code and
  templates, the difference between their two columns isolates the
  effect of the data-reduction changes from \texttt{Fuji} to
  \texttt{Iron}, which increase the scatter only modestly.

Finally, the SFRs in earlier VACs are not directly comparable to
  those in \texttt{Loa/v2.0}. Starting with \fastspecfit{}~3.5.0, we
  report the SFR averaged over the most recent 100~Myr
  (equation~\ref{eq:sfr}). Earlier versions instead reported the
  instantaneous SFR of the best-fitting model, to which only the
  youngest (0--30~Myr) age bin contributes. We attribute the large SFR
  differences in Table~\ref{table:vac-compare} (median offsets up to
  1.2~dex and NMAD values of 0.5--1.9~dex) primarily to this change in
  definition, and we recommend against combining SFRs from
  \texttt{Loa/v2.0} with those from any earlier VAC.

\end{appendix}

\bibliographystyle{aasjournalv7.1}
\bibliography{refs}

\end{document}

%% file: table-emlines.tex
\startlongtable
\begin{deluxetable}{llccc}
\tablecaption{Emission Lines\label{table:emlines}}
\tablewidth{0pt}
\tablehead{
\colhead{Emission} &
\colhead{Line} &
\colhead{Rest} &
\colhead{Initial} &
\colhead{} \\
\colhead{Line} &
\colhead{Prefix\tablenotemark{a}} &
\colhead{Wavelength\tablenotemark{b}} &
\colhead{Patch\tablenotemark{c}} &
\colhead{``Strong''\tablenotemark{d}}
}
\startdata
\cutinhead{UV/QSO Lines}
Ly$\alpha$                        & \texttt{LYALPHA}        & 1215.670   & a & \checkmark \\
{\ion{N}{5}}~$\lambda$1240      & \texttt{NV\_1240}       & 1240.14    & a & \checkmark \\
{\ion{O}{1}}~$\lambda$1304        & \texttt{OI\_1304}       & 1304.35    & b &            \\
{\ion{Si}{4}}~$\lambda$1396       & \texttt{SILIV\_1396}    & 1396.76    & c &            \\
{\ion{C}{4}}~$\lambda$1549        & \texttt{CIV\_1549}      & 1549.06    & d & \checkmark \\
{\ion{He}{2}}~$\lambda$1640       & \texttt{HEII\_1640}     & 1640.42    & e &            \\
{\ion{Al}{3}}~$\lambda$1857       & \texttt{ALIII\_1857}    & 1857.40    & f &            \\
{\ion{Si}{3}}]~$\lambda$1892      & \texttt{SILIII\_1892}   & 1892.030   & f & \checkmark \\
{\ion{C}{3}}]~$\lambda$1908       & \texttt{CIII\_1908}     & 1908.734   & f & \checkmark \\
{\ion{Mg}{2}}~$\lambda$2796       & \texttt{MGII\_2796}     & 2796.3511  & g & \checkmark \\
{\ion{Mg}{2}}~$\lambda$2803       & \texttt{MGII\_2803}     & 2803.5324  & g & \checkmark \\
{\ion{He}{2}}~$\lambda$4686       & \texttt{HEII\_4686}     & 4687.02    & o &            \\
\cutinhead{Narrow Balmer/Helium Lines}
H6                                & \texttt{H6}             & 3890.1576  & j &            \\
H$\epsilon$                       & \texttt{HEPSILON}       & 3971.195   & k &            \\
H$\delta$                         & \texttt{HDELTA}         & 4102.892   & l &            \\
H$\gamma$                         & \texttt{HGAMMA}         & 4341.684   & m & \checkmark \\
{\ion{He}{1}}~$\lambda$4471       & \texttt{HEI\_4471}      & 4472.7290  & n &            \\
H$\beta$                          & \texttt{HBETA}          & 4862.683   & p & \checkmark \\
{\ion{He}{1}}~$\lambda$5876       & \texttt{HEI\_5876}      & 5877.2690  & r &            \\
H$\alpha$                         & \texttt{HALPHA}         & 6564.613   & t & \checkmark \\
\cutinhead{Broad Balmer Lines}
H$6_\textrm{b}$                      & \texttt{H6\_BROAD}       & 3890.1576  & j &            \\
H$\epsilon_\textrm{b}$               & \texttt{HEPSILON\_BROAD} & 3971.195   & k &            \\
H$\delta_\textrm{b}$                 & \texttt{HDELTA\_BROAD}   & 4102.892   & l &            \\
H$\gamma_\textrm{b}$                 & \texttt{HGAMMA\_BROAD}   & 4341.684   & m & \checkmark \\
H$\beta_\textrm{b}$                  & \texttt{HBETA\_BROAD}    & 4862.683   & p & \checkmark \\
H$\alpha_\textrm{b}$                 & \texttt{HALPHA\_BROAD}   & 6564.613   & t & \checkmark \\
\cutinhead{Narrow Forbidden Lines}
{[\ion{Ne}{5}]}~$\lambda$3346     & \texttt{NEV\_3346}      & 3346.7828  & h &            \\
{[\ion{Ne}{5}]}~$\lambda$3426     & \texttt{NEV\_3426}      & 3426.8637  & h &            \\
{[\ion{O}{2}]}~$\lambda$3726      & \texttt{OII\_3726}      & 3727.0919  & i & \checkmark \\
{[\ion{O}{2}]}~$\lambda$3729      & \texttt{OII\_3729}      & 3729.8750  & i & \checkmark \\
{[\ion{Ne}{3}]}~$\lambda$3869     & \texttt{NEIII\_3869}    & 3869.8611  & j &            \\
{[\ion{O}{3}]}~$\lambda$4363      & \texttt{OIII\_4363}     & 4364.4351  & m &            \\
{[\ion{O}{3}]}~$\lambda$4959      & \texttt{OIII\_4959}     & 4960.2937  & p & \checkmark \\
{[\ion{O}{3}]}~$\lambda$5007      & \texttt{OIII\_5007}     & 5008.2383  & p & \checkmark \\
{[\ion{N}{2}]}~$\lambda$5755      & \texttt{NII\_5755}      & 5756.1887  & q &            \\
{[\ion{O}{1}]}~$\lambda$6300      & \texttt{OI\_6300}       & 6302.0435  & s &            \\
{[\ion{S}{3}]}~$\lambda$6312      & \texttt{SIII\_6312}     & 6313.8062  & s &            \\
{[\ion{N}{2}]}~$\lambda$6548      & \texttt{NII\_6548}      & 6549.8578  & t & \checkmark \\
{[\ion{N}{2}]}~$\lambda$6584      & \texttt{NII\_6584}      & 6585.2696  & t & \checkmark \\
{[\ion{S}{2}]}~$\lambda$6716      & \texttt{SII\_6716}      & 6718.2913  & u & \checkmark \\
{[\ion{S}{2}]}~$\lambda$6731      & \texttt{SII\_6731}      & 6732.6705  & u & \checkmark \\
{[\ion{Ar}{3}]}~$\lambda$7135     & \texttt{ARIII\_7135}    & 7137.77    & v &            \\
{[\ion{O}{2}]}~$\lambda$7320      & \texttt{OII\_7320}      & 7321.731   & v &            \\
{[\ion{O}{2}]}~$\lambda$7330      & \texttt{OII\_7330}      & 7332.199   & v &            \\
{[\ion{S}{3}]}~$\lambda$9069      & \texttt{SIII\_9069}     & 9071.103   & w &            \\
{[\ion{S}{3}]}~$\lambda$9532      & \texttt{SIII\_9532}     & 9533.227   & w &            \\
\enddata

\tablenotetext{a}{Data-model prefix for every output quantity we
  measure for this line (\S\ref{sssec:lines}), e.g.,
  \texttt{OIII\_5007\_FLUX}}
\tablenotetext{b}{Vacuum wavelengths in \AA{} come from the NIST Atomic Spectra
Database, the CHIANTI atomic database, or from \citet{vanden-berk01a};
see \S\ref{ssec:emlinelist} for more details.}
\tablenotetext{c}{Initial assignment of spectrally adjacent lines which are jointly fit by the line-masker (\S\ref{ssec:patches}).}
\tablenotetext{d}{Lines fit in the line-masker's initial patch-fitting pass (\S\ref{ssec:patches}) to estimate robust kinematics; if present, these are the lines most likely to be well detected.}
\end{deluxetable}

%% file: table-kinematic-bounds.tex
\setlength{\tabcolsep}{4pt}
\begin{deluxetable*}{llllcc}
\tablecaption{Emission-Line Kinematic Groups and Amplitude Constraints\label{table:kinematic}}
\tablewidth{0pt}
\tablehead{
  \colhead{Profile} &
  \colhead{Group} &
  \colhead{Anchor} &
  \colhead{$\sigma$ (km\,s$^{-1}$)} &
  \colhead{$|\Delta v|$ (km\,s$^{-1}$)} &
  \colhead{$\Delta v$ released\tablenotemark{a}}
}
\startdata
\multicolumn{6}{c}{\textbf{Global (both models)}} \\
Global & Free UV/AGN & \nodata & 1--10{,}000 & $<$2{,}500 & \nodata \\
\multicolumn{6}{l}{\hskip1.5em\textit{Lines (independent):}
  Ly$\alpha$,
  [\ion{N}{5}]~$\lambda$1240,
  \ion{O}{1}~$\lambda$1304,
  \ion{Si}{4}~$\lambda$1396,
  \ion{C}{4}~$\lambda$1549,
  \ion{He}{2}~$\lambda$1640,
  \ion{Al}{3}~$\lambda$1857,
  \ion{He}{2}~$\lambda$4686} \\
Global & \ion{C}{3}] pair & \ion{C}{3}]~$\lambda$1908 & 1--10{,}000 & $<$2{,}500 & N \\
\multicolumn{6}{l}{\hskip1.5em\textit{Member:} \ion{Si}{3}]~$\lambda$1892} \\
Global & \ion{Mg}{2} pair & \ion{Mg}{2}~$\lambda$2803 & 1--10{,}000 & $<$2{,}500 & N \\
\multicolumn{6}{l}{\hskip1.5em\textit{Member:} \ion{Mg}{2}~$\lambda$2796 (amplitude ratio free; see below)} \\
\hline
\multicolumn{6}{c}{\textbf{Narrow-only profile}} \\
Narrow-only & Forbidden & [\ion{O}{3}]~$\lambda$5007 & 1--750 & $<$500 & Y \\
\multicolumn{6}{l}{\hskip1.5em\textit{Members:}
  [\ion{Ne}{5}]~$\lambda\lambda$3346,3426,
  [\ion{O}{2}]~$\lambda\lambda$3726,3729,
  [\ion{Ne}{3}]~$\lambda$3869,
  [\ion{O}{3}]~$\lambda\lambda$4363,4959,} \\
\multicolumn{6}{l}{\hskip1.5em
  [\ion{N}{2}]~$\lambda$5755,
  [\ion{O}{1}]~$\lambda$6300,
  [\ion{S}{3}]~$\lambda$6312,
  [\ion{N}{2}]~$\lambda\lambda$6548,6584,
  [\ion{S}{2}]~$\lambda\lambda$6716,6731,} \\
\multicolumn{6}{l}{\hskip1.5em
  [\ion{Ar}{3}]~$\lambda$7135,
  [\ion{O}{2}]~$\lambda\lambda$7320,7330,
  [\ion{S}{3}]~$\lambda\lambda$9069,9532} \\
Narrow-only & Narrow Balmer\,+\,\ion{He}{1} & H$\alpha$ & 1--750 & $<$500 & Y \\
\multicolumn{6}{l}{\hskip1.5em\textit{Members:}
  H6,
  H$\epsilon$,
  H$\delta$,
  H$\gamma$,
  H$\beta$,
  \ion{He}{1}~$\lambda$4471,
  \ion{He}{1}~$\lambda$5876} \\
Narrow-only & Fixed ($A = 0$) & \nodata & \nodata & \nodata & \nodata \\
\multicolumn{6}{l}{\hskip1.5em\textit{Lines:}
  H$\alpha_\textrm{b}$,
  H$\beta_\textrm{b}$,
  H$\gamma_\textrm{b}$,
  H$\delta_\textrm{b}$,
  H$\epsilon_\textrm{b}$,
  H$6_\textrm{b}$} \\
\hline
\multicolumn{6}{c}{\textbf{Narrow+broad profile}} \\
Narrow+broad & All narrow & H$\alpha$ & 1--750 & $<$500 & N \\
\multicolumn{6}{l}{\hskip1.5em\textit{Members:}
  [\ion{Ne}{5}]~$\lambda\lambda$3346,3426,
  [\ion{O}{2}]~$\lambda\lambda$3726,3729,
  [\ion{Ne}{3}]~$\lambda$3869,
  [\ion{O}{3}]~$\lambda$4363,} \\
\multicolumn{6}{l}{\hskip1.5em
  [\ion{N}{2}]~$\lambda$5755,
  [\ion{O}{1}]~$\lambda$6300,
  [\ion{S}{3}]~$\lambda$6312,
  [\ion{N}{2}]~$\lambda\lambda$6548,6584,
  [\ion{S}{2}]~$\lambda\lambda$6716,6731,} \\
\multicolumn{6}{l}{\hskip1.5em
  [\ion{Ar}{3}]~$\lambda$7135,
  [\ion{O}{2}]~$\lambda\lambda$7320,7330,
  [\ion{S}{3}]~$\lambda\lambda$9069,9532,} \\
\multicolumn{6}{l}{\hskip1.5em
  H6,
  H$\epsilon$,
  H$\delta$,
  H$\gamma$,
  H$\beta$,
  \ion{He}{1}~$\lambda$4471,
  \ion{He}{1}~$\lambda$5876} \\
Narrow+broad & [\ion{O}{3}] doublet & [\ion{O}{3}]~$\lambda$5007 & 1--750 & $<$500 & N \\
\multicolumn{6}{l}{\hskip1.5em\textit{Member:} [\ion{O}{3}]~$\lambda$4959} \\
Narrow+broad & Broad Balmer & H$\alpha_\textrm{b}$ & 1--10{,}000 & $<$2{,}500 & N \\
\multicolumn{6}{l}{\hskip1.5em\textit{Members:}
  H$\beta_\textrm{b}$,
  H$\gamma_\textrm{b}$,
  H$\delta_\textrm{b}$,
  H$\epsilon_\textrm{b}$,
  H$6_\textrm{b}$} \\
\hline
\multicolumn{6}{c}{\textbf{Amplitude constraints}} \\
\multicolumn{3}{c}{\textit{Fixed ratios} ($A_\mathrm{line} = f \cdot A_\mathrm{ref}$)} &
\multicolumn{3}{c}{\textit{Free doublet ratios} ($r \equiv A_\mathrm{line}/A_\mathrm{ref}$)} \\
\relax[\ion{N}{2}]~$\lambda$6548\,/\,$\lambda$6584 & $f = 1/3.033$ & & [\ion{O}{2}]~$\lambda$3726\,/\,$\lambda$3729 & $r_0 = 1$ & $r \in [0,2]$ \\
\relax[\ion{O}{3}]~$\lambda$4959\,/\,$\lambda$5007 & $f = 1/2.964$ & & [\ion{S}{2}]~$\lambda$6731\,/\,$\lambda$6716  & $r_0 = 1$ & $r \in [0,2]$ \\
\relax[\ion{O}{2}]~$\lambda$7330\,/\,$\lambda$7320 & $f = 1/1.225$ & & \ion{Mg}{2}~$\lambda$2796\,/\,$\lambda$2803   & $r_0 = 1$ & $r \in [0,10]$ \\
\enddata
\tablenotetext{a}{Whether individual velocity shifts are freed in the final
  optimization pass (\S\ref{sssec:kinematic}).  In groups marked Y, $\Delta v$
  is released per member while $\sigma$ remains tied.  Global groups and the
  narrow+broad profile perform no kinematic relaxation.}
\tablecomments{All kinematic constraints, optimizer bounds, and amplitude ratios
  are read from \texttt{data/emline-constraints.yaml} at startup.  Lines within
  a group share a common velocity shift $\Delta v$ and width $\sigma$; the
  anchor carries the free kinematic parameters.  Subscript~b denotes the broad
  component of a Balmer line.}
\end{deluxetable*}
\setlength{\tabcolsep}{6pt}  

%% file: table-linesummary.tex
\begin{deluxetable}{lccc}
\tablecaption{Per-Line Measurements\label{table:linesummary}}
\tablewidth{0pt}
\tablehead{
  \colhead{Symbol} &
  \colhead{Column\tablenotemark{a}} &
  \colhead{Units} &
  \colhead{Inverse variance?}
}
\startdata
$A$                       & \texttt{MODELAMP}    & \fluxdensity & \nodata \\
$A_\mathrm{obs}$          & \texttt{AMP}         & \fluxdensity & Y \\
$\Delta v$                & \texttt{VSHIFT}      & \kms         & Y \\
$\sigma$                  & \texttt{SIGMA}       & \kms         & Y \\
$\Phi$                    & \texttt{FLUX}        & \flux        & Y \\
$\Phi_\mathrm{box}$       & \texttt{BOXFLUX}     & \flux        & Y \\
$F_c$                     & \texttt{CONT}        & \fluxdensity & Y \\
EW                        & \texttt{EW}          & \AA          & Y \\
$\Phi_\mathrm{lim}$       & \texttt{FLUX\_LIMIT} & \flux        & \nodata \\
$\chi^2_\mathrm{line}$    & \texttt{CHI2}        & \nodata      & \nodata \\
$n_\mathrm{pix}^\mathrm{line}$ & \texttt{NPIX}   & \nodata      & \nodata \\
\enddata
\tablenotetext{a}{Each column is prefixed by the line name (e.g.,
\texttt{OIII\_5007\_FLUX}); Table~\ref{table:emlines} lists the full
set of modeled lines.}
\tablecomments{Columns marked ``Y'' have a companion \texttt{\_IVAR}
column storing the inverse variance, estimated via the same
per-object Monte Carlo procedure described in \S\ref{ssec:uncertainties}.
$A$, $\Delta v$, and $\sigma$ are defined in \S\ref{sssec:emline_model};
$A_\mathrm{obs}$, $\Phi$, $\Phi_\mathrm{box}$, $F_c$, EW, $\Phi_\mathrm{lim}$,
$\chi^2_\mathrm{line}$, and $n_\mathrm{pix}^\mathrm{line}$ are defined in
\S\ref{sssec:lines}.}
\end{deluxetable}

%% file: table-absmag-bands.tex
\begin{deluxetable}{llcc}
\tablecaption{Rest-Frame Bandpasses for Absolute Magnitudes and K-Corrections\label{table:absmag-bands}}
\tablewidth{0pt}
\tablehead{
\colhead{Band} &
\colhead{Filter Curve} &
\colhead{$\lambda_\mathrm{eff}$\tablenotemark{a}} &
\colhead{Band-shift $z_0$}
}
\startdata
DECam $g$    & \texttt{decam2014-g} & 4890  & 1.0 \\
DECam $r$    & \texttt{decam2014-r} & 6470  & 1.0 \\
DECam $z$    & \texttt{decam2014-z} & 9196  & 1.0 \\
Johnson $U$  & \texttt{bessell-U}   & 3618  & 0.0 \\
Johnson $B$  & \texttt{bessell-B}   & 4442  & 0.0 \\
Johnson $V$  & \texttt{bessell-V}   & 5536  & 0.0 \\
2MASS $J$    & \texttt{twomass-J}   & 12408 & 0.0 \\
SDSS $u$     & \texttt{sdss2010-u}  & 3614  & 0.1 \\
SDSS $g$     & \texttt{sdss2010-g}  & 4749  & 0.1 \\
SDSS $r$     & \texttt{sdss2010-r}  & 6206  & 0.1 \\
SDSS $i$     & \texttt{sdss2010-i}  & 7526  & 0.1 \\
SDSS $z$     & \texttt{sdss2010-z}  & 8947  & 0.1 \\
WISE $W1$    & \texttt{wise2010-W1} & 34003 & 0.1 \\
\enddata
\tablecomments{Default rest-frame bandpasses, band-shifts $z_0$, and
  \texttt{speclite} filter curves used to infer K-corrections and
  absolute magnitudes (\S\ref{ssec:quantities}), as configured in the
  bundled \texttt{data/legacysurvey-dr9.yaml} parameter file.}
\tablenotetext{a}{Effective filter wavelengths in \AA{} are reported
  for the unshifted filter curve.}
\end{deluxetable}

%% file: table-sample.tex
\begin{deluxetable*}{lrr}
\tablecaption{Summary of the DR2/\texttt{Loa} \fastspecfit{} Value-Added Catalog\label{table:sample}}
\tablewidth{0pt}
\tablehead{
  \colhead{Program} &
  \colhead{\shortstack{Number of\\Objects}} &
  \colhead{\shortstack{Number of Corrected\\QSO Redshifts\tablenotemark{a}}}
}
\startdata
\cutinhead{Main Survey}
backup &     60,372 &        0 \\
bright & 13,185,499 &    1,161 \\
dark   & 23,273,230 &  157,118 \\
total  & 36,519,101 &  158,279 \\
\cutinhead{Special Survey}
backup &        512 &        0 \\
bright &    105,172 &        1 \\
dark   &    261,706 &      232 \\
total  &    367,390 &      233 \\
\cutinhead{Survey Validation\tablenotemark{b}}
backup &      2,825 &       17 \\
bright &    438,456 &       75 \\
dark   &    880,642 &    6,009 \\
other  &     36,769 &      106 \\
total  &  1,358,692 &    6,207 \\
\tableline
Total & 38,245,183 &  164,719 \\
\enddata
\tablenotetext{a}{Described in \S\ref{ssec:catalog}.}
\tablenotetext{b}{Sum of CMX, SV1, SV2, and SV3 surveys.}

\end{deluxetable*}

%% file: table-vac-history.tex
\startlongtable
\begin{deluxetable}{cccc}
\tablecaption{\fastspecfit{} VAC Revision History\label{table:vac-history}}
\tablewidth{0pt}
\tablehead{
  \colhead{VAC Version} &
  \colhead{Release Date} &
  \colhead{Software Tag(s)\tablenotemark{a}} &
  \colhead{Templates\tablenotemark{b}}
}
\startdata
\cutinhead{\texttt{Loa} (DR2) --- \fastspec}
v2.0 & 2027 Apr\tablenotemark{c} & 3.6.1 & chabrier-2.2.0 \\
v1.0 & 2027 Apr\tablenotemark{c} & 3.1.4 & chabrier-2.0.0 \\
\cutinhead{\texttt{Loa} (DR2) --- \fastphot}
v2.0 & 2027 Apr\tablenotemark{c} & 3.6.1 & chabrier-2.2.0 \\
v1.0 & 2027 Apr\tablenotemark{c} & 3.2.0 & chabrier-2.0.0 \\
\cutinhead{\texttt{Iron} (DR1) --- \fastspec}
v4.0 & 2026 Dec & 3.6.1 & chabrier-2.2.0 \\
v3.0 & 2025 Jun & 3.1.5 & chabrier-2.0.0 \\
v2.1 & 2024 Jan & 2.5.0 & chabrier-1.3.0 \\
v2.0 & 2023 Aug & 2.4.1, 2.4.2 & chabrier-1.1.0 \\
v1.0 & 2023 Feb & 2.1.0, 2.1.1 & chabrier-1.0.0 \\
\cutinhead{\texttt{Iron} (DR1) --- \fastphot}
v2.0 & 2026 Dec & 3.6.1 & chabrier-2.2.0 \\
v1.0 & 2026 Jun & 3.2.0 & chabrier-2.0.0 \\
\cutinhead{\texttt{Fuji} (EDR) --- \fastspec}
v3.2 & 2024 Jan & 2.5.0 & chabrier-1.3.0 \\
v3.1 & 2023 Dec & 2.4.3 & chabrier-1.2.0 \\
v3.0 & 2023 Aug & 2.4.1, 2.4.2 & chabrier-1.1.0 \\
v2.0 & 2023 Feb & 2.1.0, 2.1.1 & chabrier-1.0.0 \\
v1.0 & 2022 Aug & 1.0.0, 1.0.1 & SSP-CKC14z~v1.0 \\
\enddata
\tablenotetext{a}{\fastspecfit{} software tag(s) used to produce the VAC.}
\tablenotetext{b}{SPS template library, \texttt{ftemplates-\{name\}.fits} unless otherwise noted (\S\ref{ssec:templates}).}
\tablenotetext{c}{To be released in 2027 as part of DESI/DR2.}
\tablecomments{No \fastphot{} VAC was produced for the \texttt{Fuji} (EDR) specprod. Release notes, known issues, and the full change history for every version are maintained in the \fastspecfit{} online change log at \url{https://github.com/desihub/fastspecfit/blob/main/doc/changes.rst}.}
\end{deluxetable}

%% file: table-vac-compare.tex
\begin{deluxetable*}{llrrrrrr}
\tablecaption{Comparison of \texttt{Loa/v2.0} with Earlier \fastspec{} VACs\label{table:vac-compare}}
\tablewidth{0pt}
\tablehead{
  \colhead{} & \colhead{} & \multicolumn2c{\texttt{Iron/v3.0}} & \multicolumn2c{\texttt{Iron/v2.1}} & \multicolumn2c{\texttt{Fuji/v3.2}} \\
  \cline{3-4} \cline{5-6} \cline{7-8}
  \colhead{Quantity} & \colhead{Class} & \colhead{$\Delta$} & \colhead{$\sigma$} & \colhead{$\Delta$} & \colhead{$\sigma$} & \colhead{$\Delta$} & \colhead{$\sigma$}
}
\startdata
$\Delta\log_{10}\Mstar$ & BGS & $-0.002$ & $0.008$ & $+0.009$ & $0.031$ & $+0.003$ & $0.044$ \\
 & LRG & $0.000$ & $0.009$ & $+0.004$ & $0.047$ & $-0.003$ & $0.059$ \\
 & ELG & $+0.003$ & $0.065$ & $+0.012$ & $0.214$ & $-0.003$ & $0.279$ \\
\hline
$\Delta\log_{10}\mathrm{SFR}$\tablenotemark{a} & BGS & $+1.243$ & $0.841$ & $+0.618$ & $0.593$ & $+0.609$ & $0.594$ \\
 & LRG & $+0.838$ & $1.905$ & $-0.180$ & $1.264$ & $-0.250$ & $1.421$ \\
 & ELG & $+0.578$ & $0.550$ & $+0.262$ & $0.507$ & $+0.252$ & $0.531$ \\
\hline
$\Delta D_n(4000)$ & BGS & $+0.002$ & $0.023$ & $+0.004$ & $0.028$ & $+0.003$ & $0.048$ \\
 & LRG & $0.000$ & $0.028$ & $+0.002$ & $0.036$ & $+0.001$ & $0.060$ \\
 & ELG & $+0.004$ & $0.048$ & $+0.004$ & $0.052$ & $+0.004$ & $0.080$ \\
\hline
$\Delta\,{}^{0.1}\!M_r$ & BGS & $+0.001$ & $0.002$ & $-0.001$ & $0.005$ & $-0.001$ & $0.008$ \\
 & LRG & $0.000$ & $0.001$ & $-0.001$ & $0.005$ & $-0.001$ & $0.007$ \\
 & ELG & $+0.002$ & $0.019$ & $-0.005$ & $0.043$ & $-0.005$ & $0.055$ \\
\hline
$\Delta\log_{10}\vdisp$ & BGS & $+0.007$ & $0.008$ & $+0.001$ & $0.052$ & $+0.001$ & $0.053$ \\
 & LRG & $+0.004$ & $0.007$ & $-0.003$ & $0.052$ & $-0.002$ & $0.054$ \\
\hline
$\Delta\log_{10}F$(\ha) & BGS & $+0.006$ & $0.006$ & $+0.011$ & $0.023$ & $+0.009$ & $0.023$ \\
\hline
$\Delta\log_{10}F$(\oiiilam) & BGS & $+0.001$ & $0.009$ & $+0.004$ & $0.032$ & $+0.004$ & $0.034$ \\
 & ELG & $+0.001$ & $0.005$ & $+0.004$ & $0.021$ & $+0.002$ & $0.023$ \\
\hline
$\Delta\log_{10}F$(\oiidoublet) & ELG & $-0.002$ & $0.004$ & $-0.001$ & $0.021$ & $-0.002$ & $0.023$ \\
\enddata
\tablenotetext{a}{All three earlier VACs predate \fastspecfit{}~3.5.0, which changed the definition of \texttt{SFR} (see Appendix~\ref{sec:appendix_vacs}).}
\tablecomments{Median difference ($\Delta$; \texttt{Loa/v2.0} minus the earlier VAC) and normalized median absolute deviation ($\sigma$) for SV3 targets with a good redshift (\S\ref{ssec:cuts}) in both catalogs and redshifts that agree to within 1000~\kms{}. Line fluxes and \vdisp{} require flux $\mathrm{S/N}>3$ in both catalogs.}
\end{deluxetable*}